\documentclass{article}
\usepackage{jheppub}
\usepackage[utf8x, utf8]{inputenc}	
\usepackage[english]{babel}

\usepackage[ddmmyyyy,hhmmss]{datetime}

\usepackage{multirow}

\catcode`@=11
\allowdisplaybreaks
\usepackage{amsxtra}
\usepackage{mathtools}
\usepackage{float}
\usepackage{xcolor}
\usepackage{tikz}
\usepackage{macros}
\usepackage{hyperref}
\usepackage{amsmath}
\usepackage{graphicx}
\usepackage{rotating}  

\usepackage{placeins}
\usetikzlibrary{calc}
\usetikzlibrary{arrows.meta, bending, patterns}
\usepackage{soul}
\usetikzlibrary{hobby,intersections} 
\usetikzlibrary{shapes.geometric}
\usetikzlibrary{shapes.misc}
\usepackage{bm}

\tikzset{flavour/.style={draw=none,minimum size=0.3mm,fill=white, regular polygon,regular polygon sides=4,draw}}
\tikzset{flavourr/.style={draw=none,minimum size=0.3mm,fill=red, regular polygon,regular polygon sides=4,draw}}
\tikzset{flavourb/.style={draw=none,minimum size=0.3mm,fill=blue, regular polygon,regular polygon sides=4,draw}}
\tikzset{flavourblack/.style={draw=none,minimum size=0.3mm,fill=black, regular polygon,regular polygon sides=4,draw}}
\tikzset{gaugeBig/.style={inner sep=1mm,draw=none,fill=white,minimum size=2mm,circle, draw}}
\tikzset{bd/.style={circle, draw=black, inner sep=0pt, fill=black, minimum size=2mm}}
\tikzset{wd/.style={circle, draw=black, inner sep=0pt, fill=white, minimum size=2mm}}
\tikzset{Dynkin/.style={circle, draw=black, inner sep=0pt, fill=white, minimum size=2mm}}
\tikzstyle{ligne}=[draw, very thick] 
\tikzstyle{gridline}=[draw, gray] 
\tikzset{gauge/.style={circle, draw,inner sep=2.5pt}}
\tikzset{gaugeo/.style={circle, draw,inner sep=2.5pt,fill=orange}}
\tikzset{gauger/.style={circle, draw,inner sep=2.5pt,fill=red}}
\tikzset{gaugeb/.style={circle, draw,inner sep=2.5pt,fill=blue}}
\tikzset{gaugeg/.style={circle, draw,inner sep=2.5pt,fill=teal}}
\tikzset{gaugegoodteal/.style={circle, draw,inner sep=2.5pt,fill=goodteal}}
\tikzset{gaugem/.style={circle, draw,inner sep=2.5pt,fill=magenta}}
\tikzset{hasse/.style={circle, fill,inner sep=2pt}}
\tikzset{d2/.style={circle, fill,inner sep=1.3pt}}
\tikzset{shrinky/.style={circle, fill,inner sep=1pt}}
\tikzset{sized/.style={circle, draw, inner sep=1.5pt}}
\tikzset{seven/.style={circle, draw,inner sep=3pt}}

\tikzset{gaugebl/.style={circle,draw=black,fill=black,inner sep=1.5pt}}
\tikzset{gaugeblnormal/.style={circle,draw=black,fill=black,inner sep=2.5pt}}
\tikzset{hasse/.style={circle, fill,inner sep=2pt}}
\tikzstyle{dashed_brane}=[thick, dashed]
\tikzstyle{dotted_brane}=[thick, dotted]
\tikzstyle{O3plus}=[thick, color=teal]
\tikzstyle{O3minustilde}=[thick, color=blue]
\tikzstyle{O3plustilde}=[thick, color=red]
\tikzset{D5/.style={cross out, draw=black, minimum size=7, inner sep=0pt, outer sep=0pt}, cross/.default={1pt}}
\tikzset{flavor/.style={regular polygon,regular polygon sides=4,inner sep=2.5pt, label = {}, draw}}
\tikzset{redflavor/.style={regular polygon,regular polygon sides=4,inner sep=2.5pt, color=red, label = {}, draw}}
\tikzset{redgauge/.style={inner sep=1mm,color=red,draw=none,minimum size=2mm,circle, draw}}
\tikzset{blueflavor/.style={regular polygon,regular polygon sides=4,inner sep=2.5pt, color=blue, label = {}, draw}}
\tikzset{bluegauge/.style={inner sep=1mm,color=blue,draw=none,minimum size=2mm,circle, draw}}

\makeatletter
\DeclareRobustCommand{\rvdots}{%
  \vbox{
    \baselineskip4\p@\lineskiplimit\z@
    \kern-\p@
    \hbox{.}\hbox{.}\hbox{.}
  }}
\makeatother

\newcommand{\Figref}[1]{Figure~\ref{#1}}
\newcommand{\Quiver}[1]{$\mathcal Q_{\ref{#1}}$}
\newcommand{\surm}{\mathrm{SU}}

\newcommand{\urm}{\mathrm{U}}
\newcommand{\sorm}{\mathrm{SO}}
\newcommand{\slrm}{\mathrm{SL}}
\newcommand{\orm}{\mathrm{O}}
\newcommand{\sprm}{\mathrm{Sp}}
\newcommand{\hs}{\mathrm{HS}}
\newcommand{\hsc}{\mathrm{HS}^{\mathcal{C}}}
\newcommand{\hsh}{\mathrm{HS}^{\mathcal{H}}}

\newcommand{\hsC}[1]{\hs\left[\mathcal C\left(\text{\Quiver{#1}}\right)\right]}

\usepackage{amsmath}
\usepackage{array}
\usepackage{booktabs}
\usepackage{tabularx}
\usepackage{tikz}

\newcolumntype{C}[1]{>{\centering\arraybackslash}m{#1}}
\newcolumntype{Y}{>{\raggedright\arraybackslash}X}

\newcommand{\HSfrac}[2]{%
  \(\displaystyle
  \frac{\begin{aligned}#1\end{aligned}}{#2}\)}

\newcommand{\HSsep}{%
  \specialrule{0.35pt}{0.7em}{0.7em}}

\tikzset{gaugebl/.style={circle,draw=black,fill=black,inner sep=1.5pt}}
\tikzset{gaugeblnormal/.style={circle,draw=black,fill=black,inner sep=2.5pt}}
\tikzset{hasse/.style={circle, fill,inner sep=2pt}}
\tikzstyle{dashed_brane}=[thick, dashed]
\tikzstyle{dotted_brane}=[thick, dotted]
\tikzstyle{O3plus}=[thick, color=teal]
\tikzstyle{O3minustilde}=[thick, color=blue]
\tikzstyle{O3plustilde}=[thick, color=red]
\tikzset{D5/.style={cross out, draw=black, minimum size=9, inner sep=0pt, outer sep=0pt}, cross/.default={1pt}}
\tikzset{O5circle/.style={circle, minimum size=4mm,draw=black}}
\tikzset{flavor/.style={regular polygon,regular polygon sides=4,inner sep=2.5pt, label = {}, draw}}
\tikzset{redflavor/.style={regular polygon,regular polygon sides=4,inner sep=2.5pt, color=red, label = {}, draw}}
\tikzset{redgauge/.style={inner sep=1mm,color=red,draw=none,minimum size=2mm,circle, draw}}
\tikzset{blueflavor/.style={regular polygon,regular polygon sides=4,inner sep=2.5pt, color=blue, label = {}, draw}}
\tikzset{bluegauge/.style={inner sep=1mm,color=blue,draw=none,minimum size=2mm,circle, draw}}
\allowdisplaybreaks

\usepackage{todonotes}

\title{Quotient Quiver Subtraction -- Product Groups}
\author{Sam Bennett,}
\author{Amihay Hanany,}
\author{Guhesh Kumaran}

\affiliation{Abdus Salam Centre for Theoretical Physics, Imperial College London,\\ Prince Consort Road
London, SW7 2AZ, UK}
\emailAdd{samuel.bennett18@imperial.ac.uk}
\emailAdd{amihay.hanany@imperial.ac.uk}
\emailAdd{guhesh.kumaran18@alumni.imperial.ac.uk}

\preprint{Imperial/TP/26/AH/09}
\abstract{Quotient quiver subtraction translates the gauging of Coulomb branch global symmetries into a combinatoric operation on 3d $\mathcal{N}=4$ gauge theories. This work uses brane systems to introduce a systematic method to gauge $\urm(n)\times\urm(n')$ and $\surm(n)\times\surm(n')$ emergent Coulomb branch global symmetries. Similar to other recently-introduced techniques, the gauging operation is a two-step process; in this case, it involves the symplectisation of unitary gauge factors into symplectic groups. A host of explicit examples are used to verify the procedure, and connections are made to Class $\mathcal{S}$ theories and three-dimensional theories realising maximal implosion spaces.}
\begin{document}
\maketitle
\section{Introduction}
One of the more useful qualities of quiver gauge theories is their tendency to sublimate the tools of extended supersymmetry into simple combinatorics. This is particularly true for $3d\;\mathcal{N}=4$ theories, for which a host of powerful combinatorial operations have been discovered in recent years. One such tool, termed `quiver subtraction', has been successfully used to study the structure of both the Coulomb branch and the Higgs branch (sometimes termed `symplectic stratification') \cite{Cabrera:2018ann,Bourget:2019aer,Bennett:2024loi,Bourget:2022ehw,Bourget:2022tmw,Bourget:2024mgn,Bourget:2023dkj,Lawrie:2024wan,Lawrie:2023uiu,Bourget:2021siw,Henderson_2014}, as well as their FI deformations \cite{vanBeest:2021xyt,Bourget:2023uhe,Giacomelli:2024ycb,Giacomelli:2024dbd}. This paper is part of a series exploring a closely-related technique called \emph{quotient} quiver subtraction \cite{Hanany:2023tvn,Bennett:2024llh,Bennett:2025edk,Bennett:2026gtm}, which translates the question of gauging global symmetries acting on the Coulomb branch of a 3d $\mathcal{N}=4$ quiver (containing much non-perturbative physics) into a straightforward game involving little more than integer subtraction. Broadly, the technique exploits S-duality in Type IIB brane systems to identify the fate of flavour gauging under the 3d-mirror map \cite{Intriligator:1996ex, Hanany:1996ie, Aharony:1997bx, Seiberg:1996nz}; it turns out that it becomes a variety of `subtraction' (that is, the subtraction of the gauge node ranks of one quiver from another, possibly with additional steps). Importantly, the algorithm bypasses complications from strongly coupled and non-perturbative physics that arise in the IR; as such, Coulomb branch gauging becomes, for a large class of theories, as simple to perform as flavour gauging, which is well-known to enjoy protections from non-renormalisation theorems \cite{Argyres:1996eh}.

Quotient quiver subtraction is not merely a rephrasing of flavour gauging. For Lagrangian quiver gauge theories with unknown quiver 3d mirror duals, such as the affine Dynkin $E^{(1)}_{6,7,8}$ theories \cite{Dasgupta:1996ij,Minahan_1997}, there is no known dual theory for which one can classically gauge flavour symmetry subgroups. In these cases one must confront the Coulomb branch global symmetry gauging directly.

Several of the theories considered in this work arise as magnetic quivers for theories in four, five and six dimensions \cite{Cabrera:2019izd,Cabrera:2019dob,Bourget:2020asf,Bourget:2020mez,Bourget:2021csg,Lawrie:2023uiu,Mansi:2023faa,Lawrie:2024zon}. A magnetic quiver is an auxiliary  construction of the Higgs branch of a $d=4,5,6$ theory with eight supercharges as a moduli space of dressed monopoles. The magnetic quiver description becomes powerful for theories in dimensions $d=4,5,6$, where certain phases may not admit a Lagrangian description. In four dimensions, examples include Argyres-Douglas points, in five dimensions there are massless instantons at infinite coupling, and in six dimensions SCFTs contain tensionless strings. Previous work demonstrates that flavour symmetry gauging in these higher-dimensional theories may be described by quotient quiver subtraction or related operations on the magnetic quiver \cite{Dancer:2024lra,Hanany:2024fqf,Bennett:2024llh,Bennett:2026gtm,Bourget:2022tmw}.


Quotient quiver subtraction algorithms gauge $\surm(n)$, $\sprm(n)$ and $\sorm(n)$ Coulomb branch global symmetry subgroups of unitary quiver theories \cite{Hanany:2023tvn,Bennett:2026gtm}, $\sorm(n)$ and $\sprm(n)$ subgroups of framed orthosymplectic quivers \cite{Bennett:2025edk}, $\surm(2),\;\surm(3),\;G_2,\;$and $\sorm(7)$ subgroups of unframed orthosymplectic quivers \cite{Bennett:2024llh}. For unitary quivers, gaugings of $\urm(n)\times\urm(n')$ and $\surm(n)\times\surm(n')$ Coulomb branch global symmetries were, to date, unrealised. The purpose of this paper is thus to introduce a prescription for this process, using intuition from brane systems and the pairs of 3d mirror SQFTs they give rise to. 

As with previous entries in this series, the quotient quiver subtraction techniques introduced in this work lead to new IR dualities between a handful of theories, new magnetic quivers for four- and six-dimensional theories, as well as giving explicit conjectures for the identities of various Coulomb branch hyper-K\"ahler quotients. In some sense, quotient quiver subtraction presents the opportunity to build magnetic quivers from the `bottom up', by leveraging control over the Coulomb branch geometries of three-dimensional theories.

The structure of this paper is as follows.

Section \ref{sec:U(n)U(n')QQS} introduces $\urm(n)\times\urm(n')$ quotient quiver subtraction, both as an SQFT algorithm and as a brane construction, and gives various examples of its action in a selection of cases wherein the results can be verified explicitly. This uncovers a set of new results on the Coulomb branches of 3d SCFTs.

Section \ref{sec:SUnSUn'_QQS} extends the algorithm to act on unframed quivers, where it becomes an $\surm(n)\times\surm(n')$ gauging. This expands the set of possibilities to include action on affine exceptional quivers, and maximal implosions of the regular $A$-type nilpotent orbit closure. Several connections are made with four- and six-dimensional theories using the magnetic quiver correspondence, and a handful of new dualities are uncovered, some of which have equivalent derivations as degeneration limits of Class $\mathcal{S}$ theories.

Section \ref{sec:connec_class_s} makes the connection to Class $\mathcal{S}$ theories explicit, showing that the four-dimensional gluing of twisted $A$-type theories is a subset of the quotient quiver subtraction procedure in three dimensions.

Section \ref{sec:outlook} summarises the main findings of this work and expands on future directions.

Appendix \ref{app:O2O2_QQS} makes some progress on the analogous algorithm for orthosymplectic quivers. However, this becomes highly sensitive to gauge group global forms, and a general prescription does not currently exist.

Appendix \ref{sec:NSLGauge} investigates the quotient quiver subtraction construction when the $\mathrm{ON}^-$ plane used in the main body of the paper is replaced in turn by $\mathrm{ON}^+,\widetilde{\mathrm{ON}^+},$ and $\widetilde{\mathrm{ON}^-}$ planes. On the electric quiver, these three orientifolds introduce non-simply laced edges. The analogous quotient quiver subtraction algorithms on the magnetic quiver would then be associated to ``gauging with non-simply laced edges'' which does not have a clear interpretation in gauge theory.
\section{\boldmath$\urm(n)\times\urm(n')\;$\unboldmath Gauging}
\label{sec:U(n)U(n')QQS}
The construction in \cite{Bennett:2026gtm} used inspiration from orientifolds in brane systems to construct quotient quivers for $BCD$-type gaugings. By replacing the $\textrm{O5}$ planes with an $\textrm{ON}$ plane, a similar logic leads to quotient quivers for products $\urm(n)\times\urm(n')$ of two unitary gauge groups. This algorithm starts with a framed quiver as input and returns a framed quiver. The quotient quiver subtraction for products of $\surm(n)\times\surm(n')$ follow naturally by starting with unframed quivers instead, and will be explored in more detail in Section \ref{sec:SUnSUn'_QQS}.

\subsection{$n=n'$}
It is simplest to first consider the case where $n=n'$. Take the $\urm(N)$ SQCD with $2N+k$ flavours in \eqref{quiv:SQCD_3dMirror}. Gauging a $\urm(n)\times\urm(n)$ product subgroup of the $\surm(2N+k)$ flavour symmetry straightforwardly returns the quiver in \Figref{fig:Un_Un_deriv_quiv}, which admits the brane system in \Figref{fig:Un_Un_deriv_brane}. \footnote{The brane system for \eqref{quiv:SQCD_3dMirror} is straightforward and is not given here.}
\begin{figure}[h!]
\centering
    \begin{subfigure}{0.49\textwidth}
    \centering
    \begin{tikzpicture}
    \node[gauge, label=below:$N$] (n)at (0,0){};
    \node[gauge, label=left:$n$] (m1)at (-1,0.5){};
    \node[gauge, label=left:$n$] (m2)at (-1,-0.5){};
    \node[flavour, label=above:$2N+k-2n$] (flav) at (0,1){};
    \draw[-] (n)--(flav) (n)--(m1) (n)--(m2);
    \end{tikzpicture}
\caption{}
\label{fig:Un_Un_deriv_quiv}
\end{subfigure}
\begin{subfigure}{0.49\textwidth}
    \centering
    \begin{tikzpicture}
         \draw[-](0,0)--(0,2) (-2,0)--(-2,2) (-4,0)--(-4,2);
        \draw[dashed] (-6,0)--(-6,2);
        \draw[-](0,1)--(-2,1)node[midway, below]{$N$} (-2,0.5)--(-4,0.5) (-2,1.5)--(-6,1.5)--(-4,1.25);
        
         \node[D5] at (-0.5,1.75){};
        \node[D5] at (-1.5,1.75){};
        \node at (-1,1.75){$\cdots$};
        \draw [decorate, decoration = {brace, raise=10pt, 
         amplitude=5pt}] (-1.5,1.75) --  (-0.5,1.75) node[pos=0.5,above=15pt,black]{$2N+k-2n$};  
        \node at (-3,0.75) {$n$};
        \node at (-3,1.75) {$n$};
        \node at (-6,2.5) {$\text{ON}^{-}$};
    \end{tikzpicture}
\caption{}
\label{fig:Un_Un_deriv_brane}
\end{subfigure}
\begin{subfigure}{0.9\textwidth}
\centering
    \begin{tikzpicture}
    \node[gaugeb, label=below:$C_n$] (1l) at (0,0){};
    \node[gauge, label=below:$2n+1$] (2l) at (1,0){};
    \node[gauge, label=above:$2n+2$] (3l) at (2,0){};
    \node[] (cdotsl) at (3,0){$\cdots$};
    \node[gauge, label=below:$N-1$] (n1l) at (4,0){};
    \node[gauge, label=below:$N$] (n2l) at (5,0){};
    \node[] (cdotsm) at (6,0){$\cdots$};
    \node[gauge, label=below:$N$] (n1r) at (7,0){};
    \node[gauge, label=below:$N-1$] (n2r) at (8,0){};
    \node[] (cdotsr) at (9,0){$\cdots$};
    \node[gauge, label=below:$2$] (2r) at (10,0){};
    \node[gauge, label=below:$1$] (1r) at (11,0){};
    \node[flavour, label=above:$1$] (1fl) at (5,1){};
    \node[flavour, label=above:$1$] (1fr) at (7,1){};
    \node[flavourr, label=below:$D_1$] (Dl) at (-1,0){};
    \draw[-] (Dl)--(1l)--(2l)--(3l)--(cdotsl)--(n1l)--(n2l)--(cdotsm)--(n1r)--(n2r)--(cdotsr)--(2r)--(1r) (1fl)--(n2l) (1fr)--(n1r);
    \draw [decorate, decoration = {brace, raise=15pt, amplitude=5pt}] (7.25,0) --  (4.75,0) node[pos=0.5,below=20pt,black]{$k+1$};
    \end{tikzpicture}
    \caption{}
    \label{quiv:UN_QQS_mirror}
\end{subfigure}
\begin{subfigure}[t]{0.95\textwidth}
\centering
\begin{tikzpicture}
    \draw[-] (0,-1)--(0,1) (-1,-1)--(-1,1) (-2,-1)--(-2,1) (-3,-1)--(-3,1) (-4,-1)--(-4,1) (-5,-1)--(-5,1) (-6,-1)--(-6,1) (-7,-1)--(-7,1) (-8,-1)--(-8,1) (-9,-1)--(-9,1) (-10,-1)--(-10,1) (-11.5,-1)--(-11.5,1) (-13,-1)--(-13,1);
    \node[O5circle, label=above:$\mathrm{O5}^-$]at  (-14.5,0){};
    \draw[-] (0,0)--(-1,0)node[midway, above]{$1$}--(-2,0)node[midway, above]{$2$};
    \draw[-] (-2,0)--(-2.2,0) (-2.8,0)--(-3,0);
    \node at (-2.5,0){$\cdots$};
    \draw[-] (-3,0)--(-4,0)node[midway, above]{$N$};
    \node[D5] at (-3.5,0.75){};
    \node[D5] at (-13.75,0.75){};
    \draw[-] (-4,0)--(-5,0)node[midway, above]{$N$};
    \draw[-] (-5,0)--(-5.2,0) (-5.8,0)--(-6,0);
    \node at (-5.5,0){$\cdots$};
    \draw[-] (-6,0)--(-7,0)node[midway,above]{$N$}--(-8,0)node[midway,above]{$N$}--(-9,0)node[midway, above]{$N-1$};
    \node[D5] at (-7.5,0.75){};
    \draw[-] (-9,0)--(-9.2,0) (-9.8,0)--(-10,0);
    \node at (-9.5,0){$\cdots$};
    \draw[-] (-10,0)--(-11.5,0)node[midway, above]{$2n+2$}--(-13,0)node[midway,above]{$2n+1$}--(-14.3,0)node[midway, above]{$2n$};
    \draw [decorate, decoration = {brace, raise=5pt, amplitude=5pt}] (-8,1) --  (-3,1) node[pos=0.5,above=10pt,black]{$k+2$};
    \end{tikzpicture}
    \caption{}
    \label{fig:UnSOMirrorBrane}
\end{subfigure}
\caption{\Figref{fig:Un_Un_deriv_quiv} results from the $\urm(n)\times\urm(n)$ gauging of the SQCD in \eqref{quiv:SQCD_3dMirror}. It has a brane realisation using D3 branes, D5 branes, NS5 branes and an $\textrm{ON}^{0}$ orientifold plane in \Figref{fig:Un_Un_deriv_brane}, which under S-duality is exchanged for \Figref{fig:UnSOMirrorBrane} and the 3d-mirror theory in \Figref{quiv:UN_QQS_mirror}. Note that the S-duality operation between \Figref{fig:Un_Un_deriv_brane} and \Figref{fig:UnSOMirrorBrane} exchanges the $\textrm{ON}^{0}$ plane, which consists of a $\textrm{ON}^{-}$ plane and an NS5 brane, for an $\textrm{O5}^{-}$ plane and a D5 brane. Note too that the single D5 brane between the $\textrm{O5}^{-}$ plane and the leftmost NS5 brane in \Figref{fig:UnSOMirrorBrane} gives rise to the $\sorm(2)$ flavour node in \Figref{quiv:UN_QQS_mirror}.}
\label{fig:UN_QQS}
\end{figure}

Performing 3d mirror symmetry on \Figref{fig:Un_Un_deriv_quiv} results in the theory in \eqref{quiv:UN_QQS_mirror}, where now the effect of the gauging appears on the leftmost long leg. A simple inspection of the difference between the mirror theories before and after the gauging leads to a conjecture for the appropriate quotient quiver subtraction operation; in this case, the algorithm must include both the subtraction of a string of gauge nodes and the symplectisation of a unitary gauge group factor into a symplectic gauge group factor. Given these considerations, \eqref{quiv:U(n)U(n)_product_quotient_quiver} emerges as the candidate quotient quiver. Morally, the symplectisation procedure follows from the fact that an $O\text{N}^{0}$ plane is S-dual to an $O5^{-}$ plane and a single D5-brane, the latter of which sources the $\sorm(2)$ flavour. Note that this is entirely in accord with the intuition of \cite{Bennett:2026gtm}.
\begin{equation}
\raisebox{-0.5\height}{
\begin{tikzpicture}

\node[gauge, label=below:$N$] (L1) at (0,0){};
\node[flavour, label=above:$2N+k$] (L2) at (0,1){};

\draw[-] (L1)--(L2);

\draw[<->, very thick]
    (1.0,0) -- (3.0,0)
    node[midway, fill=white, inner sep=3pt, align=center]
    {\small\textit{Mirror}};

\node[gauge, label=below:$1$] (R1) at (4,0){};
\node[gauge, label=below:$2$] (R2) at (5,0){};
\node (cdotsl) at (6,0) {$\cdots$};
\node[gauge, label=below:$N$] (R3) at (7,0){};
\node (cdotsm) at (8,0) {$\cdots$};
\node[gauge, label=below:$N$] (R4) at (9,0){};
\node (cdotsr) at (10,0) {$\cdots$};
\node[gauge, label=below:$2$] (R5) at (11,0){};
\node[gauge, label=below:$1$] (R6) at (12,0){};

\node[flavour, label=above:$1$] (R7) at (7,1){};
\node[flavour, label=above:$1$] (R8) at (9,1){};

\draw [decorate, decoration = {brace, amplitude=5pt}] (9.25,-0.6) --  (6.75,-0.6) node[pos=0.5,below=2pt,black]{$k+1$};

\draw[-] (R1)--(R2)--(cdotsl)--(R3)--(cdotsm)--(R4)--(cdotsr)--(R5)--(R6);
\draw[-] (R3)--(R7);
\draw[-] (R4)--(R8);

\end{tikzpicture}
}
\label{quiv:SQCD_3dMirror}
\end{equation}
Inspection of the quotient quiver \eqref{quiv:U(n)U(n)_product_quotient_quiver} shows that the total rank of its gauge group is $2n^2-n$, less than the dimension of $\urm(n)\times\urm(n)$. The further $n$ moduli are removed precisely by the procedure of symplectising the remaining gauge node, matching the $2n^{2}$ expectation. As framed quivers are rebalanced with flavours, the net reduction of the Coulomb branch dimension after subtraction is indicative of complete Higgsing. In this case, Weyl integration is sufficient to compute the hyper-Kähler quotient of the moduli space and so the predictions of the algorithm can be verified against calculations using the nilpotent orbit localisation formula in cases in which the target theory does not have a Lagrangian 3d mirror.

Note that the quotient quiver below is identical to that for $\sorm(2n)$ \cite{Bennett:2026gtm}. Nevertheless, the second stages of the algorithms significantly differ.
\begin{equation}
    \raisebox{-0.5\height}{\begin{tikzpicture}
        \node[] (A) at (-3,-0.1){$\urm(n)\times\urm(n) \; \text{Quotient Quiver}$};
        \node[gauge, label=below:$1$] (1) at (1,0){};
        \node[gauge, label=below:$2$] (2) at (2,0){};
        \node[] (cdots) at (3,0){$\cdots$};
        \node[gauge, label=below:$2n-1$] (nm1) at (4,0){};
        \draw[-] (1)--(2)--(cdots)--(nm1);
        \end{tikzpicture}}
\label{quiv:U(n)U(n)_product_quotient_quiver}
\end{equation}
\paragraph{Formal Statement of the Rule}
The $\urm(n)\times\urm(n)$ quotient quiver subtraction algorithm is given below.
\begin{enumerate}
    \item Take a framed target quiver with a long leg of unitary gauge nodes of length at least $2n$, shown below.
    \begin{equation}
        \raisebox{-0.5\height}{\begin{tikzpicture}
        \node[gauge, label=below:$1$] (1) at (0,0){};
        \node[gauge, label=below:$2$] (2) at (1,0){};
        \node[] (cdots) (cdots) at (2,0){$\cdots$};
        \node[gauge, label=below:$2n-1$] (2r) at (3,0){};
        \node[gauge, label=below:$2n$] (2rp1) at (4,0){};
        \node[gauge] (Q) at (5,0){$Q$};
        \draw[-] (1)--(2)--(cdots)--(2r)--(2rp1);
        \draw[teal,very thick] (2rp1)--(Q);
    \end{tikzpicture}}
    \end{equation}
    \item Align the quotient quiver \eqref{quiv:U(n)U(n)_product_quotient_quiver} along the long leg of the target quiver and perform a rank-wise subtraction. Rebalance the $\urm(2n)$ gauge node with a single  $\urm(1)$ flavour. This is shown below, where the edge coloured in teal represents generic connection via bifundamental hypermultiplets to the rest of the quiver $\mathcal{Q}$.
    \begin{equation}\raisebox{-0.5\height}{\begin{tikzpicture}
    \node[font=\bfseries] at (2.5,1) {Subtraction};
    \node[gauge, label=below:$1$] (1) at (0,0){};
    \node[gauge, label=below:$2$] (2) at (1,0){};
    \node (cdots) at (2,0) {$\cdots$};
    \node[gauge, label=below:$2n-1$] (2r) at (3,0){};
    \node[gauge, label=below:$2n$] (2rp1) at (4,0){};
    \node[gauge] (Q) at (5,0) {$Q$};
    \draw[-] (1)--(2)--(cdots)--(2r)--(2rp1);
    \draw[teal, very thick] (2rp1)--(Q);

    \node[font=\Large] at (-0.8,-1.35) {$-$};
    \node[gauge, label=below:$1$] (1sub) at (0,-1){};
    \node[gauge, label=below:$2$] (2sub) at (1,-1){};
    \node (cdotssub) at (2,-1) {$\cdots$};
    \node[gauge, label=below:$2n-1$] (2rsub) at (3,-1){};
    \draw[-] (1sub)--(2sub)--(cdotssub)--(2rsub);
    
    \draw (-0.15,-1.8) -- (5.15,-1.8);
    
    \node[font=\Large] at (-0.8,-2.5) {$=$};
    \node[flavour, label=below:$1$] (f1) at (0,-2.5){};
    \node[gauge, label=below:$2n$] (2rp1r) at (1,-2.5){};
    \node[gauge] (Qr) at (2,-2.5) {$Q$};
    \draw[-] (f1)--(2rp1r);
    \draw[teal, very thick] (2rp1r)--(Qr);
    \end{tikzpicture}}
    \end{equation}
    \item Symplectise the node whose balance has changed using the projection $\urm(2n)\rightarrow\sprm(n)$. This removes a further $n$ moduli. Note that the $\urm(1)$ flavour node above used for rebalancing is now written in red as $\sorm(2)$.
    \begin{equation}\raisebox{-0.5\height}{\begin{tikzpicture}
    \node[font=\bfseries] at (5.25,1) {Symplectisation};
    \node[flavour, label=below:$1$] (f1) at (0,0){};
    \node[gauge, label=below:$2n$] (2rp1r) at (1,0){};
    \node[gauge] (Qr) at (2,0) {$Q$};
    \draw[-] (f1)--(2rp1r);
    \draw[teal, very thick] (2rp1r)--(Qr);

    \draw[->, thick]
    (2.5,0) -- (8.0,0)
    node[midway, fill=white, inner sep=3pt, align=center]
    {\small\textit{Replace}
     $\urm(2n)\;\text{\emph{with}}\;\sprm(n)$};

    \node[flavourr, label=below:$D_{1}$] (1f) at (8.5,0){};
    \node[gaugeb, label=below:$C_n$] (Cnr) at (9.5,0){};
    \node[gauge] (Qrr) at (10.5,0){$Q$};
    \draw[-] (1f)--(Cnr);
    \draw[teal, very thick] (Cnr)--(Qrr);
\end{tikzpicture}}
\end{equation}
\end{enumerate}
The following comments cover some technical aspects of the algorithm. Firstly, note that the long leg $\urm(1)-\urm(2)-\cdots-\urm(2n)-$ contributes at least $\urm(2n)$ to the Coulomb branch global symmetry, as per the Coulomb branch global symmetry algorithm. The balanced subset of nodes from $\urm(1)$ to $\urm(2n-1)$ inclusive contributes $\surm(2n-1)$; the $\urm(2n)$ gauge node is not necessarily balanced and so provides at least an additional $\urm(1)$ factor. Hence for any $n$ the long leg provides a symmetry factor large enough to perform the gauging. The embedding of $\urm(2n)\supset\urm(n)\times\urm(n)$ decomposes the adjoint as, \begin{equation}
    (\underline{4n^2})\rightarrow (\underline{n^2},\underline{1})+(\underline{1},\underline{n^2})+(\underline{n},\underline{\bar n})+(\underline{\bar n},\underline{n}),
\end{equation}

\subsection{Generic Case -- $n\neq n'$}
\label{sec:Un_Un'_generic_case}
Now consider $n\neq n'$. The following takes $n'\leq n$ without loss of generality. Consider again the $\urm(N)$ SQCD with $2N+k$ flavours in \eqref{quiv:SQCD_3dMirror}. Gauging $\urm(n)\times\urm(n')\subset\surm(2N+k)$ inside SQCD flavour symmetry in \eqref{quiv:SQCD_3dMirror} returns \Figref{fig:Un_Un'_deriv_quiv}, where $2n$, $2n'<N$. Its brane realisation is given in \Figref{fig:Un_Un'_deriv_brane}.
In the dual theory, given in \Figref{fig:UN_n_n'_QQS_mirror}, the gauging appears as the subtraction of the quotient quiver \eqref{quiv:U(n)U(n')_quotient_quiver} along with a symplectisation of the unitary node that undergoes a change in rank.
\begin{figure}[h!]
    \centering
    \begin{subfigure}{0.49\textwidth}
    \centering
    \begin{tikzpicture}
    \node[gauge, label=below:$N$] (n)at (0,0){};
    \node[gauge, label=left:$n$] (m1)at (-1,0.5){};
    \node[gauge, label=left:$n'$] (m2)at (-1,-0.5){};
    \node[flavour, label=above:$2N+k-n-n'$] (flav) at (0,1){};
    \draw[-] (n)--(flav) (n)--(m1) (n)--(m2);
    \end{tikzpicture}
\caption{}
\label{fig:Un_Un'_deriv_quiv}
\end{subfigure}
\begin{subfigure}{0.49\textwidth}
    \centering
    \begin{tikzpicture}
        \draw[-](0,0)--(0,2) (-2,0)--(-2,2) (-4,0)--(-4,2);
        \draw[dashed] (-6,0)--(-6,2);
        \draw[-](0,1)--(-2,1)node[midway, below]{$N$} (-2,0.5)--(-4,0.5) (-2,1.5)--(-6,1.5)--(-4,1.25);
        
         \node[D5] at (-0.5,1.75){};
        \node[D5] at (-1.5,1.75){};
        \node at (-1,1.75){$\cdots$};
        \draw [decorate, decoration = {brace, raise=10pt, 
         amplitude=5pt}] (-1.5,1.75) --  (-0.5,1.75) node[pos=0.5,above=15pt,black]{$2N+k-n-n'$};  
        \node at (-3,0.75) {$n'$};
        \node at (-3,1.75) {$n$};
        \node at (-6,2.5) {$\text{ON}^{-}$};
    \end{tikzpicture}
\caption{}
\label{fig:Un_Un'_deriv_brane}
\end{subfigure}
\begin{subfigure}{0.9\textwidth}
    \begin{tikzpicture}
    \node[gaugeb, label=below:$C_{n'}$] (0) at (-1,0){};
    \node[gauge, label=above:$2n'+2$] (1) at (0,0){};
    \node[gauge, label=below:$2n'+4$] (2) at (1,0){};
    \node[] (cdotsa) at (2,0){$\cdots$};
    \node[gauge, label=below:$2n-4$] (3) at (3,0){};
    \node[gauge, label=above:$2n-2$] (4) at (4,0){};
    \node[gauge, label=below:$2n$] (5) at (5,0){};
    \node[gauge, label=below:$2n+1$] (6) at (6,0){};
    \node[] (cdotsb) at (7,0){$\cdots$};
    \node[gauge, label=below:$N$] (7) at (8,0){};
    \node[] (cdotsc) at (9,0){$\cdots$};
    \node[gauge, label=below:$N$] (8) at (10,0){};
    \node[] (cdotsd) at (11,0){$\cdots$};
    \node[gauge, label=below:$1$] (9) at (12,0){};
    \node[flavour, label=above:$1$] (10) at (5,1){};
    \node[flavour, label=above:$1$] (11) at (8,1){};
    \node[flavour, label=above:$1$] (12) at (10,1){};
    \draw[-] (0)--(1)--(2)--(cdotsa)--(3)--(4)--(5)--(6)--(cdotsb)--(7)--(cdotsc)--(8)--(cdotsd)--(9) (5)--(10) (7)--(11) (8)--(12);
    \draw [decorate, decoration = {brace, raise=15pt, amplitude=5pt}] (10.25,0) --  (7.75,0) node[pos=0.5,below=20pt,black]{$k+1$};
    \end{tikzpicture}
\caption{}
\label{fig:UN_n_n'_QQS_mirror}
\end{subfigure}
\begin{subfigure}[t]{0.95\textwidth}
\centering
\begin{tikzpicture}
    \draw[-] (0,-1)--(0,1) (-1,-1)--(-1,1) (-2,-1)--(-2,1) (-3,-1)--(-3,1) (-4,-1)--(-4,1) (-5,-1)--(-5,1) (-6,-1)--(-6,1) (-7,-1)--(-7,1) (-8,-1)--(-8,1) (-9,-1)--(-9,1) (-10,-1)--(-10,1) (-11.5,-1)--(-11.5,1) (-13,-1)--(-13,1);
    \node[O5circle, label=above:$\mathrm{O5}^-$] at (-14.5,0){};
    \draw[-] (0,0)--(-1,0)node[midway, above]{$1$}--(-2,0)node[midway, above]{$2$};
    \draw[-] (-2,0)--(-2.2,0) (-2.8,0)--(-3,0);
    \node at (-2.5,0){$\cdots$};
    \draw[-] (-3,0)--(-4,0)node[midway, above]{$N$};
    \node[D5] at (-3.5,0.75){};
    \draw[-] (-4,0)--(-5,0)node[midway, above]{$N$};
    \draw[-] (-5,0)--(-5.2,0) (-5.8,0)--(-6,0);
    \node at (-5.5,0){$\cdots$};
    \draw[-] (-6,0)--(-7,0)node[midway,above]{$N$}--(-8,0)node[midway,above]{$N$}--(-9,0)node[midway, above]{$N+1$};
    \node[D5] at (-7.5,0.75){};
    \draw[-] (-9,0)--(-9.2,0) (-9.8,0)--(-10,0);
    \node at (-9.5,0){$\cdots$};
    \draw[-] (-10,0)--(-11.5,0)node[midway, above]{$2n'+4$}--(-13,0)node[midway,above]{$2n'+2$}--(-14.3,0)node[midway, above]{$2n'$};
    \draw [decorate, decoration = {brace, raise=5pt, amplitude=5pt}] (-8,1) --  (-3,1) node[pos=0.5,above=10pt,black]{$k+1$};
    \end{tikzpicture}
    \caption{}
    \label{fig:UnUn'MirrorBrane}
\end{subfigure}
\caption{Gauging a $\urm(n)\times\urm(n')$ subgroup of the flavour symmetry in \eqref{quiv:SQCD_3dMirror} returns the theory in \Figref{fig:Un_Un'_deriv_quiv}, the low-energy theory of the brane system in \Figref{fig:Un_Un'_deriv_brane}. The quiver in \Figref{fig:UN_n_n'_QQS_mirror} is the 3d mirror of \Figref{fig:Un_Un'_deriv_quiv}, with brane system given in \Figref{fig:UnUn'MirrorBrane}.}
\label{}
\end{figure}
In this case, the $\sprm(n')$ gauge node is balanced without the need for a $D_1$ flavour. The $\urm(1)$ flavour for rebalancing instead appears on the $\urm(2n)$ gauge node. The sum of the ranks of the quotient quiver \eqref{quiv:U(n)U(n')_quotient_quiver} is $n^{2}+(n')^{2}-n'$. As before, a further $n'$ moduli are removed in the symplectisation process, ensuring that the Coulomb branch dimension decreases by $n^{2}+{n'}^{2}$. Note that for $n=n'$ the quotient quiver in \eqref{quiv:U(n)U(n')_quotient_quiver} reduces to that of \eqref{quiv:U(n)U(n)_product_quotient_quiver}.
\begin{equation}
    \raisebox{-0.5\height}{\begin{tikzpicture}
    \node[] (A) at (-3,-0.1){$\urm(n)\times\urm(n') \; \text{Quotient Quiver}$};
    \node[gauge, label=below:\small$1$] (1) at (7,0){};
    \node[gauge, label=below:\small$2$] (2) at (6,0){};
    \node[] (cdots) at (5,0){$\cdots$};
    \node[gauge, label=above:\small$n-n'$] (nm2) at (4,0){};
    \node[gauge, label=below:\small$n+n'-1$] (nm3) at (3,0){};
    \node[] (cdots2) at (2,0){$\cdots$};
    \node[gauge, label=below:\small$2$] (nm5) at (1,0){};
    \node[gauge, label=below:\small$1$] (nm6) at (0,0){};
    \draw[-] (1)--(2)--(cdots)--(nm2)--(nm3)--(cdots2)--(nm5)--(nm6);
    \end{tikzpicture}}
\label{quiv:U(n)U(n')_quotient_quiver}
\end{equation}
\paragraph{Formal Statement of the Rule}
The quotient quiver subtraction process for a $\urm(n)\times\urm(n')$ subgroup is given below.
\begin{enumerate}
    \item Take a framed target quiver \eqref{quiv:UnUn'_step1} with a long leg of unitary gauge nodes up to at least $\urm(2n)$.
    \begin{equation}
    \raisebox{-0.5\height}{\begin{tikzpicture}
    \node[gauge, label=below:$1$] (1) at (0,0){};
    \node[gauge, label=below:$2$] (2) at (1,0){};
    \node[] (cdotsa) at (2,0){$\cdots$};
    \node[gauge, label=below:$n+n'-1$] (3) at (3,0){};
    \node[gauge, label=above:$n+n'$] (4) at (4,0){};
    \node[gauge, label=below:$n+n'+1$] (5) at (5,0){};
    \node[] (cdotsb) at (6,0){$\cdots$};
    \node[gauge, label=below:$2n-1$] (6) at (7,0){};
    \node[gauge, label=below:$2n$] (7) at (8,0){};
    \node[gauge] (8) at (9,0){$Q$};
    \draw[-] (1)--(2)--(cdotsa)--(3)--(4)--(5)--(cdotsb)--(6)--(7);
    \draw[teal,very thick] (7)--(8);
    \end{tikzpicture}}   
    \label{quiv:UnUn'_step1}
    \end{equation}
    \item Subtract the ranks of the nodes of the quotient quiver from those of the target quiver.
    \begin{equation}\raisebox{-0.5\height}{\begin{tikzpicture}
    \node[font=\bfseries] at (4.1,1) {Subtraction};
    \node[gauge, label=below:$1$] (t1) at (0,0){};
    \node[gauge, label=below:$2$] (t2) at (1,0){};
    \node (tdotsa) at (2,0) {$\cdots$};
    \node[gauge, label=below:$n+n'-1$] (t3) at (3,0){};
    \node[gauge, label=above:$n+n'$] (t4) at (4,0){};
    \node[gauge, label=below:$n+n'+1$] (t5) at (5,0){};
    \node (tdotsb) at (6,0) {$\cdots$};
    \node[gauge, label=below:$2n-1$] (t6) at (7,0){};
    \node[gauge, label=below:$2n$] (t7) at (8,0){};
    \node[gauge] (tQ) at (9,0) {$Q$};
    \draw[-] (t1)--(t2)--(tdotsa)--(t3)--(t4)--(t5)--(tdotsb)--(t6)--(t7);
    \draw[teal, very thick] (t7)--(tQ);
    
    \node[font=\Large] at (-0.8,-1.6) {$-$};
    \node[gauge, label=below:\small$1$] (s1) at (0,-1.6){};
    \node[gauge, label=below:\small$2$] (s2) at (1,-1.6){};
    \node (sdotsa) at (2,-1.6) {$\cdots$};
    \node[gauge, label=above:\small$n+n'-1$] (s3) at (3,-1.6){};
    \node[gauge, label=below:\small$n-n'$] (s4) at (4,-1.6){};
    \node (sdotsb) at (5,-1.6) {$\cdots$};
    \node[gauge, label=below:\small$2$] (s5) at (6,-1.6){};
    \node[gauge, label=below:\small$1$] (s6) at (7,-1.6){};
    \draw[-] (s1)--(s2)--(sdotsa)--(s3)--(s4)--(sdotsb)--(s5)--(s6);
    
    \draw (-0.15,-2.45) -- (8.15,-2.45);

    \node[font=\Large] at (-0.8,-4.2) {$=$};
    \node[gauge, label=below:$2n'$] (r1) at (0,-4.2){};
    \node[gauge, label=below:$2n'+2$] (r2) at (1,-4.2){};
    \node (rdots) at (2,-4.2) {$\cdots$};
    \node[gauge, label=below:$2n-2$] (r3) at (3,-4.2){};
    \node[gauge, label=below:$2n$] (r4) at (4,-4.2){};
    \node[flavour, label=left:$1$] (f) at (4,-3.2){};
    \node[gauge] (rQ) at (5,-4.2) {$Q$};
    \draw[-] (r1)--(r2)--(rdots)--(r3)--(r4)--(f);
    \draw[teal, very thick] (r4)--(rQ);

\end{tikzpicture}}
\end{equation}
    \item `Symplectise' the node that has undergone a change in balance using the projection $\urm(2p)\rightarrow\sprm(p)$. Rebalance the node of rank $2n$ with a $\urm(1)$ flavour.
    \begin{equation}\raisebox{-0.5\height}{\begin{tikzpicture}
    \node[font=\bfseries] at (2.5,2) {Symplectisation};
    \node[gauge, label=below:$2n'$] (L1) at (0,0){};
    \node[gauge, label=below:$2n'+2$] (L2) at (1,0){};
    \node (Ldots) at (2,0) {$\cdots$};
    \node[gauge, label=below:$2n-2$] (L3) at (3,0){};
    \node[gauge, label=below:$2n$] (L4) at (4,0){};
    \node[gauge] (LQ) at (5,0) {$Q$};
    \node[flavour, label=left:$1$] (Rf1) at (4,1){};
    \draw[-] (L1)--(L2)--(Ldots)--(L3)--(L4)--(Rf1);
    \draw[teal, very thick] (L4)--(LQ);

    \draw[->, thick]
    (2.5,-0.9) -- (2.5,-3.1)
    node[midway, fill=white, inner sep=3pt, align=center]
    {\small\textit{Replace}\\[-2pt]
     $\urm(2n')\;\text{\emph{with}}\;\sprm(n')$};

    \node[gaugeb, label=below:$C_{n'}$] (R1) at (0,-4){};
    \node[gauge, label=below:$2n'+2$] (R2) at (1,-4){};
    \node (Rdots) at (2,-4) {$\cdots$};
    \node[gauge, label=below:$2n-2$] (R3) at (3,-4){};
    \node[gauge, label=below:$2n$] (R4) at (4,-4){};
    \node[flavour, label=left:$1$] (Rf) at (4,-3){};
    \node[gauge] (RQ) at (5,-4) {$Q$};
    \draw[-] (R1)--(R2)--(Rdots)--(R3)--(R4)--(Rf);
    \draw[teal, very thick] (R4)--(RQ);
\end{tikzpicture}}
\end{equation}
\end{enumerate}

The long leg of gauge nodes from $\urm(1)$ up to $\urm(2n)$ contributes at least a $\urm(2n)$ factor to the global symmetry. The $\urm(n)\times\urm(n')$ is embedded into $\urm(2n)$ with a continuous commutant of $\urm(n-n')$, which is sourced by the gauge nodes from $C_{n'}$ up to $\urm(2n)$ after the quotient quiver subtraction. The embedding of $\urm(2n)\supset\urm(n)\times\urm(n')\times\urm(n-n')$ decomposes the adjoint as three adjoints and all combinations of bifundamentals.
\begin{align}
    (\underline{4n^2})&\rightarrow (\underline{n^2},\underline{1},\underline{1})+(\underline{1},\underline{n'^2},\underline{1})+(\underline{1},\underline{1},\underline{(n-n')^2})+(\underline{n},\underline{\overline{n'}},\underline{1})+(\underline{\overline n},\underline{n'},\underline{1})+(\underline{n},\underline{1},\underline{\overline{n-n'}})\\&+(\underline{\overline n},\underline{1},\underline{n-n'})\nonumber+(\underline{1},\underline{n'},\underline{\overline{n-n'}})+(\underline{1},\underline{\overline{n'}},\underline{n-n'})
\end{align}






        

\subsection{Examples}
The algorithms that have been proposed are now demonstrated through a collection of examples largely drawn from nilpotent orbit closures of type-A.
\subsubsection{$\overline{reg.A_{2}}///\urm(1)^{2}$}
\begin{figure}[h!]
\centering

\begin{subfigure}[t]{0.47\textwidth}
\centering
\begin{tikzpicture}

\node[font=\bfseries] at (1,0.9) {Step 1: Subtraction};

\node[gauge, label=below:$1$] (1) at (0,0){};
\node[gauge, label=below:$2$] (2) at (1,0){};
\node[flavour, label=below:$3$] (3f) at (2,0){};

\draw[-] (1)--(2)--(3f);

\node[font=\Large] at (-0.8,-1.35) {$-$};

\node[gauge, label=below:$1$] (1sub) at (0,-1.35){};

\draw (-0.15,-2.15) -- (2.15,-2.15);

\node[font=\Large] at (-0.8,-2.75) {$=$};

\node[gauge, label=below:$2$] (2res) at (0,-2.75){};
\node[flavour, label=below:$4$] (4fres) at (1,-2.75){};

\draw[-] (2res)--(4fres);

\end{tikzpicture}
\caption{}
\label{fig:A2U1U1Sub}
\end{subfigure}
\hfill
\begin{subfigure}[t]{0.47\textwidth}
\centering
\begin{tikzpicture}

\node[font=\bfseries] at (0.5,0.9) {Step 2: Symplectisation};

\node[gauge, label=below:$2$] (2res) at (0,0){};
\node[flavour, label=below:$4$] (4fres) at (1,0){};

\draw[-] (2res)--(4fres);

\draw[->, thick]
    (0.5,-.8) -- (0.5,-2.25)
    node[midway, fill=white, inner sep=2pt, align=center]
    {\small\textit{Replace}\\[-2pt]
     $\urm(2)\;\text{\emph{with}}\;\sprm(1)$};

\node[gaugeb, label=below:$C_1$] (C1res) at (0,-2.75){};
\node[flavourr, label=below:$D_4$] (D4res) at (1,-2.75){};

\draw[-] (C1res)--(D4res);
\end{tikzpicture}
\caption{}
\label{fig:A2U1U1Symp}
\end{subfigure}
\caption{$\urm(1)\times\urm(1)$ quotient quiver subtraction on the $T[\surm(3)]$ which occurs in two steps. The first step involves the subtraction of the $\urm(1)\times\urm(1)$ quotient quiver (\eqref{quiv:U(n)U(n)_product_quotient_quiver} with $n=1$) and the rebalancing of the $\urm(2)$ gauge node, which promotes the $\surm(3)$ flavour symmetry into $\surm(4)$. The second step is the symplectisation of the $\urm(2)$ node into an $\sprm(1)$ gauge node, while also changing the flavour symmetry group, from $\surm(4)$ acting on four full hypermultiplets to $\sorm(8)$ acting on eight half-hypermultiplets.}
\label{fig:A2U1U1QQS}
\end{figure}
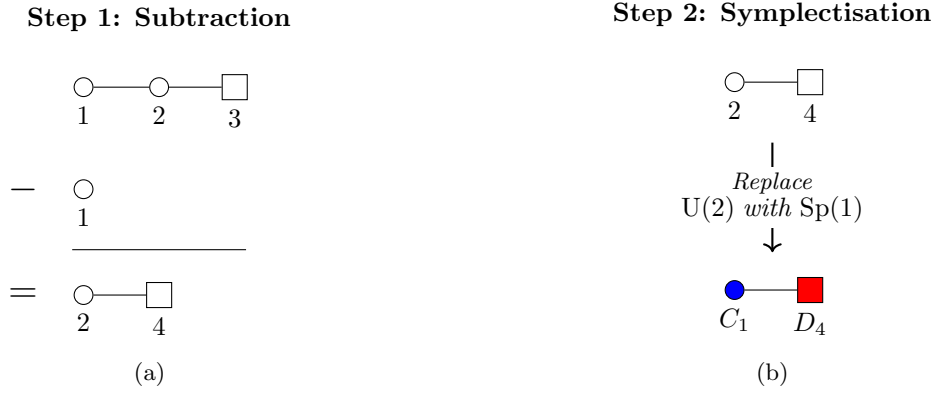

One of the simplest examples of the $\urm(n)\times\urm(n)$ quotient quiver subtraction algorithm concerns the $T[\surm(3)]$ theory, which is self-dual under 3d mirror symmetry. Gauging a $\urm(1)^{2}$ flavour symmetry subgroup returns the affine $D_4$ theory, dual to $\sprm(1)$ with four flavours. 
The same result can be constructed using the prescription above on the 3d mirror theory, which is also $T[\surm(3)]$. Performing a $\urm(1)^{2}$ quotient quiver subtraction on $T[\surm(3)]$ involves subtracting the $\urm(1)\times\urm(1)$ quotient quiver and rebalancing with a flavour, shown in \Figref{fig:A2U1U1Sub}, and then symplectising the remaining $\urm(2)$ into $\sprm(1)$, shown in \Figref{fig:A2U1U1Symp}. Note that the flavour symmetry group changes from $\urm(4)$ to $\sorm(8)$.
At the level of moduli spaces, the flavour symmetry gauging of $T[\surm(3)]$ acts on its Higgs branch via the hyper-K\"ahler quotient below \cite{Kobak1996CLASSICALNO}.\begin{equation}
    \overline{reg. A_2}///\left(\urm(1)\times\urm(1)\right)= D_4
\end{equation}
The embedding decomposes the Coulomb branch symmetry as,
\begin{equation}
    \nu \rightarrow wq+\frac{w}{q}+\frac{1}{w^{2}},
\end{equation}
where $\nu$ is an $\surm(3)$ fugacity, and $w$ and $q$ are $\urm(1)$ fugacities.
It is simple to check that this is the same as the action on the Coulomb branch of $T[\surm(3)]$ by the quotient quiver and subsequent symplectisation. Note that in this example the rebalancing process increases the rank of the flavour group by one. \footnote{In the case that the gauge node undergoing rebalancing already sees non-zero flavours, the rank of the flavour group increases by one. This is best thought of in terms of the brane system -- the number of half D5-branes in the NS5-brane interval giving rise to the gauge group increases by two.}
\subsubsection{$\overline{reg.A_{4}}///\urm(2)^{2}$}
\begin{figure}[h!]
\centering

\begin{subfigure}[t]{0.47\textwidth}
\centering
\begin{tikzpicture}

\node[font=\bfseries] at (1.5,0.9) {Step 1: Subtraction};

\node[gauge, label=below:$1$] (1) at (0,0){};
\node[gauge, label=below:$2$] (2) at (1,0){};
\node[gauge, label=below:$3$] (3) at (2,0){};
\node[gauge, label=below:$4$] (4) at (3,0){};
\node[flavour, label=below:$5$] (5f) at (4,0){};

\draw[-] (1)--(2)--(3)--(4)--(5f);

\node[font=\Large] at (-0.8,-1.35) {$-$};

\node[gauge, label=below:$1$] (1sub) at (0,-1.35){};
\node[gauge, label=below:$2$] (2sub) at (1,-1.35){};
\node[gauge, label=below:$3$] (3sub) at (2,-1.35){};

\draw[-] (1sub)--(2sub)--(3sub);

\draw (-0.15,-2.15) -- (3.15,-2.15);

\node[font=\Large] at (-0.8,-2.75) {$=$};

\node[gauge, label=below:$4$] (4res) at (0,-2.75){};
\node[flavour, label=below:$6$] (6fres) at (1,-2.75){};

\draw[-] (4res)--(6fres);

\end{tikzpicture}
\caption{}
\label{fig:A4U2U2Sub}
\end{subfigure}
\hfill
\begin{subfigure}[t]{0.47\textwidth}
\centering
\begin{tikzpicture}

\node[font=\bfseries] at (1,0.9) {Step 2: Symplectisation};

\node[gauge, label=below:$4$] (4res) at (0,0){};
\node[flavour, label=below:$6$] (6fres) at (1,0){};

\draw[-] (4res)--(6fres);

\draw[->, thick]
    (0.5,-.8) -- (0.5,-2.25)
    node[midway, fill=white, inner sep=2pt, align=center]
    {\small\textit{Replace}\\[-2pt]
     $\urm(4)\;\text{\emph{with}}\;\sprm(2)$};

\node[gaugeb, label=below:$C_2$] (C2res) at (0,-2.75){};
\node[flavourr, label=below:$D_6$] (D6res) at (1,-2.75){};

\draw[-] (C2res)--(D6res);

\end{tikzpicture}
\caption{}
\label{fig:A4U2U2Symp}
\end{subfigure}

\caption{$\urm(2)\times\urm(2)$ quotient quiver subtraction on the $T[\surm(5)]$ which occurs in two steps. The first step involves the subtraction of the $\urm(2)\times\urm(2)$ quotient quiver and the rebalancing of the $\urm(4)$ gauge node, which promotes the flavour symmetry from $\surm(5)$ to $\surm(6)$. The second step is the symplectisation of the $\urm(4)$ node into an $\sprm(2)$ gauge node.}
\label{fig:A4U2U2QQS}
\end{figure}
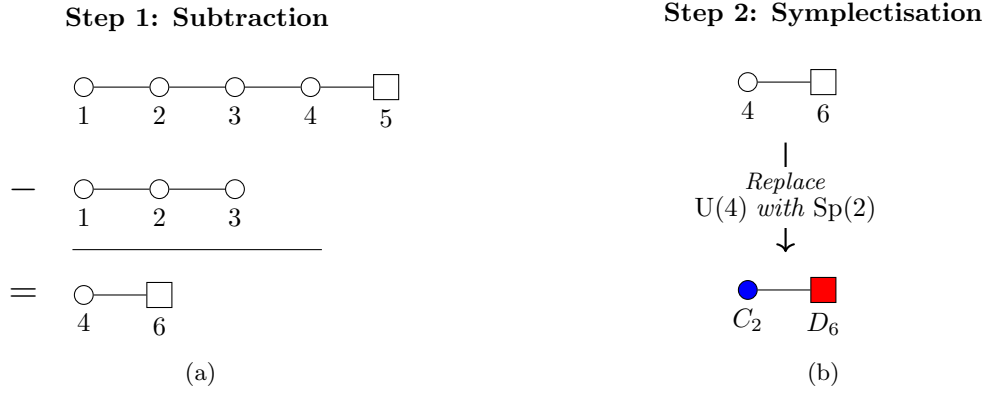

The simplest non-Abelian example concerns the $\urm(2)^{2}$ gauging on the Coulomb branch of $T[\surm(5)]$, shown in \Figref{fig:A4U2U2QQS}.
The unrefined Hilbert series of the resulting $\sprm(2)$ SQCD in \Figref{fig:A4U2U2QQS}, given in \eqref{eq:A4U2U2QQSHS}, matches that obtained using Weyl integration under the embedding $A_1\times A_1\times\urm(1)\times\urm(1)$ under which the $[1,0,0,0]$ decomposes as
\begin{equation}
    \mu_{1}\rightarrow\nu_{1}q_1q_2+\frac{\sigma_{1}q_2}{q_1}+\frac{1}{q_{2}^{4}}
\label{eq:A4_A1A1u1u1_embedding}
\end{equation}
Where $\mu$, $\nu$ and $\sigma$ are fugacities for $A_4$, $A_1$ and $A_1$ respectively and $q_{1,2}$ label the $\mathfrak{u}_1$ charges. The integration is performed over all fugacities.
\begin{equation}
    \hsC{fig:A4U2U2QQS}=\frac{(1-t^{16})(1-t^{20})}{(1-t^4)^2(1-t^6)(1-t^8)^2(1-t^{10})}\label{eq:A4U2U2QQSHS}
\end{equation}
The moduli space is identified as the slice, $\mathcal S^{D_6}_{\mathcal N,(7,5)}$, \cite{Cabrera:2018ldc} to the nilcone of $D_6$ from the orbit labelled by partition $(7,5)$.

This straightforwardly extends to the general identification given below
\begin{equation}
\overline{reg.A_{2n}}///\left(\urm(n)\times\urm(n)\right)=\mathcal{C}\left[
\raisebox{-0.5\height}{\begin{tikzpicture}
    \node[gaugeb, label=left:$C_{n}$] (a) at (0,0){};
    \node[flavourr, label=right:$D_{2n+2}$] (b) at (1,0){};
    \draw[-] (a)--(b);
\end{tikzpicture}}
\right]=\mathcal S^{D_{2n+2}}_{\mathcal N,(2n+3,2n+1)}
\label{eq:1}
\end{equation}
The general embedding is,
\begin{equation}
    \mu_{1}\rightarrow\nu_1 q_1 q_2+\sigma_{1}\frac{q_1}{q_2}+\frac{1}{q_2^{2n}},
\end{equation}
where $\mu$ is an $\surm(2n+1)$ fugacity, $\nu$ and $\sigma$ are $\surm(n)$ fugacities and $q_1$, $q_2$ are $\urm(1)$ fugacities. In general the Hilbert series is,
\begin{equation}
    \hs\left[\mathcal S^{D_{2n+2}}_{\mathcal N,(2n+3,2n+1)}\right] = \prod_{i=1}^{n}\frac{1-t^{4n+4+4i}}{(1-t^{4i})^2(1-t^{4i+2})}.
\end{equation}
\subsubsection{$\overline{reg.A_{4}}$///$\urm(2)\times\urm(1)$} An instructive example consists of gauging a $\urm(2)\times\urm(1)$ subgroup of the $T[\surm(5)]$ theory.
In the magnetic theory, the gauging involves the two-step process shown in \Figref{fig:regA4_U2U1_QQS}. After subtracting the quotient quiver, the $\urm(2)$ gauge node is symplectised into an $\sprm(1)$. Notice that the quotient quiver in \Figref{fig:regA4_U2U1_QQS} is identical to that used to gauge a $\urm(2)$ global symmetry under the algorithm developed in \cite{Hanany:2023tvn}. The symplectisation process augments the gauging to $\urm(2)\times\urm(1)$. This behaviour is generic for $\urm(n)\times\urm(1)$ since $\surm(n)$ quotient quiver subtraction always leaves a $\urm(2)$ gauge node and gauging a $\urm(1)$ inside this $\urm(2)$ turns it into $\sprm(1)$.
\begin{figure}[h!]
\centering

\begin{subfigure}[t]{0.47\textwidth}
\centering
\begin{tikzpicture}

\node[font=\bfseries] at (1.5,0.9) {Step 1: Subtraction};

\node[gauge, label=below:$1$] (1) at (0,0){};
\node[gauge, label=below:$2$] (2) at (1,0){};
\node[gauge, label=below:$3$] (3) at (2,0){};
\node[gauge, label=below:$4$] (4) at (3,0){};
\node[flavour, label=below:$5$] (5f) at (4,0){};

\draw[-] (1)--(2)--(3)--(4)--(5f);

\node[font=\Large] at (-0.8,-1.35) {$-$};

\node[gauge, label=below:$1$] (1sub) at (0,-1.35){};
\node[gauge, label=below:$2$] (2sub) at (1,-1.35){};
\node[gauge, label=below:$1$] (1subr) at (2,-1.35){};

\draw[-] (1sub)--(2sub)--(1subr);

\draw (-0.15,-2.15) -- (3.15,-2.15);

\node[font=\Large] at (-0.8,-2.75) {$=$};

\node[gauge, label=below:$2$] (2res) at (0,-2.75){};
\node[gauge, label=below:$4$] (4res) at (1,-2.75){};
\node[flavour, label=below:$6$] (6fres) at (2,-2.75){};

\draw[-] (2res)--(4res)--(6fres);


\end{tikzpicture}
\caption{}
\label{fig}
\end{subfigure}
\hfill
\begin{subfigure}[t]{0.47\textwidth}
\centering
\begin{tikzpicture}

\node[font=\bfseries] at (1,0.9) {Step 2: Symplectisation};

\node[gauge, label=below:$2$] (2res) at (0,0){};
\node[gauge, label=below:$4$] (4res) at (1,0){};
\node[flavour, label=below:$6$] (6fres) at (2,0){};

\draw[-] (2res)--(4res)--(6fres);



\draw[->, thick]
    (1,-.8) -- (1,-2.25)
    node[midway, fill=white, inner sep=2pt, align=center]
    {\small\textit{Replace}\\[-2pt]
     $\urm(2)\;\text{\emph{with}}\;\sprm(1)$};

\node[gaugeb, label=below:$C_1$] (C1res) at (0,-2.75){};
\node[gauge, label=below:$4$] (4resb) at (1,-2.75){};
\node[flavour, label=below:$6$] (6fresb) at (2,-2.75){};

\draw[-] (C1res)--(4resb)--(6fresb);


\end{tikzpicture}
\caption{}
\label{fig}
\end{subfigure}
\caption{$\urm(2)\times\urm(1)$ quotient quiver subtraction on
$T[\mathrm{SL}(5)]$, which occurs in two steps. The first step involves the subtraction of the $\urm(2)\times\urm(1)$ quotient quiver and the rebalancing of the $\urm(4)$ gauge node, which promotes its flavour symmetry from $\surm(5)$ to $\surm(6)$. The second step is the symplectisation of the $\urm(2)$ node into an $\sprm(1)$ gauge node.}
\label{fig:regA4_U2U1_QQS}
\end{figure}
At the level of varieties, this procedure constructs the hyper-K\"ahler quotient of $\overline{reg.A_4}$ by $\urm(2)\times\urm(1)$. The unrefined Coulomb branch Hilbert series of \Quiver{fig:regA4_U2U1_QQS} is given in \eqref{hs:A4U2U1QQS}, which matches the result of Weyl integration under the embedding of \eqref{eq:A4_A1A1u1u1_embedding}. Now, the integration is over one $A_1$ fugacity and both $\mathfrak{u}_1$ fugacities.
\begin{equation}
    \hsC{fig:regA4_U2U1_QQS}=\frac{1-2t^2+6t^4-4t^6+8t^8-3t^{10}-8 t^{14}+ 12 t^{16} - \cdots - t^{34}}{(1-t^2)^5(1-t^4)^3(1-t^6)(1-t^8)^2}
\label{hs:A4U2U1QQS}
\end{equation}
It is straightforward to extend this to a generic $\urm(n)\times\urm(n')$ quotient quiver subtraction on the (self-dual) $T[\surm(N)]$ theory given at the top of \Figref{fig:TSUN_gauging_diagram} (see also \cite{Gaiotto:2008ak}). In this case, the two (dual) operations of gauging flavour and Coulomb-branch global symmetries take place on the same target quiver. Applying the algorithm described above yields the unitary-symplectic quiver at the bottom right of \Figref{fig:TSUN_gauging_diagram}, which under 3d mirror symmetry is dual to the flavour-gauged result on the bottom left of \Figref{fig:TSUN_gauging_diagram}.
\subsubsection{$\overline{reg. A_6}///\urm(3)\times\urm(2)$}
The simplest case where a product of two non-abelian unitary Coulomb branch isometry subgroups of different ranks are gauged takes the $T[\mathrm{SL}(7)]$ theory and gauges the Coulomb branch product subgroup $\urm(3)\times\urm(2)$. This is shown in \Figref{fig:A6U3U2QQS} to produce \Quiver{fig:A6U3U2QQS}. Like in the previous examples, the rebalancing increases the rank of the explicit flavour symmetry $\surm(7)$ without augmenting it via Abelian factors.

\begin{figure}[h!]
\centering

\begin{subfigure}[t]{0.47\textwidth}
\centering
\begin{tikzpicture}

\node[font=\bfseries] at (2.5,0.9) {Step 1: Subtraction};

\node[gauge, label=below:$1$] (1) at (0,0){};
\node[gauge, label=below:$2$] (2) at (1,0){};
\node[gauge, label=below:$3$] (3) at (2,0){};
\node[gauge, label=below:$4$] (4) at (3,0){};
\node[gauge, label=below:$5$] (5) at (4,0){};
\node[gauge, label=below:$6$] (6) at (5,0){};
\node[flavour, label=below:$7$] (7f) at (6,0){};

\draw[-] (1)--(2)--(3)--(4)--(5)--(6)--(7f);

\node[font=\Large] at (-0.8,-1.35) {$-$};

\node[gauge, label=below:$1$] (1sub) at (0,-1.35){};
\node[gauge, label=below:$2$] (2sub) at (1,-1.35){};
\node[gauge, label=below:$3$] (3sub) at (2,-1.35){};
\node[gauge, label=below:$4$] (4sub) at (3,-1.35){};
\node[gauge, label=below:$1$] (1subr) at (4,-1.35){};

\draw[-] (1sub)--(2sub)--(3sub)--(4sub)--(1subr);

\draw (-0.15,-2.15) -- (5.15,-2.15);

\node[font=\Large] at (-0.8,-2.75) {$=$};

\node[gauge, label=below:$4$] (4res) at (0,-2.75){};
\node[gauge, label=below:$6$] (6res) at (1,-2.75){};
\node[flavour, label=below:$8$] (8fres) at (2,-2.75){};

\draw[-] (4res)--(6res)--(8fres);

\end{tikzpicture}
\caption{}
\label{fig:A6U3U2Sub}
\end{subfigure}
\hfill
\begin{subfigure}[t]{0.47\textwidth}
\centering
\begin{tikzpicture}

\node[font=\bfseries] at (1,0.9) {Step 2: Symplectisation};

\node[gauge, label=below:$4$] (4res) at (0,0){};
\node[gauge, label=below:$6$] (6res) at (1,0){};
\node[flavour, label=below:$8$] (8fres) at (2,0){};

\draw[-] (4res)--(6res)--(8fres);

\draw[->, thick]
    (1,-.8) -- (1,-2.25)
    node[midway, fill=white, inner sep=2pt, align=center]
    {\small\textit{Replace}\\[-2pt]
     $\urm(4)\;\text{\emph{with}}\;\sprm(2)$};

\node[gaugeb, label=below:$C_2$] (Sp2res) at (0,-2.75){};
\node[gauge, label=below:$6$] (6resb) at (1,-2.75){};
\node[flavour, label=below:$8$] (8fresb) at (2,-2.75){};

\draw[-] (Sp2res)--(6resb)--(8fresb);

\end{tikzpicture}
\caption{}
\label{fig:A6U3U2Symp}
\end{subfigure}

\caption{$\urm(3)\times\urm(2)$ quotient quiver subtraction on the $T[\mathrm{SL}(7)]$ which occurs in two steps. The first involves the subtraction of the $\urm(3)\times\urm(2)$ quotient quiver and rebalances the $\urm(6)$ gauge node, promoting its flavour symmetry from $\surm(7)$ to $\surm(8)$. The second step is the symplectisation of the $\urm(4)$ node into an $\sprm(2)$ gauge node.}
\label{fig:A6U3U2QQS}
\end{figure}
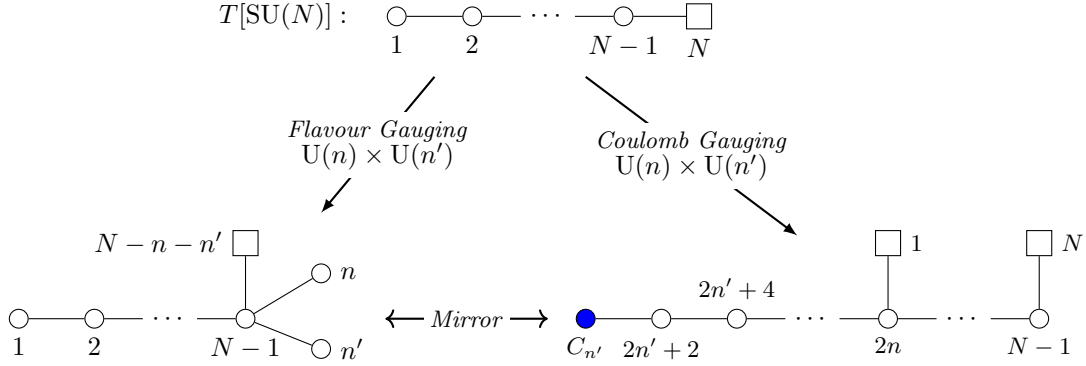
\begin{figure}[h!]
\centering
\begin{tikzpicture}


\node[gauge, label=below:$1$] (t1) at (0,3){};
\node[gauge, label=below:$2$] (t2) at (1,3){};
\node (tdots) at (2,3) {$\cdots$};
\node[gauge, label=below:$N-1$] (tnm1) at (3,3){};
\node[flavour, label=below:$N$] (tnf) at (4,3){};

\draw[-] (t1)--(t2)--(tdots)--(tnm1)--(tnf);

\node[font=\bfseries] at (-1.5,3) {$T[\surm(N)]:$};


\node[gauge, label=below:$1$] (l1) at (-5,-1){};
\node[gauge, label=below:$2$] (l2) at (-4,-1){};
\node (ldots) at (-3,-1) {$\cdots$};
\node[gauge, label=below:$N-1$] (lnm1) at (-2,-1){};
\node[flavour, label=left:$N-n-n'$] (lnf) at (-2,0){};

\node[gauge, label=right:$n$] (lr1) at (-1, -0.4){};
\node[gauge, label=right:$n'$] (lr2) at (-1, -1.4){};

\draw[-] (l1)--(l2)--(ldots)--(lnm1)--(lnf);
\draw[-] (lnm1)--(lr1);
\draw[-] (lnm1)--(lr2);


\node[gaugeb, label=below:\small$C_{n'}$] (r1) at (2.5,-1){};
\node[gauge, label=below:\small$2n'+2$] (r2) at (3.5,-1){};
\node[gauge, label=above:\small$2n'+4$] (r3) at (4.5,-1){};
\node (rdotsa) at (5.5,-1) {$\cdots$};
\node[gauge, label=below:\small$2n$] (r4) at (6.5,-1){};
\node (rdotsb) at (7.5,-1) {$\cdots$};
\node[gauge, label=below:\small$N-1$] (r5) at (8.5,-1){};

\node[flavour, label=right:\small$N$] (rnf) at (8.5,0){};
\node[flavour, label=right:\small$1$] (rnf2) at (6.5,0){};

\draw[-] (r1)--(r2)--(r3)--(rdotsa)--(r4)--(rdotsb)--(r5)--(rnf);
\draw[-] (r4)--(rnf2);


\draw[-latex, thick]
    (0.5,2.2) -- (-1,0.4)
    node[midway, fill=white, inner sep=3pt, align=center]
    {\small\textit{Flavour Gauging}\\[-2pt]
     $\urm(n)\times\urm(n')$};

\draw[-latex, thick]
    (2.5,2.2) -- (5.3,0.1)
    node[midway, fill=white, inner sep=3pt, align=center]
    {\small\textit{Coulomb Gauging}\\[-2pt]
     $\urm(n)\times\urm(n')$};

\draw[<->, thick]
    (-0.15,-1) -- (2,-1)
    node[midway, fill=white, inner sep=2pt]
    {\small\textit{Mirror}};

\end{tikzpicture}

\caption{Another illustration of the two gauging operations on the moduli spaces of the $T[\surm(N)]$ theory in the upper row. The two theories in the lower row are a well-known 3d mirror pair, which forces the two gauging operations to be dual to each other. Hence realising a Coulomb branch gauging using quotient quiver subtraction, before subsequently performing 3d mirror symmetry, is equivalent to classical flavour gauging.}
\label{fig:TSUN_gauging_diagram}
\end{figure}

It is straightfoward to calculate the Coulomb branch Hilbert series of \Quiver{fig:A6U3U2QQS} perturbatively to $\mathcal{O}\left(t^{20}\right)$,
    \begin{align} \hsC{fig:A6U3U2QQS}= 1 +& 3 t^2 + 14 t^4 + 41 t^6 + 128 t^8 + 338 t^{10} + 885 t^{12} + 2124 t^{14} + 4974 t^{16} + 11033 t^{18}\\ +& 23803 t^{20} + \mathcal{O}(t^{21})\nonumber\end{align}
which matches the Hilbert series computed exactly through Weyl integration under the embedding \eqref{emb:A6_U3U2}. For brevity, only the unrefined Hilbert series is presented below,
\begin{equation}
\hs\left[
\frac{\overline{\mathrm{reg.}\,\mathrm{SL}(7)}}
{\urm(3)\times\urm(2)}
\right]
=
\frac{(1-t^{14})(1-t^{16})\,P_{\ref{hs:SL7_U3U2}}(t)}
{(1-t^2)^4(1-t^4)^4(1-t^6)^3(1-t^8)^3
 (1-t^{10})^2(1-t^{12})^2},
 \label{hs:SL7_U3U2}
\end{equation}

where
\begin{align}
P_{\ref{hs:SL7_U3U2}}(t)={}&1-t^2+4t^4+11t^8+6t^{10}+29t^{12}
        +22t^{14}+65t^{16} \nonumber\\
&+62t^{18}+115t^{20}+116t^{22}+183t^{24}
  +168t^{26}+237t^{28} \nonumber\\
&+208t^{30}+255t^{32}+\cdots+t^{64}.\nonumber
\end{align}
The emebedding is,
\begin{equation}
    \mu_{1} \rightarrow \sigma q_1q_2^3+\tilde{\sigma}\frac{q_2^3}{q_1}+\frac{\rho_{1}}{q_2^4}
\label{emb:A6_U3U2}
\end{equation}
Where $\mu$ is an $A_6$ fugacity, $\sigma$, $\tilde{\sigma}$ are $A_1$ fugacities and $q_1$, $q_2$ are $\urm(1)$ charges. As with the previous example, this is an instance of the general prescription in \Figref{fig:TSUN_gauging_diagram}.
\section{\boldmath$\surm(n)\times\surm(n')\;$\unboldmath  Gauging}
\label{sec:SUnSUn'_QQS}
A small adjustment to the rules above yields an algorithm for $\surm(n)\times\surm(n')$ gauging. In this case, the target quiver must be unframed and rebalancing occurs using a single $\urm(1)$ gauge node. An unframed target quiver of course ensures that the $\urm(1)_{\text{COM}}$ is present, which together with the $\urm(1)$ rebalancing node reduces the action of the quotient quiver from a $\urm(n)\times\urm(n')$ gauging to $\surm(n)\times\surm(n')$.  The resulting quiver need not be framed, as the $\urm(2n')\rightarrow\sprm(n')$ symplectisation explicitly breaks the $\urm(1)_{\text{COM}}$ shift symmetry. The following examples explore the actions of $\surm(2)\times\surm(2)$, $\surm(2)\times\surm(3)$ and $\surm(3)\times\surm(3)$ gauging on the 3d-mirrors of the 3d affine $E_7$ and $E_8$ theories. Since these theories have non-Lagrangian duals, it is not possible to check these results against a Higgs branch gauging, and hence is a set of non-trivial examples of the algorithm gauging symmetries that cannot be gauged in any other way in 3d.

Recall that for unframed unitary-symplectic quivers contributions to the monopole formula arise from both the integer and half-integer magnetic lattices. All Coulomb branch Hilbert series presented here include contributions from both lattices, when relevant.

The following examples of $\surm(n)\times\surm(n')$ quotient quiver subtraction focus on exceptional nilpotent orbits and the maximal implosion of $\overline{reg.\slrm(N)}$.
\subsection{$\surm(2)\times\surm(2)$ Gauging of the $E_7$ Theory}
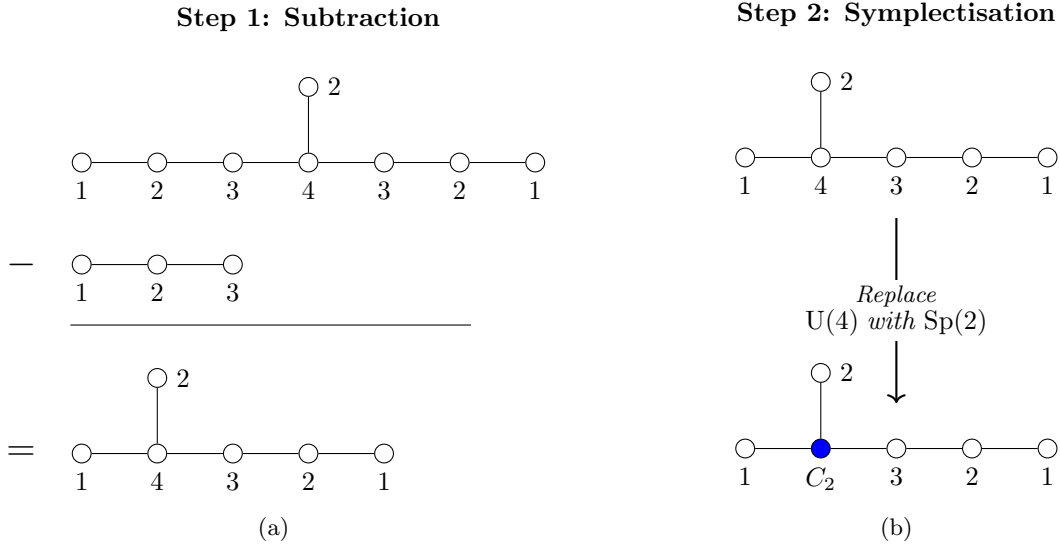
\begin{figure}[h!]
\centering

\begin{subfigure}[t]{0.47\textwidth}
\centering
\begin{tikzpicture}

\node[font=\bfseries] at (3,1.9) {Step 1: Subtraction};

\node[gauge, label=below:$1$] (1) at (0,0){};
\node[gauge, label=below:$2$] (2) at (1,0){};
\node[gauge, label=below:$3$] (3) at (2,0){};
\node[gauge, label=below:$4$] (4) at (3,0){};
\node[gauge, label=below:$3$] (5) at (4,0){};
\node[gauge, label=below:$2$] (6) at (5,0){};
\node[gauge, label=below:$1$] (7) at (6,0){};
\node[gauge, label=right:$2$] (8) at (3,1){};

\draw[-] (1)--(2)--(3)--(4)--(5)--(6)--(7);
\draw[-] (4)--(8);

\node[font=\Large] at (-0.8,-1.35) {$-$};

\node[gauge, label=below:$1$] (1sub) at (0,-1.35){};
\node[gauge, label=below:$2$] (2sub) at (1,-1.35){};
\node[gauge, label=below:$3$] (3sub) at (2,-1.35){};

\draw[-] (1sub)--(2sub)--(3sub);

\draw (-0.15,-2.15) -- (5.15,-2.15);

\node[font=\Large] at (-0.8,-3.85) {$=$};

\node[gauge, label=below:$1$] (1a) at (0,-3.85){};
\node[gauge, label=below:$4$] (2a) at (1,-3.85){};
\node[gauge, label=below:$3$] (3a) at (2,-3.85){};
\node[gauge, label=below:$2$] (4a) at (3,-3.85){};
\node[gauge, label=below:$1$] (5a) at (4,-3.85){};
\node[gauge, label=right:$2$] (6a) at (1,-2.85){};

\draw[-] (1a)--(2a)--(3a)--(4a)--(5a);
\draw[-] (6a)--(2a);

\end{tikzpicture}
\caption{}
\label{fig:E8SU3SU3Sub}
\end{subfigure}
\hfill
\begin{subfigure}[t]{0.47\textwidth}
\centering
\begin{tikzpicture}

\node[font=\bfseries] at (2,1.9) {Step 2: Symplectisation};

\node[gauge, label=below:$1$] (1a) at (0,0){};
\node[gauge, label=below:$4$] (2a) at (1,0){};
\node[gauge, label=below:$3$] (3a) at (2,0){};
\node[gauge, label=below:$2$] (4a) at (3,0){};
\node[gauge, label=below:$1$] (5a) at (4,0){};
\node[gauge, label=right:$2$] (6a) at (1,1){};

\draw[-] (1a)--(2a)--(3a)--(4a)--(5a);
\draw[-] (6a)--(2a);

\draw[->, thick]
    (2,-.8) -- (2,-3.25)
    node[midway, fill=white, inner sep=2pt, align=center]
    {\small\textit{Replace}\\[-2pt]
     $\urm(4)\;\text{\emph{with}}\;\sprm(2)$};

\node[gauge, label=below:$1$] (1b) at (0,-3.85){};
\node[gaugeb, label=below:$C_2$] (2b) at (1,-3.85){};
\node[gauge, label=below:$3$] (3b) at (2,-3.85){};
\node[gauge, label=below:$2$] (4b) at (3,-3.85){};
\node[gauge, label=below:$1$] (5b) at (4,-3.85){};
\node[gauge, label=right:$2$] (6b) at (1,-2.85){};

\draw[-] (1b)--(2b)--(3b)--(4b)--(5b);
\draw[-] (6b)--(2b);

\end{tikzpicture}
\caption{}
\label{fig:E7SU2SU2Symp}
\end{subfigure}
\caption{This presents the $\surm(2)\times\surm(2)$ gauging of the $E_7$ Coulomb branch symmetry of the 3d-mirror to the 3d affine $E_7$ theory. Note that this theory is star-shaped before and after the gauging (which in Class $\mathcal{S}$ terms means the gauging relates a 3-punctured sphere to a 5-punctured sphere).}
\label{fig:E7SU2SU2QQS}
\end{figure}

Consider the $\surm(2)\times\surm(2)$ Coulomb branch gauging of the 3d mirror of the 3d $E_7$ theory, shown in \Figref{fig:E7SU2SU2QQS}. This reduces the Coulomb branch dimension by six and changes the quiver from unitary star-shaped to unitary-symplectic star-shaped, with a central $\sprm(2)$ gauge node. It is straightforward to calculate the Coulomb branch Hilbert series of the gauged theory which agrees exactly with the result obtained from an explicit $\surm(2)\times\surm(2)$ hyper-K\"ahler quotient.
\begin{equation}
\hs\left[
\frac{\overline{\mathrm{min.}\,E_7}}
{\surm(2)\times\surm(2)}
\right]
=
\frac{P_{\ref{hs_e7_A1A1}}(t)}
{(1-t^2)^{11}(1-t^4)^{11}},
\label{hs_e7_A1A1}
\end{equation}
where
\begin{align}
P_{\ref{hs_e7_A1A1}}(t)={}&1+20t^2+282t^4+2494t^6+15510t^8
        +70186t^{10}+241225t^{12} +643075t^{14}\nonumber\\
&+1357618t^{16}+2294386t^{18}
  +3135012t^{20} +3475614t^{22}+\cdots+t^{44}.\nonumber
\end{align}
This is associated with the embedding,
\begin{equation}
    \mu_{} \rightarrow 
    \nu_{1}\alpha+\nu_{3}\beta +\nu_{4}\gamma+\alpha\beta\gamma
\end{equation}
where $\mu$ is an $E_7$ fugacity, $\nu$ is a $D_4$ fugacity and $\alpha$, $\beta$, $\gamma$ are $A_1$ fugacities. Note that this is expected to be degeneration $\mathrm{F}$ in \cite{Chacaltana:2012ch}, Section 6.3.

It's also straightforward to perform an $\surm(2)\times\surm(2)$ quotient quiver subtraction on the orthosymplectic quiver dual to the $E_7$ theory \cite{Hanany:2023tvn,Bourget:2020xdz}. In this case, the resulting theory \eqref{quiv:OS_E7A1A1_QQS} has the same Coulomb branch as that of \Figref{fig:E7SU2SU2QQS}. Interestingly, the theory \eqref{quiv:OS_E7A1A1_QQS} has an $S_2\times S_3$ outer automorphism symmetry, as well as various choices of $\mathbb{Z}_{2}^{k}$ orbifolds from the choice of global form of $\mathfrak{so}_{2n}$ gauge nodes. It is unclear how these are related to the quiver in \Figref{fig:E7SU2SU2QQS}.
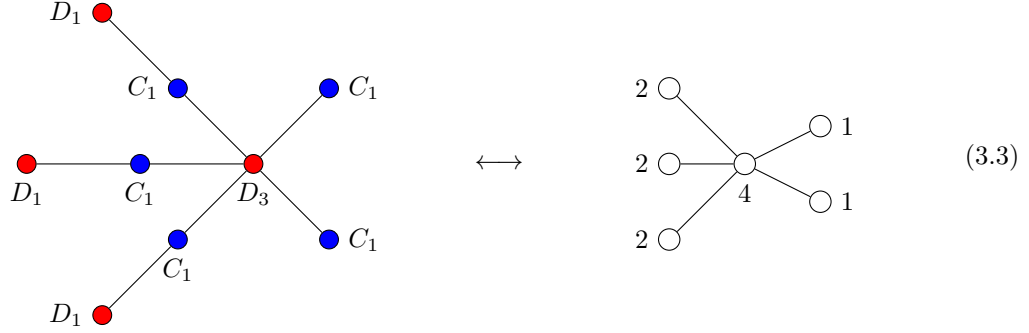
\begin{equation}
\raisebox{-0.5\height}{\begin{tikzpicture}
    \node[gauger, label=below:$D_1$] (1) at (-3,0){};
    \node[gaugeb, label=below:$C_1$] (2) at (-1.5,0){};
    \node[gauger, label=below:$D_3$] (3) at (0,0){};
    \node[gauger, label=left:$D_1$] (4) at (-2,-2){};
    \node[gaugeb, label=below:$C_1$] (5) at (-1,-1){};
    \node[gauger, label=left:$D_1$] (6) at (-2,2){};
    \node[gaugeb, label=left:$C_1$] (7) at (-1,1){};
    \node[gaugeb, label=right:$C_1$] (8) at (1,1){};
    \node[gaugeb, label=right:$C_1$] (9) at (1,-1){};
    \draw[-] (1)--(2)--(3)--(5)--(4) (3)--(7)--(6) (8)--(3)--(9);
    \node (arr) at (3.25,0){$\longleftrightarrow$};
    \node[gaugeBig, label=left:$2$] (1a) at (5.5,1){};
    \node[gaugeBig, label=left:$2$] (2a) at (5.5,0){};
    \node[gaugeBig, label=left:$2$] (3a) at (5.5,-1){};
    \node[gaugeBig, label=right:$1$] (4a) at (7.5,0.5){};
    \node[gaugeBig, label=right:$1$] (5a) at (7.5,-0.5){};
    \node[gaugeBig, label=below:$4$] (6a) at (6.5,0){};
    \draw[-] (1a)--(6a)--(2a) (3a)--(6a)--(4a) (5a)--(6a);
\end{tikzpicture}}
\label{quiv:OS_E7A1A1_QQS}
\end{equation}
The Higgs branch Hilbert series of \Figref{fig:E7SU2SU2QQS} and \eqref{quiv:OS_E7A1A1_QQS} also match. This is given in \eqref{hs:e7A1A1_higgs_check}. Hence there is strong expectation that these theories are IR dual.
\begin{equation}
    \hsh\left[{\ref{fig:E7SU2SU2QQS}}\right] = \hsh\left[\eqref{quiv:OS_E7A1A1_QQS}\right] = \frac{1+t^{4}+3t^{6}+4t^{8}+4t^{10}+2t^{12}+4t^{14}+4t^{16}+3t^{18}+t^{20}+t^{24}}{(1-t^{4})^{4}(1-t^6)(1-t^8)}
\label{hs:e7A1A1_higgs_check}
\end{equation}
The theory in \Figref{fig:E7SU2SU2QQS} that results from the $\surm(2)\times\surm(2)$ gauging appears as the 3d mirror for the three dimensional reduction of the theory given in entry 19 of [\cite{Argyres:2007tq}, Table 3] under the embedding given in [\cite{Argyres:2007tq}, Table 4]. This is an $\surm(2)\times\surm(2)$ flavour gauging of the $\mathcal{N}=2$ $E_7$ SCFT \cite{Minahan_1997,Dasgupta:1996ij}, which is dual to $\surm(2)\times\sprm(2)$ gauge theory with a single full hypermultiplet in $\left(\bm{2,4}\right)$ and four full hypermultiplets in $\left(\bm{1,4}\right)$. Both the Higgs and Coulomb branch Hilbert series match those of \Figref{fig:E7SU2SU2QQS} and \eqref{quiv:OS_E7A1A1_QQS} under the mirror map (the Coulomb branch Hilbert series of Argyres-Wittig's entry 19 is \eqref{hs:e7A1A1_higgs_check}). Hence there appear to be two distinct three-dimensional mirrors for the $S^{1}$ reduction of the four-dimensional theory, with markedly different presentations. Interestingly, there is an $S_3\times S_2$ outer-automorphism symmetry in \eqref{quiv:OS_E7A1A1_QQS}, along with various possible $S_2$ orbifolds associated with a choice of global form of the $\mathfrak{so}_{2n}$ gauge algebra. \footnote{All quivers in this work are chosen to have $\sorm(2n)$ global form. Promoting this to $\orm(2n)$ acts as a $\mathbb{Z}_{2}$ gauging with the corresponding orbifold action on the Coulomb branch \cite{Bennett:2025zor,Cabrera:2017njm,Cabrera:2018ldc,Cabrera:2017ucb,Harding:2025vov}.} It would be interesting to understand whether these gaugings have a consistent interpretation in the four-dimensional theory.

The theory in \Figref{fig:E7SU2SU2Symp} presents a further opportunity for $\surm(2)$ Coulomb branch gauging, arising from the long leg untouched by the initial subtraction. In this case, an $\surm(2)$ quotient quiver subtraction returns the theory on the right of \eqref{quiv:extra_A1_gauging_in_E7}. It's straightforward to compute its Coulomb and Higgs branch Hilbert series, which match the Higgs and Coulomb branch Hilbert series (respectively) of the theory on the left of \eqref{quiv:extra_A1_gauging_in_E7}. The theory on the left of \eqref{quiv:extra_A1_gauging_in_E7} is the $\surm(2)$ gauging of the $\sorm(8)$ flavour symmetry in Argyres-Wittig's entry 19 -- note that this embedding uses an $\surm(2)$ inside the $\sorm(4)^{2}$ decomposition of $\sorm(8)$ -- as such the resulting flavour symmetry is $\sorm(4)$, as the other $\surm(2)$ becomes the flavour symmetry of the $\left(\bm{2,4}\right)$ full hypermultiplet of $\sprm(1)\times\sprm(2)$ that emerges under the gauging. This further gauging exposes an asymmetry in the presentations of the IR dual theories in \Figref{fig:E7SU2SU2Symp} and \eqref{quiv:OS_E7A1A1_QQS} -- while the extra $\surm(2)$ can be gauged in \Figref{fig:E7SU2SU2Symp}, it is unclear in \eqref{quiv:OS_E7A1A1_QQS}.
\begin{equation}
\raisebox{-0.5\height}{\begin{tikzpicture}
    \node[gaugeb, label=right:$C_1$] (1) at (2,-0.5){};
    \node[gaugeb, label=below:$C_2$] (2) at (1,0){};
    \node[gaugeb, label=right:$C_1$] (3) at (2,0.5){};
    \node[flavourr, label=below:$D_2$] (4) at (0,0){};
    \draw[-] (1)--(2)--(3) (2)--(4);
    \draw[<->, thick] (3,0) -- (6,0) node[midway, fill=white, inner sep=2pt]{\small\textit{Mirror}};
    \node[gaugeBig, label=below:$2$] (1a) at (7,-0.15){};
    \node[gaugeb, label=below:$C_2$] (1b) at (8,-0.15){};
    \node[gaugeBig, label=below:$2$] (1c) at (9,-0.15){};
    \node[gaugeBig, label=left:$1$] (1d) at (7.5,0.6){};
    \node[gaugeBig, label=right:$1$] (1e) at (8.5,0.6){};
    \draw[-] (1a)--(1b)--(1c) (1d)--(1b)--(1e);
\end{tikzpicture}}
\label{quiv:extra_A1_gauging_in_E7}
\end{equation}
\begin{align}
    \hsh\left[\ref{quiv:extra_A1_gauging_in_E7}|_{\text{Left}}\right] = \hsc\left[\ref{quiv:extra_A1_gauging_in_E7}|_{\text{Right}}\right] &= \frac{P_{\ref{hs:e7_other_quivs_i}}(t)}{(1-t^2)^3(1-t^4)^8(1-t^6)^5},\label{hs:e7_other_quivs_i}\\
    \hsc\left[\ref{quiv:extra_A1_gauging_in_E7}|_{\text{Left}}\right] = \hsh\left[\ref{quiv:extra_A1_gauging_in_E7}|_{\text{Right}}\right] &= \frac{P_{\ref{hs:e7_other_quivs_ii}}(t)}{(1-t^2)^3(1-t^4)^3(1-t^6)(1-t^8)}.
\label{hs:e7_other_quivs_ii}
\end{align}
where,
\begin{align}
    P_{\ref{hs:e7_other_quivs_i}}(t)={}&1 + 9 t^2 + 91 t^4 + 619 t^6 + 3129 t^8 + 12451 t^{10} + 40117 t^{12} + 106516 t^{14} + 237745 t^{16}\\ &+ 451833 t^{18} + 737839 t^{20} + 1042257 t^{22} + 1280262 t^{24} + 1370774 t^{26}+\cdots+t^{52},\\
    P_{\ref{hs:e7_other_quivs_ii}}(t)={}&1-3t^{2}+10t^4-10t^6+23t^8-15t^{10}+32t^{12}-15t^{14}+23t^{16}-10t^{18}+10t^{20}\\&-3t^{22}+t^{24}.
\end{align}
\subsection{Dualities from $\overline{min. E_8}$}
In \cite{Bennett:2024llh}, various gauging operations were performed on orthosymplectic 3d $\mathcal{N}=4$ quivers with quotient quivers derived from 6d $\mathcal{N}=(1,0)$ theories. Amongst the four admissible quotient quivers were those for $\surm(2)$ and $\surm(3)$, which were sequentially applied to the two long legs of the orthosymplectic theory conjecturally mirror to the 3d affine $E_8$ theory. Naturally, these gaugings realised $\surm(2)\times\surm(2)$, $\surm(2)\times\surm(3)$ and $\surm(3)\times\surm(3)$ hyper-K\"ahler quotients on the Coulomb branch. Given the $\surm(n)\times\surm(n')$ quotient quiver subtraction algorithm introduced in Section \ref{sec:SUnSUn'_QQS}, it is interesting to examine whether they provide the same hyper-K\"ahler quotient on the Coulomb branch of the unitary mirror to affine $E_8$. Interestingly, this indeed turns out to be the case. As such, it is possible to conjecture that the two differently-presented gaugings lead to IR dual field theories.

\subsubsection*{\underline{$\overline{min. E_8}///\surm(2)\times\surm(2)$}}
\begin{figure}[h!]
\centering
\begin{subfigure}[t]{0.47\textwidth}
\centering
\begin{tikzpicture}
\node[font=\bfseries] at (3.5,1.9) {Step 1: Subtraction};
\node[gauge, label=below:$1$] (1) at (0,0){};
\node[gauge, label=below:$2$] (2) at (1,0){};
\node[gauge, label=below:$3$] (3) at (2,0){};
\node[gauge, label=below:$4$] (4) at (3,0){};
\node[gauge, label=below:$5$] (5) at (4,0){};
\node[gauge, label=below:$6$] (6) at (5,0){};
\node[gauge, label=below:$4$] (4r) at (6,0){};
\node[gauge, label=below:$2$] (2r) at (7,0){};
\node[gauge, label=right:$3$] (3t) at (5,1){};

\draw[-] (1)--(2)--(3)--(4)--(5)--(6)--(4r)--(2r);
\draw[-] (6)--(3t);

\node[font=\Large] at (-0.8,-1.35) {$-$};

\node[gauge, label=below:$1$] (1sub) at (0,-1.35){};
\node[gauge, label=below:$2$] (2sub) at (1,-1.35){};
\node[gauge, label=below:$3$] (3sub) at (2,-1.35){};

\draw[-] (1sub)--(2sub)--(3sub);

\draw (-0.15,-2.15) -- (6.15,-2.15);

\node[font=\Large] at (-0.8,-3.85) {$=$};
\node[gauge, label=below:$1$] (0res) at (0,-3.85){};
\node[gauge, label=below:$4$] (1ares) at (1,-3.85){};
\node[gauge, label=below:$5$] (1bres) at (2,-3.85){};
\node[gauge, label=below:$6$] (6res) at (3,-3.85){};
\node[gauge, label=below:$4$] (4res) at (4,-3.85){};
\node[gauge, label=below:$2$] (2res) at (5,-3.85){};
\node[gauge, label=right:$3$] (3res) at (3,-2.85){};

\draw[-] (0res)--(1ares)--(1bres)--(6res)--(4res)--(2res);
\draw[-] (6res)--(3res);

\end{tikzpicture}
\caption{}
\label{}
\end{subfigure}
\hfill
\begin{subfigure}[t]{0.47\textwidth}
\centering
\begin{tikzpicture}

\node[font=\bfseries] at (1.5,1.9) {Step 2: Symplectisation};

\node[gauge, label=below:$1$] (0res) at (0,0){};
\node[gauge, label=below:$4$] (1ares) at (1,0){};
\node[gauge, label=below:$5$] (1bres) at (2,0){};
\node[gauge, label=below:$6$] (6res) at (3,0){};
\node[gauge, label=below:$4$] (4res) at (4,0){};
\node[gauge, label=below:$2$] (2res) at (5,0){};
\node[gauge, label=right:$3$] (3res) at (3,1){};

\draw[-] (0res)--(1ares)--(1bres)--(6res)--(4res)--(2res);
\draw[-] (6res)--(3res);
\draw[->, thick]
    (1.5,-.8) -- (1.5,-2.25)
    node[midway, fill=white, inner sep=2pt, align=center]
    {\small\textit{Replace}\\[-2pt]
     $\urm(4)\;\text{\emph{with}}\;\sprm(2)$};

\node[gauger, label=below:$D_1$] (D1res) at (0,-3.85){};
\node[gaugeb, label=below:$C_2$] (C2res) at (1,-3.85){};
\node[gauge, label=below:$5$] (1bres) at (2,-3.85){};
\node[gauge, label=below:$6$] (6resb) at (3,-3.85){};
\node[gauge, label=below:$4$] (4resb) at (4,-3.85){};
\node[gauge, label=below:$2$] (2resb) at (5,-3.85){};
\node[gauge, label=right:$3$] (3resb) at (3,-2.85){};

\draw[-] (D1res)--(C2res)--(1bres)--(6resb)--(4resb)--(2resb);
\draw[-] (6resb)--(3resb);

\end{tikzpicture}
\caption{}
\label{}
\end{subfigure}

\caption{This shows the $\surm(2)\times\surm(2)$ gauging of the $E_8$ Coulomb branch symmetry of the mirror to the affine $E_8$ theory, which takes the form of the affine $E_8$ Dynkin diagram. Note that the sympelctisation affects only the $C_2$ gauge node, and the $D_1$ gauge node appearing is identical to the previous $\urm(1)$.}
\label{fig:E8SU2SU2Symp}
\end{figure}
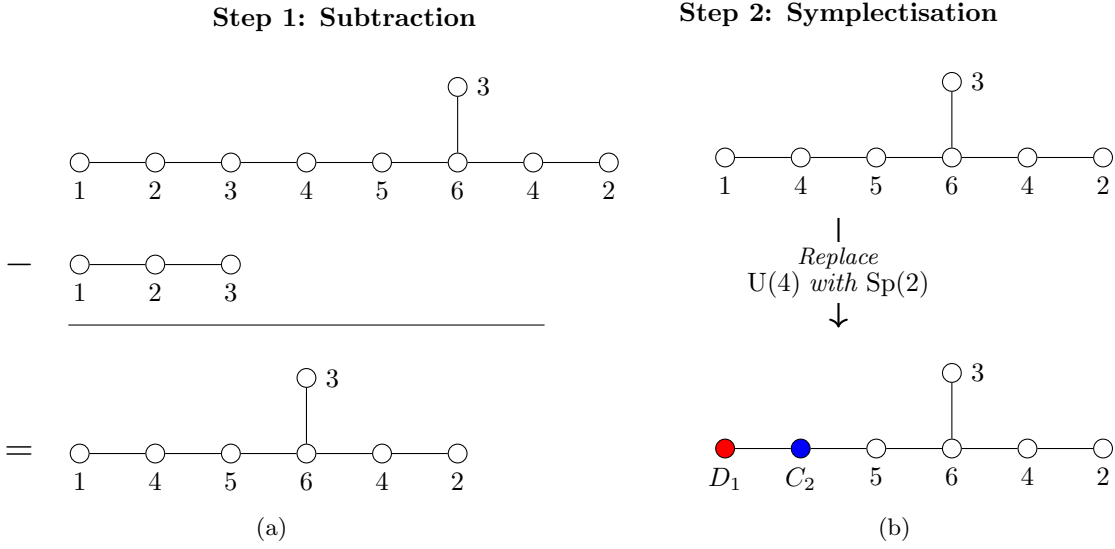

First consider the $\surm(2)\times\surm(2)$ quotient quiver subtraction on the unitary quiver 3d mirror to affine $E_8$, in the shape of an affine $E_8$ Dynkin diagram and realising $\overline{min.E_8}$ on its Coulomb branch and the Klein $E_8$ singularity on its Higgs branch. Using the rules in Section \ref{sec:SUnSUn'_QQS} it is straightforward to apply an $\surm(2)\times\surm(2)$ quotient quiver subtraction, shown in \Figref{fig:E8SU2SU2Symp}.
It's illustrative to compare this to the result obtained in \cite{Bennett:2024llh} after $\surm(2)\times\surm(2)$ gauging and given in \eqref{quiv:OS_E8A1A1_QQS}.
\begin{equation}
\raisebox{-0.5\height}{\begin{tikzpicture}
    \node[gauger, label=below:$D_1$] (1) at (0,0){};
    \node[gaugeb, label=below:$C_1$] (2) at (1,0){};
    \node[gauger, label=below:$D_3$] (3) at (2,0){};
    \node[gaugeb, label=below:$C_3$] (4) at (3,0){};
    \node[gauger, label=below:$D_4$] (5) at (4,0){};
    \node[gaugeb, label=below:$C_3$] (6) at (5,0){};
    \node[gauger, label=below:$D_3$] (7) at (6,0){};
    \node[gaugeb, label=below:$C_1$] (8) at (7,0){};
    \node[gauger, label=below:$D_1$] (9) at (8,0){};
    \node[gaugeb, label=left:$C_1$] (10) at (2,1){};
    \node[gaugeb, label=right:$C_1$] (11) at (4,1){};
    \node[gaugeb, label=right:$C_1$] (12) at (6,1){};
    \draw[-] (1)--(2)--(3)--(4)--(5)--(6)--(7)--(8)--(9) (3)--(10) (5)--(11) (7)--(12);
\end{tikzpicture}}
\label{quiv:OS_E8A1A1_QQS}
\end{equation}
Direct computation shows that the two theories' Coulomb branch Hilbert series match up to $\mathcal{O}(t^{10})$, which suggests that they are the same algebraic variety. Since these theories are expected to be IR dual, they are also conjectured to have the same Higgs branch; this is not computationally testable under current techniques. Interestingly, the purely orthosymplectic quiver \eqref{quiv:OS_E8A1A1_QQS} displays a $\mathbb{Z}_{2}$ outer automorphism symmetry about the central $D_4$ gauge node that does not appear in the unitary-orthosymplectic quiver of \Figref{fig:E8SU2SU2Symp}. At present there does not appear to be a candidate $\mathbb{Z}_{2}$ in the quiver of \Figref{fig:E8SU2SU2Symp} that maps to this outer automorphism of \eqref{quiv:OS_E8A1A1_QQS} across the duality -- the only obvious $\mathbb{Z}_{2}$ orbifold action on the Coulomb branch relates to the global form of the $D_1$ gauge node. Note also that the resulting theory in \Figref{fig:E8SU2SU2Symp} does not have a known brane construction, nor does it appear to descend from Class $\mathcal{S}$.

Neither is there an obvious higher-dimensional theory for which either \Figref{fig:E8SU2SU2Symp} or \eqref{quiv:OS_E8A1A1_QQS} might serve as a magnetic quiver. One near-miss concerns the 6d $\mathcal{N}=(1,0)$ little string theory with curve skeleton $2-1-2$ supporting the gauge algebra $\mathfrak{su}_2\oplus\mathfrak{sp}_0\oplus\mathfrak{su}_2$. In order to respect anomaly cancellation, each $\mathfrak{su}_2$ gauge factor must see an additional eight half-hypers rotated under $\sorm(8)$. Hence the chiral ring is schematically $\overline{min.E_8}\times\mathbb{H}^{16}///\surm(2)\times\surm(2)$, with global symmetry $\sorm(8)^{3}\times\surm(2)^{2}$ as given in Table 5 of \cite{Baume_2022}. The construction of $\overline{min.E_8}///\surm(2)\times\surm(2)$ here is thus of too small dimension to lead to a candidate magnetic quiver for this theory.
\subsubsection*{$\underline{\overline{min. E_8}///\surm(2)\times\surm(3)}$}
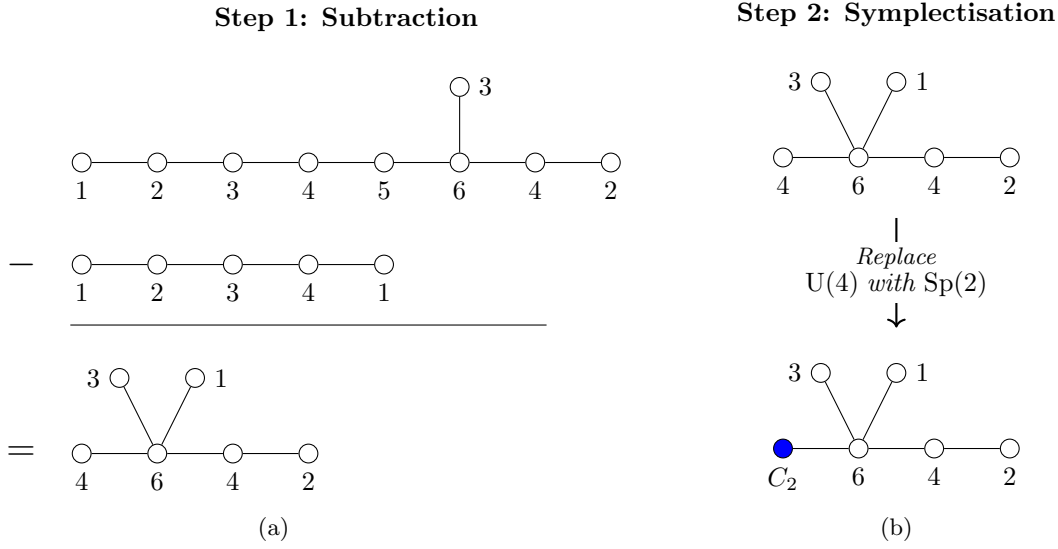
\begin{figure}[h!]
\centering
\begin{subfigure}[t]{0.47\textwidth}
\centering
\begin{tikzpicture}
\node[font=\bfseries] at (3.5,1.9) {Step 1: Subtraction};
\node[gauge, label=below:$1$] (1) at (0,0){};
\node[gauge, label=below:$2$] (2) at (1,0){};
\node[gauge, label=below:$3$] (3) at (2,0){};
\node[gauge, label=below:$4$] (4) at (3,0){};
\node[gauge, label=below:$5$] (5) at (4,0){};
\node[gauge, label=below:$6$] (6) at (5,0){};
\node[gauge, label=below:$4$] (4r) at (6,0){};
\node[gauge, label=below:$2$] (2r) at (7,0){};
\node[gauge, label=right:$3$] (3t) at (5,1){};

\draw[-] (1)--(2)--(3)--(4)--(5)--(6)--(4r)--(2r);
\draw[-] (6)--(3t);

\node[font=\Large] at (-0.8,-1.35) {$-$};

\node[gauge, label=below:$1$] (1sub) at (0,-1.35){};
\node[gauge, label=below:$2$] (2sub) at (1,-1.35){};
\node[gauge, label=below:$3$] (3sub) at (2,-1.35){};
\node[gauge, label=below:$4$] (4sub) at (3,-1.35){};
\node[gauge, label=below:$1$] (5sub) at (4,-1.35){};

\draw[-] (1sub)--(2sub)--(3sub)--(4sub)--(5sub);

\draw (-0.15,-2.15) -- (6.15,-2.15);

\node[font=\Large] at (-0.8,-3.85) {$=$};

\node[gauge, label=below:$4$] (1res) at (0,-3.85){};
\node[gauge, label=below:$6$] (6res) at (1,-3.85){};
\node[gauge, label=below:$4$] (4res) at (2,-3.85){};
\node[gauge, label=below:$2$] (2res) at (3,-3.85){};
\node[gauge, label=left:$3$] (3res) at (0.5,-2.85){};
\node[gauge, label=right:$1$] (1tres) at (1.5,-2.85){};

\draw[-] (1res)--(6res)--(4res)--(2res);
\draw[-] (6res)--(3res);
\draw[-] (6res)--(1tres);

\end{tikzpicture}
\caption{}
\label{}
\end{subfigure}
\hfill
\begin{subfigure}[t]{0.47\textwidth}
\centering
\begin{tikzpicture}

\node[font=\bfseries] at (1.5,1.9) {Step 2: Symplectisation};

\node[gauge, label=below:$4$] (1res) at (0,0){};
\node[gauge, label=below:$6$] (6res) at (1,0){};
\node[gauge, label=below:$4$] (4res) at (2,0){};
\node[gauge, label=below:$2$] (2res) at (3,0){};
\node[gauge, label=left:$3$] (3res) at (0.5,1){};
\node[gauge, label=right:$1$] (1tres) at (1.5,1){};

\draw[-] (1res)--(6res)--(4res)--(2res);
\draw[-] (6res)--(3res);
\draw[-] (6res)--(1tres);

\draw[->, thick]
    (1.5,-.8) -- (1.5,-2.25)
    node[midway, fill=white, inner sep=2pt, align=center]
    {\small\textit{Replace}\\[-2pt]
     $\urm(4)\;\text{\emph{with}}\;\sprm(2)$};

\node[gaugeb, label=below:$C_2$] (C2res) at (0,-3.85){};
\node[gauge, label=below:$6$] (6resb) at (1,-3.85){};
\node[gauge, label=below:$4$] (4resb) at (2,-3.85){};
\node[gauge, label=below:$2$] (2resb) at (3,-3.85){};
\node[gauge, label=left:$3$] (3resb) at (0.5,-2.85){};
\node[gauge, label=right:$1$] (1tresb) at (1.5,-2.85){};

\draw[-] (C2res)--(6resb)--(4resb)--(2resb);
\draw[-] (6resb)--(3resb);
\draw[-] (6resb)--(1tresb);

\end{tikzpicture}
\caption{}
\label{}
\end{subfigure}

\caption{This shows the $\surm(2)\times\surm(3)$ gauging of the $E_8$ Coulomb branch symmetry of the mirror to the affine $E_8$ theory, which takes the form of the affine $E_8$ Dynkin diagram. Note that the symplectic node appears in a leg.}
\label{fig:E8SU2SU3Symp}
\end{figure}

The next smallest non-Abelian gauging consisting of two factors is the $\surm(2)\times\surm(3)$ subgroup inside $E_8$, shown in \Figref{fig:E8SU2SU3Symp}. This leads to a star-shaped quiver which, like that of \Figref{fig:E8SU2SU2Symp}, has no known brane realisation and does not appear to be of Class $\mathcal{S}$.
Like the $\surm(2)\times\surm(2)$ gauging, this quiver is also conjectured to be dual to that which appears in \cite{Bennett:2024llh}, reproduced in \eqref{quiv:OS_E8A1A2_QQS}. Direct computation of the (unrefined) Coulomb branch Hilbert series matches between the two examples to order $\mathcal{O}(t^{10})$. As with the previous example, although these theories are expected to be IR dual their Higgs branches Hilbert series are too computationally expensive to directly check. As is often the case with orthosymplectic and unitary-orthosymplectic dual quivers, the $\mathbb{Z}_{2}$ outer automorphism acting on the bouquet of $C_1$ nodes in \eqref{quiv:OS_E8A1A2_QQS} does not appear to have an obvious analogue in \Figref{fig:E8SU2SU3Symp}. 
\begin{equation}
\raisebox{-0.5\height}{\begin{tikzpicture}
    \node[gauger, label=below:$D_1$] (1) at (0,0){};
    \node[gaugeb, label=below:$C_2$] (2) at (1,0){};
    \node[gauger, label=below:$D_4$] (3) at (2,0){};
    \node[gaugeb, label=below:$C_3$] (4) at (3,0){};
    \node[gauger, label=below:$D_3$] (5) at (4,0){};
    \node[gaugeb, label=below:$C_1$] (6) at (5,0){};
    \node[gauger, label=below:$D_1$] (7) at (6,0){};
    \node[gaugeb, label=left:$C_1$] (8) at (1,1){};
    \node[gaugeb, label=right:$C_1$] (9) at (3,1){};
    \node[gaugeb, label=right:$C_1$] (10) at (5,1){};
    \draw[-] (1)--(2)--(3)--(4)--(5)--(6)--(7) (8)--(3)--(9) (6)--(10);
\end{tikzpicture}}
\label{quiv:OS_E8A1A2_QQS}
\end{equation}
Searching for a higher-dimensional theory for which \Figref{fig:E8SU2SU3Symp} and \eqref{quiv:OS_E8A1A2_QQS} both act as magnetic quivers again does not yield positive results. As with the $\surm(2)\times\surm(2)$ gauging, it's tempting to consider the little string theory on the curve skeleton $2-1-3$ with gauge algebra $\mathfrak{su}_2\oplus\mathfrak{sp}_0\oplus\mathfrak{su}_3$. However, although the $\mathfrak{su}_3$ remains anomaly-free with zero flavours, the $\mathfrak{su}_2$ again requires eight half-hypermultiplets, with means that the chiral ring of the LST is schematically of the form $\overline{min.E_8}\times\mathbb{H}^{8}///\surm(2)\times\surm(3)$, and again both \Figref{fig:E8SU2SU3Symp} \eqref{quiv:OS_E8A1A2_QQS} do not supply the required dimension. It is unclear whether \Figref{fig:E8SU2SU3Symp} and \eqref{quiv:OS_E8A1A2_QQS} are magnetic quivers for such a higher-dimensional theory. Nevertheless, it is expected that they are IR dual theories.
\subsubsection*{$\underline{\overline{min. E_8}///\surm(3)\times\surm(3)}$}
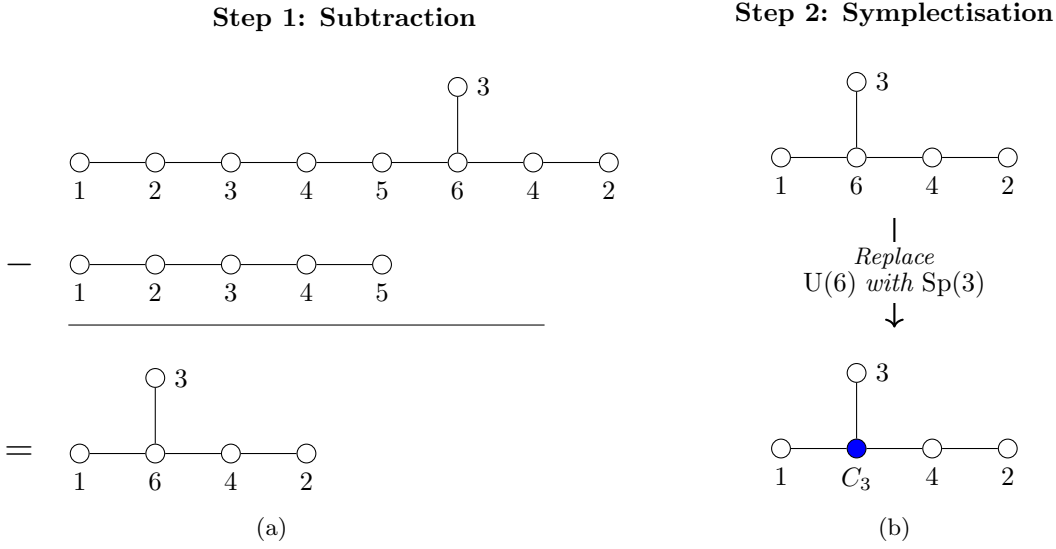
\begin{figure}[h!]
\centering

\begin{subfigure}[t]{0.47\textwidth}
\centering
\begin{tikzpicture}

\node[font=\bfseries] at (3.5,1.9) {Step 1: Subtraction};

\node[gauge, label=below:$1$] (1) at (0,0){};
\node[gauge, label=below:$2$] (2) at (1,0){};
\node[gauge, label=below:$3$] (3) at (2,0){};
\node[gauge, label=below:$4$] (4) at (3,0){};
\node[gauge, label=below:$5$] (5) at (4,0){};
\node[gauge, label=below:$6$] (6) at (5,0){};
\node[gauge, label=below:$4$] (4r) at (6,0){};
\node[gauge, label=below:$2$] (2r) at (7,0){};
\node[gauge, label=right:$3$] (3t) at (5,1){};

\draw[-] (1)--(2)--(3)--(4)--(5)--(6)--(4r)--(2r);
\draw[-] (6)--(3t);

\node[font=\Large] at (-0.8,-1.35) {$-$};

\node[gauge, label=below:$1$] (1sub) at (0,-1.35){};
\node[gauge, label=below:$2$] (2sub) at (1,-1.35){};
\node[gauge, label=below:$3$] (3sub) at (2,-1.35){};
\node[gauge, label=below:$4$] (4sub) at (3,-1.35){};
\node[gauge, label=below:$5$] (5sub) at (4,-1.35){};

\draw[-] (1sub)--(2sub)--(3sub)--(4sub)--(5sub);

\draw (-0.15,-2.15) -- (6.15,-2.15);

\node[font=\Large] at (-0.8,-3.85) {$=$};

\node[gauge, label=below:$1$] (1res) at (0,-3.85){};
\node[gauge, label=below:$6$] (6res) at (1,-3.85){};
\node[gauge, label=below:$4$] (4res) at (2,-3.85){};
\node[gauge, label=below:$2$] (2res) at (3,-3.85){};
\node[gauge, label=right:$3$] (3res) at (1,-2.85){};

\draw[-] (1res)--(6res)--(4res)--(2res);
\draw[-] (6res)--(3res);

\end{tikzpicture}
\caption{}
\label{fig:E8SU3SU3Sub}
\end{subfigure}
\hfill
\begin{subfigure}[t]{0.47\textwidth}
\centering
\begin{tikzpicture}

\node[font=\bfseries] at (1.5,1.9) {Step 2: Symplectisation};
\node[gauge, label=below:$1$] (1res) at (0,0){};
\node[gauge, label=below:$6$] (6res) at (1,0){};
\node[gauge, label=below:$4$] (4res) at (2,0){};
\node[gauge, label=below:$2$] (2res) at (3,0){};
\node[gauge, label=right:$3$] (3res) at (1,1){};

\draw[-] (1res)--(6res)--(4res)--(2res);
\draw[-] (6res)--(3res);

\draw[->, thick]
    (1.5,-.8) -- (1.5,-2.25)
    node[midway, fill=white, inner sep=2pt, align=center]
    {\small\textit{Replace}\\[-2pt]
     $\urm(6)\;\text{\emph{with}}\;\sprm(3)$};

\node[gauge, label=below:$1$] (1b) at (0,-3.85){};
\node[gaugeb, label=below:$C_3$] (C3res) at (1,-3.85){};
\node[gauge, label=below:$4$] (4b) at (2,-3.85){};
\node[gauge, label=below:$2$] (2b) at (3,-3.85){};
\node[gauge, label=right:$3$] (3b) at (1,-2.85){};

\draw[-] (1b)--(C3res)--(4b)--(2b);
\draw[-] (C3res)--(3b);

\end{tikzpicture}
\caption{}
\label{}
\end{subfigure}

\caption{$\surm(3)\times\surm(3)$ quotient quiver subtraction on the affine $E_8^{(1)}$ which occurs in two steps. The first step is shown in \ref{fig:E8SU3SU3Sub} and involves the subtraction of the $\surm(3)\times\surm(3)$ quotient quiver. The second step is the symplectisation of the $\urm(6)$ node into an $\sprm(3)$ gauge node producing the resulting quiver \Quiver{fig:E8SU3SU3QQS}.}
\label{fig:E8SU3SU3QQS}
\end{figure}

Performing an $\surm(3)\times\surm(3)$ quotient quiver subtraction on the affine $E_8$ quiver, as shown in \Figref{fig:E8SU3SU3QQS}, yields a star-shaped quiver with a central $\sprm(3)$ gauge node surrounded by legs of $\urm(1)$, $\urm(3)$ and $\urm(2)-\urm(4)$. Like the previous two examples, this quiver also has a direct analogue in the form of an orthosymplectic quiver, which can be obtained from an $\surm(3)\times\surm(3)$ quotient quiver subtraction on the orthosymplectic $E_8$ theory \cite{Bennett:2024llh}.
This dual theory, given in \eqref{quiv:E8SU3SU3Ortho}, possesses an $S_3\times S_2$ outer automorphism symmetry absent in \Figref{fig:E8SU3SU3QQS}. Nevertheless, their Coulomb branch Hilbert series are the same, and may be computed exactly by gluing $\mathbb Z_2$-twisted $A_5$ Slodowy slices following the prescription in \cite[Appendix D]{Bourget:2020xdz}. Their Higgs branches are also expected to be equal, however are too computationally expensive to check.
\begin{equation}
\raisebox{-0.5\height}{\begin{tikzpicture}
    \node[gauger, label=below:$D_1$] (d1l) at (0,0){};
    \node[gaugeb, label=below:$C_2$] (c2l) at (1,0){};
    \node[gauger, label=below:$D_4$] (d4) at (2,0){};
    \node[gaugeb, label=below:$C_2$] (c2r) at (3,0){};
    \node[gauger, label=below:$D_1$] (d1r) at (4,0){};
    \node[gaugeb, label=above:$C_1$] (c1tl) at ({2-sin(45)},{cos(45)}){};
    \node[gaugeb, label=above:$C_1$] (c1tm) at (2,1){};
    \node[gaugeb, label=above:$C_1$] (c1tr) at ({2+sin(45)},{cos(45)}){};
    \draw[-] (d1l)--(c2l)--(d4)--(c2r)--(d1r) (c1tl)--(d4)--(c1tr) (d4)--(c1tm);
    \end{tikzpicture}}
\label{quiv:E8SU3SU3Ortho}
\end{equation}
Interestingly, both the quivers \Figref{fig:E8SU3SU3QQS} and \eqref{quiv:E8SU3SU3Ortho} have direct interpretations as magnetic quivers for a higher-dimensional theory. Consider the little string theory on a $3-1-3$ curve skeleton with gauge algebra $\mathfrak{su}_3\oplus\mathfrak{sp}_0\oplus\mathfrak{su}_3$ and $\surm(3)\times\surm(3)$ global symmetry, as given in \cite{Baume_2022,Sela_2019} and written explicitly in \eqref{quiv:3-1-3_6d_theory}. In this setup, the tensormultiplet associated with the $\mathfrak{sp}_0$ gauge factor is at infinite coupling, while the two $\mathfrak{su}_3$ tensors are at finite coupling. 
\begin{equation}
\raisebox{-0.5\height}{\begin{tikzpicture}
    \node[gaugeblack] (surm3l) at (0,0) {};
    \node[align=center,anchor=north] (lab) at (surm3l.south) {$\surm(3)$\\$-3$};
    \node[gaugeb] (sprm0) at (1,0) {};
    \node[align=center,anchor=north] (lab2) at (sprm0.south) {$\sprm(0)$\\$-1$};
    \node[gaugeblack] (surm3r) at (2,0) {};
    \node[align=center,anchor=north] (lab3) at (surm3r.south) {$\surm(3)$\\$-3$};
    \node[flavourblack, label=left:$\surm(3)$] (flav1) at (0.25,0.75) {};
    \node[flavourblack, label=right:$\surm(3)$] (flav2) at (1.75,0.75) {};
    \draw[-] (surm3l)--(sprm0)--(surm3r) (sprm0)--(flav1) (sprm0)--(flav2);
\end{tikzpicture}}
\label{quiv:3-1-3_6d_theory}
\end{equation}
Taking the theory to infinite coupling is straightforward and results in the $S_3$ orbifold of the Coulomb branch given by the magnetic quiver in \eqref{quiv:E8SU3SU3Ortho_orbifold}, as derived from the Type IIA brane system in \cite{Hanany:2022itc}. It is interesting to consider whether such a transition is possible in \Figref{fig:E8SU3SU3QQS} -- clearly, there is no $S_3$ outer automorphism to wreath. Neither is there an obvious brane system that gives rise to \Figref{fig:E8SU3SU3QQS}. Hence, although the two theories \Figref{fig:E8SU3SU3QQS} and \eqref{quiv:E8SU3SU3Ortho} are conjectured to be dual, only \eqref{quiv:E8SU3SU3Ortho} appears to capture the inifnite-coupling physics of the higher-dimensional theory for which it acts as a magnetic quiver.
\begin{equation}
\raisebox{-0.5\height}{\begin{tikzpicture}
    \node[gauger, label=below:$D_1$] (d1l) at (0,0){};
    \node[gaugeb, label=below:$C_2$] (c2l) at (1,0){};
    \node[gauger, label=below:$D_4$] (d4) at (2,0){};
    \node[gaugeb, label=below:$C_2$] (c2r) at (3,0){};
    \node[gauger, label=below:$D_1$] (d1r) at (4,0){};
    \node[gaugeb, label=left:$C_3$] (c1tm) at (2,1){};
    \draw[-] (c1tm) to[out=45,in=135,looseness=8] node[pos=0.5,above]{} (c1tm);
    \draw[-] (d1l)--(c2l)--(d4)--(c2r)--(d1r) (d4)--(c1tm);
    \end{tikzpicture}}
\label{quiv:E8SU3SU3Ortho_orbifold}
\end{equation}
\subsection{Maximal Implosion of $\mathrm{T}\left[\surm(2n+1)\right]$}
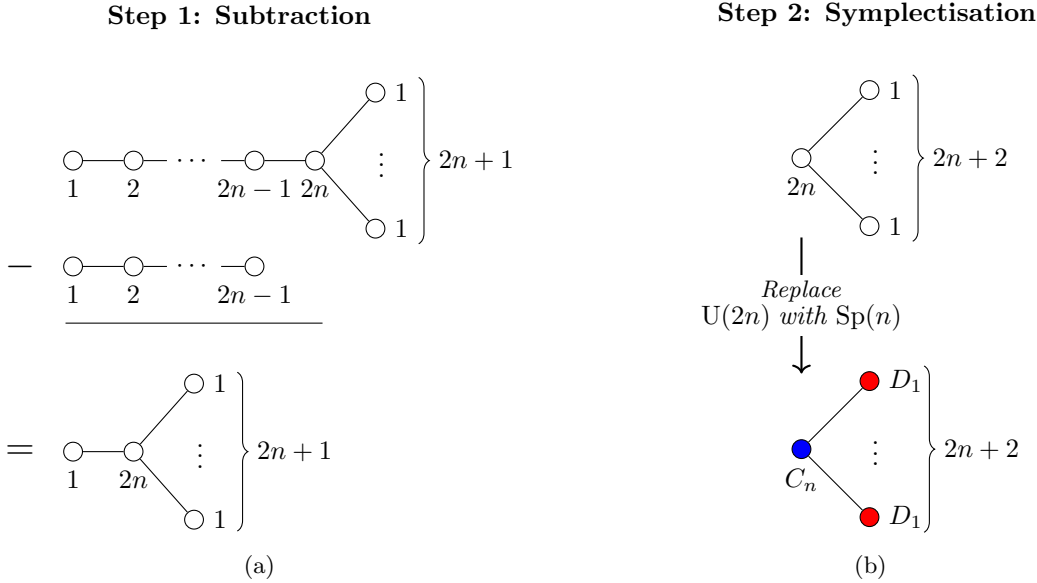
\begin{figure}[h!]
\centering

\begin{subfigure}[t]{0.48\textwidth}
\centering
\begin{tikzpicture}

\node[font=\bfseries] at (2.2,1.9) {Step 1: Subtraction};

\node[gauge, label=below:$1$] (1) at (0,0){};
\node[gauge, label=below:$2$] (2) at (0.8,0){};
\node (cdots) at (1.6,0) {$\cdots$};
\node[gauge, label=below:$2n-1$] (5) at (2.4,0){};
\node[gauge, label=below:$2n$] (6) at (3.2,0){};
\node[gauge, label=right:$1$] (4r) at (4,0.9){};
\node (vdots) at (4.08,0.05) {$\vdots$};
\node[gauge, label=right:$1$] (2r) at (4,-0.9){};

\draw[-] (1)--(2)--(cdots)--(5)--(6)--(4r);
\draw[-] (6)--(2r);

\draw[
    decorate,
    decoration={brace, raise=12pt, amplitude=4pt}
]
(4.15,1.1) -- (4.15,-1.1)
node[pos=0.5,right=16pt,black] {$2n+1$};

\node[font=\Large] at (-0.7,-1.4) {$-$};

\node[gauge, label=below:$1$] (1sub) at (0,-1.4){};
\node[gauge, label=below:$2$] (2sub) at (0.8,-1.4){};
\node (cdotssub) at (1.6,-1.4) {$\cdots$};
\node[gauge, label=below:$2n-1$] (5sub) at (2.4,-1.4){};

\draw[-] (1sub)--(2sub)--(cdotssub)--(5sub);

\draw (-0.1,-2.15) -- (3.3,-2.15);

\node[font=\Large] at (-0.7,-3.85) {$=$};

\node[gauge, label=below:$1$] (0aa) at (0,-3.85){};
\node[gauge, label=below:$2n$] (1aa) at (0.8,-3.85){};
\node[gauge, label=right:$1$] (2aa) at (1.6,-2.95){};
\node (vdotsaa) at (1.68,-3.8) {$\vdots$};
\node[gauge, label=right:$1$] (3aa) at (1.6,-4.75){};

\draw[-] (0aa)--(1aa)--(2aa);
\draw[-] (1aa)--(3aa);

\draw[
    decorate,
    decoration={brace, raise=12pt, amplitude=4pt}
]
(1.75,-2.8) -- (1.75,-4.9)
node[pos=0.5,right=16pt,black] {$2n+1$};

\end{tikzpicture}
\caption{}
\label{}
\end{subfigure}
\hfill
\begin{subfigure}[t]{0.48\textwidth}
\centering
\begin{tikzpicture}

\node[font=\bfseries] at (1,1.9) {Step 2: Symplectisation};

\node[gauge, label=below:$2n$] (1aa) at (0,0){};
\node[gauge, label=right:$1$] (2aa) at (0.9,0.9){};
\node (vdotsaa) at (0.98,0.05) {$\vdots$};
\node[gauge, label=right:$1$] (3aa) at (0.9,-0.9){};

\draw[-] (1aa)--(2aa);
\draw[-] (1aa)--(3aa);

\draw[
    decorate,
    decoration={brace, raise=12pt, amplitude=4pt}
]
(1.05,1.1) -- (1.05,-1.1)
node[pos=0.5,right=16pt,black] {$2n+2$};

\draw[->, thick]
    (0,-1.05) -- (0,-2.85)
    node[midway, fill=white, inner sep=2pt, align=center]
    {\small\textit{Replace}\\[-2pt]
     $\urm(2n)\;\text{\emph{with}}\;\sprm(n)$};

\node[gaugeb, label=below:$C_n$] (Cnaa) at (0,-3.85){};
\node[gauger, label=right:$D_1$] (D1top) at (0.9,-2.95){};
\node (vdotsb) at (0.98,-3.8) {$\vdots$};
\node[gauger, label=right:$D_1$] (D1bottom) at (0.9,-4.75){};

\draw[-] (Cnaa)--(D1top);
\draw[-] (Cnaa)--(D1bottom);

\draw[
    decorate,
    decoration={brace, raise=12pt, amplitude=4pt}
]
(1.2,-2.75) -- (1.2,-4.95)
node[pos=0.5,right=16pt,black] {$2n+2$};

\end{tikzpicture}
\caption{}
\label{}
\end{subfigure}

\caption{The action of $\surm(n)\times\surm(n)$ quotient quiver subtraction on the 3d $\mathcal{N}=4$ theory whose Coulomb branch is the maximal implosion of $\overline{reg.\surm(2n+1)}$. The resulting quiver is entirely orthosymplectic, and the initial $S_{2n+1}$ permutation symmetry between its bouqeuet of $\urm(1)$ nodes is enhanced to $S_{2n+2}$. Note that this is essentially similar to \eqref{eq:1}, which points to the fact that the implosion operation identified in \cite{Bourget:2021qpx} appears to commute with the Coulomb branch gauging.}
\label{fig:qqs_on_maximp}
\end{figure}

Another natural target for the $\surm(n)\times\surm(n)$ quotient quiver subtraction algorithm is that realising the maximal implosion of $T\left[\surm(2n+1)\right]$. In this case, the long leg of the quiver supports an $\surm(n)\times\surm(n)$ gauging, shown explicitly in \Figref{fig:qqs_on_maximp}. The resulting theory has a central gauge node of $\sprm(n)$ with a bouquet of $2n+2$ $\sorm(2)$ gauge nodes. Note that this quiver is straightforwardly the $\sorm(2)^{2n+2}$ flavour gauging of $\sprm(n)$ SQCD with $2n+2$ flavours. From the perspective of Coulomb branch geometry, this theory is interesting for the multitude of orbifolds it admits -- each $D$-type gauge factor has a $\mathbb{Z}_{2}$ related to a choice of global form, and there is an overall $S_{2n+2}$ permutation symmetry acting on the identical $D_1$ gauge nodes, which is enhanced from the initial $S_{2n+1}$ symmetry in the bouquet of the maximal implosion quiver. The Coulomb branch of the resulting theory does not in general have a simple description. \footnote{The Coulomb branch geometry was analysed to some extent in \cite{Lawrie:2024wan}.} Note that for $n=1$ the gauging action is trivial and it simply sends $\overline{min.\sorm(8)}$ to $\overline{min.\sorm(8)}$.

This example is also interesting for its interpretation in terms of the action of a Class $\mathcal{S}$ cylinder of type $\surm(n)\times\surm(n)$. As mentioned in the Introduction, quotient quiver subtracion is in general not reducible to Class $\mathcal{S}$ gluing owing to the fact that it is not restricted to act on full punctures. However, in the case of \Figref{fig:qqs_on_maximp} the $\surm(n)\times\surm(n)$ is precisely acting on a full puncture -- a $2n+2$ punctured sphere with $2n+1$ minimal punctures and one full puncture is glued to an $A_{2n-1}$ twisted fixture with one full puncture and two trivial twisted punctures and a $\left[2n-1,1\right]$ puncture. The resulting theory is a $2n+2$ punctured sphere of type twisted $A_{2n-1}$, whose 3d reduction has \Figref{fig:qqs_on_maximp} as its mirror.

\section{Discussion and Outlook}
\label{sec:outlook}
\subsection{Connection to Class $\mathcal{S}$}
\label{sec:connec_class_s}
As mentioned in the Introduction, and made clear in the previous explicit example, the action of $\surm(n)\times\surm(n)$ gauging identified in this work has in some cases a straightforward interpretation in terms of twisted Class $\mathcal{S}$ fixtures of type $A_{2n-1}$ \cite{Chacaltana:2012ch}. In fact, this is similarly true for the $\sprm(n)$ quotient quiver subtraction operation introduced in \cite{Bennett:2026gtm}. For simplicity, the following discussion considers $(2,0)$ theory on an $S^{2}$ with punctures.

\subsection*{$\sprm(N)$ Quotient Quiver Subtraction}
The action of $\sprm(n)$ quotient quiver subtraction generalises the gluing of an $\sprm(N)$ cylinder to a $A$-type $n$-punctured sphere with at least one maximal puncture. From this perspective, it is unsurprising that the quotient quiver subtraction algorithm requires node splitting. 

Consider an $A_{2N+1}$ $n$ punctured sphere with at least one maximal puncture. As explained in \cite{Chacaltana:2010ks}, this supports an $\sprm(N)$ cylinder gluing it with an $A_{2N+1}$ three punctured sphere with two regular punctures $\left[N+2,N\right]$ and $\left[N+1,N+1\right]$, and an irregular puncture $\left(\left[1^{2N+2}\right],\sprm(N)\right)$. Gluing results in an $A_{2N+1}$ $n+1$ punctured sphere with the surviving $n-1$ original punctures and the $\left[N+2,N\right]$ and $\left[N+1,N+1\right]$ punctures. This action is the Class $\mathcal{S}$ uplift of $\sprm(N)$ quotient quiver subtraction.

As given in \cite{Chacaltana:2010ks}, there are various other ways to glue an $\sprm(N)$ cylinder to a Class $\mathcal{S}$ $n$ punctured sphere. Only the method above is known to effect a pure $\sprm(N)$ hyper-K\"ahler quotient on the target theory; all other possibilities are expected to perform a diagonal $\sprm(N)$ hyper-K\"ahler quotient with some other matter sector, free or interacting. Of course, the advantage of quotient quiver subtraction in three dimensions is that it retains many of the non-conformal gaugings that disappear under RG flow in four dimensions.

\subsection*{$\sorm(N)$ Quotient Quiver Subtraction}
This case is more interesting than $\sprm(N)$. It is well known that $A$-type Class $\mathcal{S}$ theories do not admit $\sorm(N)$ cylinders \cite{Chacaltana:2010ks}. As such, there is no known direct four-dimensional uplift of $\sorm(N)$ quotient quiver subtraction. One possibility would be to perform an $\surm(N)$ gluing between two copies of the same theory, before effecting a twisted compactification under an assumed non-anomalous puncture-permutation symmetry. This rather contrived construction is engineered to directly reproduce the resulting 3d SCFT -- it does not give a purely four-dimensional interpretation to the quotient quiver subtraction operation in this instance.

\subsection*{$\surm(N)\times\surm(N)$ Quotient Quiver Subtraction}

Consider a star-shaped target quiver $\mathcal{Q}$ arising as the 3d-mirror of the $S^{1}$ reduction of a regular $A_{2N-1}$ $k$-punctured sphere with at least one full puncture, termed $T_{\mathcal{Q}}$. Following Section 4.2 of \cite{Chacaltana:2012ch}, an $\surm(N)\times\surm(N)$ cylinder can be constructed between the full puncture of $T_{\mathcal{Q}}$ and a twisted $A_{2N-1}$ fixture, with two minimal twisted punctures $\left[2N+1\right]$ and a single minimal untwisted puncture $\left[2N-1,1\right]$. Upon collapsing the cylinder, $T_{\mathcal{Q}}$ becomes a twisted $A_{2N-1}$ $k+2$ punctured sphere, inheriting the $\{\left[2N+1\right], \left[2N+1\right], \left[2N-1,1\right]\}$ punctures from the fixture. As an operation on the mirrors of the compactified theories, this is precisely $\surm(N)\times\surm(N)$ quotient quiver subtraction; the two $\left[2N+1\right]$ punctures ensure that the final theory is twisted with central gauge group $\sprm(N)$, while the $\left[2N-1,1\right]$ puncture is the $\urm(1)$ gauge node used in the algorithm of Section \ref{sec:U(n)U(n')QQS} for rebalancing.

As mentioned previously, the generalisation of quotient quiver subtraction beyond this picture is that the gauging in three dimensions need not be conformal in four dimensions. Hence, quotient quiver subtraction realises $\surm(N)\times\surm(N')$ gaugings as an extension of the Class $\mathcal{S}$ picture where these are non-conformal. Similarly, the $\surm(N)\times\surm(N)$ cylinder of \cite{Chacaltana:2012ch} applies to a star-shaped theory in the usual sense.

\subsection{Further Directions}
This paper introduces quotient quiver subtraction prescriptions for the gauging of $\urm(n)\times\urm(n')$ and $\surm(n)\times\surm(n')$ Coulomb branch global symmetry subgroups for unitary quiver theories. As with much of the quotient quiver subtraction programme, this is in some sense a three dimensional extension of Class $\mathcal{S}$ gluing in four dimensions. Unlike the Class $\mathcal{S}$ formalism, three-dimensional theories open the possibility of gauging \emph{non-full} quiver legs, resulting in theories which are not star-shaped. Equally, as shown in \cite{Bennett:2026gtm}, a larger family of quivers are easily constructed involving non-simply laced `matter'. For the present construction, two perspectives are illuminating. The first considers Type IIB brane systems with D3, D5 and NS5 branes in the presence of $\mathrm{ON}$ planes, which engineers 3d $\mathcal{N}=4$ quiver gauge theories with adjacency matrices of finite $D$-type. Across $S$-duality $\mathrm{ON}^{-}$ planes are transformed into $\mathrm{O5}^{-}$ planes, which supports $\sprm(N)$ vectormultiplets in the 3d mirror theory. Although standard in the literature, this translation leads to a novel prescription for gauging $\urm(n)\times\urm(n')$ Coulomb branch symmetry subgroups in a variety of framed quiver gauge theories. Interestingly, the method can be shown to hold for unframed theories also, at the cost of realising $\surm(n)\times\surm(n')$ gaugings instead.

Like the previous papers in this series \cite{Hanany:2023tvn,Hanany:2024fqf,Bennett:2024llh,Bennett:2025edk,Bennett:2026gtm}, this work tests the action of these gaugings against a variety of standard examples involving the closures of nilpotent orbits of classical and exceptional type. In cases in which a 3d mirror Higgs branch exists, explicit checks are made using Hilbert series techniques. For theories with non-Lagrangian mirrors, such as those with exceptional adjacency matrix, these quotient quiver subtractions yield conjectures for the action of Coulomb branch gauging.

Many concordances exist between the gauging of three dimensional Coulomb branch symmetries and actions involving theories in higher dimensions. As shown in Section \ref{sec:SUnSUn'_QQS}, this prescription leads to an independent construction of the magnetic quiver for an SCFT in the catalogue of Argyres and Wittig \cite{Argyres:2007tq}, and maps correspondences between unitary-symplectic and orthosymplectic IR-dual theories derived from the closure of the minimal nilpotent orbit of $E_8$, which are simply dual degeneration limits from the Class $\mathcal{S}$ perspective. Notably, one example leads to an alternative magnetic quiver for a six dimensional SCFT \cite{Hanany:2022itc}. In a twist on the familiar logic of magnetic quivers, it appears that such a `bottom-up' approach to studying the moduli spaces of higher-dimensional theories can yield non-trivial results.

It is worth noting an interesting feature of the $\urm(n)\times\urm(n')$ quotient quiver subtraction algorithm introduced in Section \ref{sec:U(n)U(n')QQS}. The quotient quiver subtraction algorithm introduced in \cite{Hanany:2023tvn} prohibits  sequential $\urm(n)$ gaugings on the same long leg (this is because the leg ceases to be maximal after a subtraction). The challenge of gauging $\prod_{i}\urm(n_i)$ Coulomb branch subgroups of a \emph{single} Coulomb branch global symmetry factor thus remained open. There is no a priori obstruction in the field theory to such a gauging. The problem, which is solved in this work using intuition arising from $O5$ and $O\text{N}$ planes in Type IIB brane systems, leads to a set of conjectured IR-dual theories that hence share the same Coulomb and Higgs branches.

As with previous work, these gaugings are checked against explicit hyper-K\"ahler quotients on nilpotent orbits where such examples are available. Interestingly, for the $\orm(2)\times\orm(2)$ gaugings considered in Appendix \ref{app:O2O2_QQS}, these checks show that the quotient quiver subtraction is sensitive to the global form of the resulting quivers' gauge groups. In some cases this may involve diagonal $\mathbb{Z}_2$ quotients on the magnetic lattice, as introduced in \cite{Cabrera:2017ucb}.

There are several natural extensions to the work presented here. The first is to build a bridge between these diagrammatic techniques and analytical methods of the superconformal index. Although quotient quiver subtraction is more efficient than index techniques for examples within its reach, it notably cannot yet act in all cases of incomplete Higgsing. Partition function manipulations do not generically suffer from such limitations, and thus it may be possible to derive new subtraction rules by translating between the two techniques.

Another continuation of the present work concerns rigorous proofs of the gauging procedure in terms of GIT quotients on the Coulomb branch. The mathematical literature does not currently contain satisfactory techniques for building the Coulomb branches of orthosymplectic theories -- that an $\surm(n)\times\surm(n')$ quotient promotes a purely unitary quiver into a unitary-symplectic quiver thus takes it beyond the celebrated BFN construction \cite{braverman2018coulombbranches3dmathcal}. The results contained here constitute an interesting and highly non-trivial check of a future orthosymplectic Coulomb branch formalism.

Lastly, the extension of the Coulomb branch gauging procedure to encompass products of unitary groups leaves open the question of whether a similar prescription exists for products of orthogonal and symplectic groups. Some progress on this problem is made in Appendix \ref{app:O2O2_QQS}, which surprisingly relies on a non-standard choice of global form for $D$-type gauge groups. These subtleties merit further study, although they limit the immediate applicability of such an algorithm to quivers arising as ancillaries of higher dimensional theories.


\acknowledgments
The work of SB, AH, and GK is partially supported by STFC Consolidated Grants ST/T000791/1 and ST/X000575/1. The work of SB is supported by the STFC DTP research studentship grant ST/Y509231/1. The work of GK is supported by STFC DTP research studentship grant ST/X508433/1.
\appendix
\section{\boldmath Magnetic Quivers for $\orm(2)\times\orm(2)$ Gauging\unboldmath}
\label{app:O2O2_QQS}
In principle, the techniques introduced in this paper admit an extension for gauging products of orthogonal groups in (framed) orthosymplectic quivers. Insofar as the quotient quiver \eqref{quiv:U(n)U(n)_product_quotient_quiver} emerges from intuition involving D3-/D5-/NS5-brane systems, some analogous statement is likely to arise after the introduction of $\mathrm{O}3$ orientifold planes. Unsurprisingly, it turns out that this is indeed the case. However, several subtleties regarding gauge group global forms complicate matters considerably. As such, this appendix explores a handful of examples of $\orm(n)\times\orm(n')$ quotient quiver subtraction while remaining agnostic as to the existence of an algorithm of comparable generality to \eqref{quiv:U(n)U(n)_product_quotient_quiver}.

\begin{table}[tbp]
\centering

\small
\renewcommand{\arraystretch}{1.35}
\renewcommand{\tabularxcolumn}[1]{m{#1}}
\setlength{\tabcolsep}{5pt}

\begin{tabularx}{\textwidth}{@{} C{0.17\textwidth} Y Y @{}}
\toprule

\multicolumn{3}{c}{\textbf{Electric Quiver}} \\[0.4em]

\multicolumn{3}{c}{%
  \begin{tikzpicture}[
    baseline=(current bounding box.center),
    x=1.15cm,
    y=0.9cm
  ]
    \node[gauger,   label=below:{$G_1$}] (q1) at (0,0) {};
    \node[gaugeb,   label=below:{$C_2$}] (q2) at (1,0) {};
    \node[gauger,   label=below:{$G_2$}] (q3) at (2,0) {};
    \node[flavourr, label=above:{$D_3$}] (q4) at (1,1) {};

    \draw (q1) -- (q2) -- (q3);
    \draw (q2) -- (q4);
  \end{tikzpicture}
} \\[0.7em]

\midrule

\([G_1,G_2]\)
&
\multicolumn{1}{c}{\(\mathrm{HS}[\mathcal H]\)}
&
\multicolumn{1}{c}{\(\mathrm{HS}[\mathcal C]\)}
\\

\midrule

\(\bigl[\orm(2),\orm(2)\bigr]\)
&
\HSfrac{
  &1+7t^2+40t^4+152t^6+471t^8+1061t^{10}\\[-0.2ex]
  &{}+1938t^{12}+2690t^{14}+3060t^{16}+\cdots+t^{32}
}{
  (1-t^2)^8(1-t^4)^8
}
&
\HSfrac{
  &1-2t^2+6t^4-5t^6+12t^8\\[-0.2ex]
  &{}-7t^{10}+16t^{12}+\cdots+t^{24}
}{
  (1-t^2)^3(1-t^4)^3(1-t^6)(1-t^8)
}
\\

\HSsep

\(\bigl[\sorm(2),\orm(2)\bigr]\)
&
\HSfrac{
  &1+7t^2+55t^4+241t^6+834t^8+2017t^{10}\\[-0.2ex]
  &{}+3877t^{12}+5555t^{14}+6386t^{16}+\cdots+t^{32}
}{
  (1-t^2)^8(1-t^4)^8
}
&
\HSfrac{
  (1+t^4)\bigl(1-t^2+4t^4+5t^8+\cdots+t^{16}\bigr)
}{
  (1-t^2)^3(1-t^4)^4(1-t^6)
}
\\

\HSsep

\(\bigl[\sorm(2),\sorm(2)\bigr]\)
&
\HSfrac{
  &1+7t^2+71t^4+357t^6+1426t^8+3779t^{10}\\[-0.2ex]
  &{}+7745t^{12}+11497t^{14}+13354t^{16}+\cdots+t^{32}
}{
  (1-t^2)^8(1-t^4)^8
}
&
\HSfrac{
  (1+t^4)\bigl(1+5t^4+4t^6+6t^8+\cdots+t^{16}\bigr)
}{
  (1-t^2)^3(1-t^4)^4(1-t^6)
}
\\

\midrule

\multicolumn{2}{c}{\textbf{Incorrect Magnetic Quiver}}
&
\multicolumn{1}{c}{\(\mathrm{HS}\left[\mathcal{C}\right]\)}
\\

\multicolumn{2}{c}{%
  \begin{tikzpicture}[
    baseline=(current bounding box.center),
    x=0.95cm,
    y=0.85cm
  ]
    \node[gauger,   label=below:{$D_1$}] (a1) at (0,0) {};
    \node[gaugeb,   label=below:{$C_1$}] (a2) at (1,0) {};
    \node[gauger,   label=below:{$D_2$}] (a3) at (2,0) {};
    \node[gaugeb,   label=below:{$C_2$}] (a4) at (3,0) {};
    \node[gauger,   label=below:{$D_2$}] (a5) at (4,0) {};
    \node[flavourr, label=above:{$D_1$}] (a6) at (3,1) {};

    \draw (a1) -- (a2) -- (a3) -- (a4);
    \draw (a4) -- (a6);
    \draw[transform canvas={yshift= 1.3pt}] (a4) -- (a5);
    \draw[transform canvas={yshift=-1.3pt}] (a4) -- (a5);
    \draw (3.6,0.2) -- (3.4,0) -- (3.6,-0.2);
  \end{tikzpicture}
}
&
\HSfrac{
  &1+7t^2+41t^4+179t^6+700t^8+1867t^{10}\\[-0.2ex]
  &{}+3867t^{12}+5767t^{14}+6702t^{16}+\cdots+t^{32}
}{
  (1-t^2)^8(1-t^4)^8
}
\\
\HSsep
\multicolumn{2}{c}{\textbf{Correct Magnetic Quiver}}
&
\multicolumn{1}{c}{\(\mathrm{HS}\left[\mathcal{C}\right]\)}
\\
\multicolumn{2}{c}{%
  \begin{tikzpicture}[
    baseline=(current bounding box.center),
    x=0.95cm,
    y=0.85cm
  ]
    \node[gauger,   label=below:{$D_1$}]     (b1) at (0,0) {};
    \node[gaugeb,   label=below:{$C_1$}]     (b2) at (1,0) {};
    \node[gauger,   label=below:{$D_2$}]     (b3) at (2,0) {};
    \node[gaugeb,   label=below:{$C_2$}]     (b4) at (3,0) {};
    \node[gauger,   label=below:{$\orm(4)$}] (b5) at (4,0) {};
    \node[flavourr, label=above:{$D_1$}]     (b6) at (3,1) {};

    \draw (b1) -- (b2) -- (b3) -- (b4);
    \draw (b4) -- (b6);
    \draw[transform canvas={yshift= 1.3pt}] (b4) -- (b5);
    \draw[transform canvas={yshift=-1.3pt}] (b4) -- (b5);
    \draw (3.6,0.2) -- (3.4,0) -- (3.6,-0.2);
  \end{tikzpicture}
}
&
\HSfrac{
  &1+7t^2+40t^4+152t^6+471t^8+1061t^{10}\\[-0.2ex]
  &{}+1938t^{12}+2690t^{14}+3060t^{16}+\cdots+t^{32}
}{
  (1-t^2)^8(1-t^4)^8
}
\\

\bottomrule
\end{tabularx}
\caption{Hilbert series for the different choices of \((G_1,G_2)\) for the $G_1\times G_2$ gauging of the $\sorm(10)$ flavour symmetry of $\sprm(2)$ with five full hypermultiplets. As shown by the unrefined Hilbert series, the match occurs between the $\orm(2)\times\orm(2)$ gauged electric theory and the non-simply laced $\orm(4)$ magnetic theory. All other choices are inconsisntent.}
\label{tab:appendix1}
\end{table}
Consider first the $\sprm(2)$ SQCD with ten half-hypers rotated by $\sorm(10)$. There are three obvious candidates for an $\mathfrak{so}_{2}\times\mathfrak{so}_{2}$ flavour subgroup product gauging: $\orm(2)\times\orm(2)$ and $\sorm(2)\times\orm(2)$ and $\sorm(2)\times\sorm(2)$, all given in Table \ref{tab:appendix1}. Each example effects a hyper-K\"ahler quotient on $d_5$, although Hilbert series calculations show that the resulting Higgs and Coulomb branches differ considerably. The magnetic quiver for this theory, which is non-simply laced, is sensitive to the global form of the gauge group. As shown in Table \ref{tab:appendix1}, only one choice provides an electric-magnetic pair (at the level of unrefined Hilbert series -- without refinement this identification of course remains conjectural). No other choice provides an electric-magnetic match. Note that since the magnetic quiver is non-simply laced there is no current understanding of its Higgs branch. 

The second example, in Table \ref{tab:appendix2}, concerns a product flavour subgroup gauging of the $T\left[\sorm(6)\right]\simeq T\left[\surm(4)\right]$ theory. Again, between the choices of $\orm(2)\times\orm(2)$, $\orm(2)\times\sorm(2)$ and $\sorm(2)\times\sorm(2)$ only the first can be mapped to the non-simply laced magnetic quiver shown with a distinct choice of gauge group global form. Like in Table \ref{tab:appendix1}, there is an exact match at the level of unrefined Hilbert series, which is conjectured to indicate an equivalence of Higgs and Coulomb branches across the electric-magnetic relationship.

The third example in Table \ref{tab:appendix3} is somewhat different. This concerns the $T\left[\sorm(8)\right]$ theory with product subgroup gauging $\orm(2)\times\orm(2)$, $\orm(2)\times\sorm(2)$ and $\sorm(2)\times\sorm(2)$. There are two non-simply laced magnetic quiver proposals given, and it turns out that the only consistent electric-magnetic pair is between the $\orm(2)\times\orm(2)$ electric gauging and a magnetic gauging together with a \emph{diagonal} $\mathbb{Z}_2$ action on the magnetic lattice. This diagonal action is explained in more detail in \cite{Cabrera:2017ucb} -- note that in this case its compatibility with the non-simply laced `matter' is assumed. It is unclear as to why the required $\mathbb{Z}_2$ action here becomes diagonal, as opposed to its localisation on specific nodes in Tables \ref{tab:appendix1} and \ref{tab:appendix2}. It would be interesting to understand these electric-magnetic correspondences more generally.

\begin{table}[tbp]
\centering

\small
\renewcommand{\arraystretch}{1.35}
\renewcommand{\tabularxcolumn}[1]{m{#1}}
\setlength{\tabcolsep}{2pt}

\begin{tabularx}{\textwidth}{@{} C{0.17\textwidth} Y Y @{}}
\toprule

\multicolumn{3}{c}{\textbf{Electric Quiver}} \\[0.4em]

\multicolumn{3}{c}{%
  \begin{tikzpicture}[
    baseline=(current bounding box.center),
    x=1.0cm,
    y=0.9cm
  ]
    \node[gauger,   label=below:{$D_1$}] (q1) at (0,0) {};
    \node[gaugeb,   label=below:{$C_1$}] (q2) at (1,0) {};
    \node[gauger,   label=below:{$D_2$}] (q3) at (2,0) {};
    \node[gaugeb,   label=below:{$C_2$}] (q4) at (3,0) {};
    \node[gauger,   label=below:{$G_1$}] (q5) at (4,0.5) {};
    \node[gauger,   label=below:{$G_2$}] (q6) at (4,-0.5) {};
    \node[flavourr, label=left:{$D_1$}]  (q7) at (3,1) {};

    \draw (q1) -- (q2) -- (q3) -- (q4) -- (q5);
    \draw (q6) -- (q4) -- (q7);
  \end{tikzpicture}
} \\[0.7em]

\midrule

\(\left[G_1,G_2\right]\)
&
\multicolumn{1}{c}{\(\mathrm{HS}[\mathcal H]\)}
&
\multicolumn{1}{c}{\(\mathrm{HS}[\mathcal C]\)}
\\

\midrule

\(\bigl[\orm(2),\orm(2)\bigr]\)
&
\HSfrac{
  &1-t^2+4t^4-2t^6+12t^8-3t^{10}\\[-0.2ex]
  &{}+22t^{12}-2t^{14}+30t^{16}+\cdots+t^{32}
}{
  (1-t^2)^2(1-t^4)^2(1-t^6)^2(1-t^8)^2
}
&
\HSfrac{
  &1+9t^2+54t^4+254t^6+953t^8+2932t^{10}\\[-0.2ex]
  &{}+7523t^{12}+16323t^{14}+30328t^{16}+48694t^{18}\\[-0.2ex]
  &{}+67979t^{20}+82910t^{22}+88564t^{24}+\cdots+t^{48}
}{
  (1-t^2)^6(1-t^4)^4(1-t^6)^6
}
\\

\HSsep

\(\bigl[\sorm(2),\orm(2)\bigr]\)
&
\HSfrac{
  &1-t^2+5t^4+t^6+16t^8+6t^{10}\\[-0.2ex]
  &{}+31t^{12}+12t^{14}+42t^{16}+\cdots+t^{32}
}{
  (1-t^2)^2(1-t^4)^2(1-t^6)^2(1-t^8)^2
}
&
\HSfrac{
  &1+10t^2+69t^4+371t^6+1532t^8+5058t^{10}\\[-0.2ex]
  &{}+13675t^{12}+30866t^{14}+59076t^{16}+96939t^{18}\\[-0.2ex]
  &{}+137363t^{20}+169002t^{22}+181044t^{24}+\cdots+t^{48}
}{
  (1-t^2)^6(1-t^4)^4(1-t^6)^6
}
\\

\HSsep

\(\bigl[\sorm(2),\sorm(2)\bigr]\)
&
\HSfrac{
  &1-t^2+6t^4+7t^6+16t^8+17t^{10}\\[-0.2ex]
  &{}+33t^{12}+26t^{14}+\cdots+t^{28}
}{
  (1-t^2)^2(1-t^4)^3(1-t^6)^2(1-t^8)
}
&
\HSfrac{
  &1+11t^2+85t^4+524t^6+2395t^8+8553t^{10}\\[-0.2ex]
  &{}+24540t^{12}+57935t^{14}+114653t^{16}+192732t^{18}\\[-0.2ex]
  &{}+277579t^{20}+344741t^{22}+370438t^{24}+\cdots+t^{48}
}{
  (1-t^2)^6(1-t^4)^4(1-t^6)^6
}
\\

\midrule

\multicolumn{2}{c}{\textbf{Incorrect Magnetic Quiver}}
&
\multicolumn{1}{c}{\(\mathrm{HS}\left[\mathcal{C}\right]\)}
\\

\multicolumn{2}{c}{%
  \begin{tikzpicture}[
    baseline=(current bounding box.center),
    x=1.0cm,
    y=0.85cm
  ]
    \node[gauger,   label=below:{$D_2$}] (a1) at (0,0) {};
    \node[gaugeb,   label=below:{$C_2$}] (a2) at (1,0) {};
    \node[flavourr, label=above:{$D_3$}] (a3) at (1,1) {};

    \draw (a2) -- (a3);
    \draw[transform canvas={yshift= 1.3pt}] (a1) -- (a2);
    \draw[transform canvas={yshift=-1.3pt}] (a1) -- (a2);
    \draw (0.4,0.2) -- (0.6,0) -- (0.4,-0.2);
  \end{tikzpicture}
}
&
\HSfrac{
  &1-t^2+4t^4+t^6+10t^8+5t^{10}\\[-0.2ex]
  &{}+21t^{12}+10t^{14}+\cdots+t^{28}
}{
  (1-t^2)^2(1-t^4)^3(1-t^6)^2(1-t^8)
}
\\

\HSsep
\multicolumn{2}{c}{\textbf{Correct Magnetic Quiver}}
&
\multicolumn{1}{c}{\(\mathrm{HS}\left[\mathcal{C}\right]\)}
\\

\multicolumn{2}{c}{%
  \begin{tikzpicture}[
    baseline=(current bounding box.center),
    x=1.0cm,
    y=0.85cm
  ]
    \node[gauger,   label=below:{$\orm(4)$}] (b1) at (0,0) {};
    \node[gaugeb,   label=below:{$C_2$}]     (b2) at (1,0) {};
    \node[flavourr, label=above:{$D_3$}]     (b3) at (1,1) {};

    \draw (b2) -- (b3);
    \draw[transform canvas={yshift= 1.3pt}] (b1) -- (b2);
    \draw[transform canvas={yshift=-1.3pt}] (b1) -- (b2);
    \draw (0.4,0.2) -- (0.6,0) -- (0.4,-0.2);
  \end{tikzpicture}
}
&
\HSfrac{
  &1-t^2+4t^4-2t^6+12t^8-3t^{10}\\[-0.2ex]
  &{}+22t^{12}-2t^{14}+30t^{16}+\cdots+t^{32}
}{
  (1-t^2)^2(1-t^4)^2(1-t^6)^2(1-t^8)^2
}
\\

\bottomrule
\end{tabularx}
\caption{Hilbert series for the different choices of \((G_1,G_2)\) for the $G_1\times G_2$ gauging of the $\sorm(6)$ flavour symmetry of $T\left[\sorm(6)\right]$. The only match occurs for the $\orm(2)\times\orm(2)$ gauging of the electric theory and the $\orm(4)$ choice in the magnetic theory. All other choices are inconsistent.}
\label{tab:appendix2}
\end{table}

\begin{landscape}

\begin{table}[tbp]
\centering

\small
\renewcommand{\arraystretch}{1.35}
\renewcommand{\tabularxcolumn}[1]{m{#1}}
\setlength{\tabcolsep}{5pt}

\begin{tabularx}{\linewidth}{@{} C{0.24\linewidth} Y @{}}
\toprule

\multicolumn{2}{c}{\textbf{Electric Quiver}} \\[0.4em]

\multicolumn{2}{c}{%
  \begin{tikzpicture}[
    baseline=(current bounding box.center),
    x=1.0cm,
    y=0.9cm
  ]
    \node[gauger,   label=below:{$D_1$}] (q1) at (0,0) {};
    \node[gaugeb,   label=below:{$C_1$}] (q2) at (1,0) {};
    \node[gauger,   label=below:{$D_2$}] (q3) at (2,0) {};
    \node[gaugeb,   label=below:{$C_2$}] (q4) at (3,0) {};
    \node[gauger,   label=below:{$D_3$}] (q5) at (4,0) {};
    \node[gaugeb,   label=below:{$C_3$}] (q6) at (5,0) {};
    \node[gauger,   label=below:{$G_1$}] (q7) at (6,0.5) {};
    \node[gauger,   label=below:{$G_2$}] (q8) at (6,-0.5) {};
    \node[flavourr, label=left:{$D_2$}]  (q9) at (5,1) {};

    \draw (q1) -- (q2) -- (q3) -- (q4) -- (q5) -- (q6) -- (q9);
    \draw (q7) -- (q6) -- (q8);
  \end{tikzpicture}
} \\[0.7em]

\midrule

\(\left[G_1,G_2\right]\)
&
\multicolumn{1}{c}{\(\mathrm{HS}[\mathcal H]\)}
\\

\midrule

\(\bigl[\orm(2),\orm(2)\bigr]\)
&
\HSfrac{
  &1+2t^2+17t^4+42t^6+188t^8+490t^{10}
  +1569t^{12}+3746t^{14}+9419t^{16}+19668t^{18}+40611t^{20}+73286t^{22}
  +127977t^{24}\\[-0.2ex]
  &{}+200910t^{26}+303149t^{28}+417820t^{30}+552236t^{32}+673004t^{34}
  +786318t^{36}+851352t^{38}+883750t^{40}+\cdots+t^{80}
}{
  (1-t^2)^4(1-t^4)^6(1-t^6)^6(1-t^8)^4
}
\\

\HSsep

\(\bigl[\sorm(2),\orm(2)\bigr]\)
&
\HSfrac{
  &1+2t^2+23t^4+70t^6+323t^8+936t^{10}
  +2989t^{12}+7454t^{14}+18584t^{16}+39480t^{18}+80915t^{20}+147216t^{22}
  +255689t^{24}\\[-0.2ex]
  &{}+403220t^{26}+605829t^{28}+837448t^{30}+1103175t^{32}+1347648t^{34}
  +1570228t^{36}+1703886t^{38}+1764488t^{40}+\cdots+t^{80}
}{
  (1-t^2)^4(1-t^4)^6(1-t^6)^6(1-t^8)^4
}
\\

\HSsep

\(\bigl[\sorm(2),\sorm(2)\bigr]\)
&
\HSfrac{
  &(1-t^2+t^4)\bigl(1+3t^2+30t^4+139t^6+594t^8
  +1995t^{10}+6051t^{12}+15650t^{14}+36450t^{16}
  +75352t^{18}+141860t^{20}\\[-0.2ex]
  &\phantom{(1-t^2+t^4)\bigl(}{}+241680t^{22}+378003t^{24}+540685t^{26}+713928t^{28}
  +867619t^{30}+976523t^{32}+1014234t^{34}+\cdots+t^{68}\bigr)
}{
  (1-t^2)^4(1-t^4)^8(1-t^6)^6(1-t^8)^2
}
\\

\midrule

\textbf{Incorrect Magnetic Quiver}
&
\multicolumn{1}{c}{\(\mathrm{HS}\left[\mathcal{C}\right]\)}
\\

\begin{tikzpicture}[
  baseline=(current bounding box.center),
  x=1.0cm,
  y=0.85cm
]
  \node[gauger,   label=below:{$D_2$}] (b1) at (0,0) {};
  \node[gaugeb,   label=below:{$C_2$}] (b2) at (1,0) {};
  \node[gauger,   label=below:{$D_3$}] (b3) at (2,0) {};
  \node[gaugeb,   label=below:{$C_3$}] (b4) at (3,0) {};
  \node[flavourr, label=above:{$D_4$}] (b5) at (3,1) {};

  \draw (b2) -- (b3) -- (b4) -- (b5);
  \draw[transform canvas={yshift= 1.3pt}] (b1) -- (b2);
  \draw[transform canvas={yshift=-1.3pt}] (b1) -- (b2);
  \draw (0.4,0.2) -- (0.6,0) -- (0.4,-0.2);
\end{tikzpicture}
&
\HSfrac{
  &(1-t^2+t^4)\bigl(1+3t^2+18t^4+71t^6+292t^8
  +981t^{10}+3003t^{12}+7816t^{14}+18252t^{16}
  +37744t^{18}+71002t^{20}\\[-0.2ex]
  &\phantom{(1-t^2+t^4)\bigl(}{}+120858t^{22}+188923t^{24}+270195t^{26}+356838t^{28}
  +433799t^{30}+488391t^{32}+507306t^{34}+\cdots+t^{68}\bigr)
}{
  (1-t^2)^4(1-t^4)^8(1-t^6)^6(1-t^8)^2
}
\\
\midrule

\textbf{Correct Magnetic Quiver}
&
\multicolumn{1}{c}{\(\mathrm{HS}\left[\mathcal{C}\right]\)}
\\
\begin{tikzpicture}[
  baseline=(current bounding box.center),
  x=1.0cm,
  y=0.85cm
]
  \node[gauger,   label=below:{$[D_2$}] (a1) at (0,0) {};
  \node[gaugeb,   label=below:{$C_2$}]  (a2) at (1,0) {};
  \node[gauger,   label=below:{$D_3]$}] (a3) at (2,0) {};
  \node[gaugeb,   label=below:{$C_3$}]  (a4) at (3,0) {};
  \node[flavourr, label=above:{$D_4$}]  (a5) at (3,1) {};

  \draw (a2) -- (a3) -- (a4) -- (a5);
  \draw[transform canvas={yshift= 1.3pt}] (a1) -- (a2);
  \draw[transform canvas={yshift=-1.3pt}] (a1) -- (a2);
  \draw (0.4,0.2) -- (0.6,0) -- (0.4,-0.2);
\end{tikzpicture}
&
\HSfrac{
  &1+2t^2+17t^4+42t^6+188t^8+490t^{10}
  +1569t^{12}+3746t^{14}+9419t^{16}+19668t^{18}+40611t^{20}+73286t^{22}
  +127977t^{24}\\[-0.2ex]
  &{}+200910t^{26}+303149t^{28}+417820t^{30}+552236t^{32}+673004t^{34}
  +786318t^{36}+851352t^{38}+883750t^{40}+\cdots+t^{80}
}{
  (1-t^2)^4(1-t^4)^6(1-t^6)^6(1-t^8)^4
}
\\

\bottomrule
\end{tabularx}
\caption{Hilbert series for the different choices of \((G_1,G_2)\) for the $G_1\times G_2$ gauging of the $\sorm(8)$ flavour symmetry of the $T\left[\sorm(8)\right]$ theory. Only $\orm(2)\times\orm(2)$ and the diagonal gauging of the magnetic quiver are a consistent electric-magnetic pair, as shown by the Hilbert series. Further details on diagonal gauging of Coulomb branch magnetic lattices can be found in \cite{Cabrera:2017ucb}.}
\label{tab:appendix3}
\end{table}
\end{landscape}
\section{``Gauging'' with non-simply laced edges}
\label{sec:NSLGauge}
The construction in the main body of this paper uses an $\mathrm{ON}^-$ plane which has the effect of introducing a bifurcation of gauge nodes in the shape of a $D$-type Dynkin diagram in the electric quiver. However there are three other types of orientifold plane that one can use, the $\mathrm{ON}^+, \widetilde{\mathrm{ON}^+},$ and the $\widetilde{\mathrm{ON}^-}$ planes. The first two correspond to $C$-type endings (or specifically a $C'$-type ending in the case of $\widetilde{\mathrm{ON}^+}$) and the last to a $B$-type ending in the electric quiver which add non-simply laced edges.

Although non-simply laced edges have a clear interpretation on the magnetic side in the context of the monopole formula, there is to date no such interpretation on the electric side. Nevertheless one can still follow the steps taken in this work to ``derive rules'' which correspond to ``gauging'' with non-simply laced edges.

It is important to clarify that this Appendix is largely for completeness and there are no known Higgs branch Hilbert series methods to realise ``gauging'' with non-simply laced edges. Despite this, we press on in the hope that non-simply laced edges in electric theories will be better understood in the future and the techniques outlined in this Appendix will find natural use or even shed light on the nature of non-simply laced edges in electric theories by studying the ``dual'' operation on the magnetic quiver.

\subsection{Gauging a Long $\urm(2n)$ Factor}
The starting point for this construction is the $\urm(N)$ SQCD with $2N+k$ flavours. The meaning of gauging a long $\urm(2n)$ factor refers to the following operation \begin{equation}
    \raisebox{-0.5\height}{
\begin{tikzpicture}

\node[gauge, label=below:$N$] (L1) at (0,0){};
\node[flavour, label=above:$2N+k$] (L2) at (0,1){};

\draw[-] (L1)--(L2);

\draw[->, very thick]
    (1.0,0) -- (3.0,0);

\node[gauge, label=below:$2n$] (2n) at (4,0){};
\node[gauge, label=below:$N$] (N) at (5,0){};
\node[flavour, label=above:$2N+k-2n$] (flav) at (5,1){};

\draw[transform canvas={yshift=1.3pt}](2n)--(N);
\draw[transform canvas={yshift=-1.3pt}](2n)--(N);
\draw[-] (4.4,-0.2)--(4.6,0)--(4.4,0.2) (N)--(flav);
\end{tikzpicture}
}
\end{equation}
The effect of gauging a long $\urm(2n)$ factor is summarised in \Figref{fig:ONPlus_QQS}. Note that the orientifold requires the rank of the gauged node to be even. Here the S-dual orientifold to the $\mathrm{ON}^+$ plane is the $\mathrm{O5}^+$ plane which gives a $D$-type gauge node in the magnetic quiver.
\begin{figure}[h!]
\centering
    \begin{subfigure}{0.49\textwidth}
    \centering
    \begin{tikzpicture}
    \node[gauge, label=below:$2n$] (2n) at (4,0){};
\node[gauge, label=below:$N$] (N) at (5,0){};
\node[flavour, label=above:$2N+k-2n$] (flav) at (5,1){};
\draw[transform canvas={yshift=1.3pt}](2n)--(N);
\draw[transform canvas={yshift=-1.3pt}](2n)--(N);
\draw[-] (4.4,-0.2)--(4.6,0)--(4.4,0.2) (N)--(flav);
    \end{tikzpicture}
\caption{}
\label{fig:ONPlus_deriv_quiv}
\end{subfigure}
\begin{subfigure}{0.49\textwidth}
    \centering
    \begin{tikzpicture}
         \draw[-](0,0)--(0,2) (-2,0)--(-2,2);
        \draw[dashed] (-4,0)--(-4,2);
        \draw[-](0,1)--(-2,1)node[midway, below]{$N$}-- (-4,1)node[midway, below]{$2n$};
        
         \node[D5] at (-0.5,1.75){};
        \node[D5] at (-1.5,1.75){};
        \node at (-1,1.75){$\cdots$};
        \draw [decorate, decoration = {brace, raise=10pt, 
         amplitude=5pt}] (-1.5,1.75) --  (-0.5,1.75) node[pos=0.5,above=15pt,black]{$2N+k-2n$};  
        \node at (-4,2.5) {$\text{ON}^{+}$};
    \end{tikzpicture}
\caption{}
\label{fig:ONPlus_deriv_brane}
\end{subfigure}
\begin{subfigure}{0.9\textwidth}
\centering
    \begin{tikzpicture}
    \node[gauger, label=below:$D_n$] (1l) at (0,0){};
    \node[gauge, label=below:$2n+1$] (2l) at (1,0){};
    \node[gauge, label=above:$2n+2$] (3l) at (2,0){};
    \node[] (cdotsl) at (3,0){$\cdots$};
    \node[gauge, label=below:$N-1$] (n1l) at (4,0){};
    \node[gauge, label=below:$N$] (n2l) at (5,0){};
    \node[] (cdotsm) at (6,0){$\cdots$};
    \node[gauge, label=below:$N$] (n1r) at (7,0){};
    \node[gauge, label=below:$N-1$] (n2r) at (8,0){};
    \node[] (cdotsr) at (9,0){$\cdots$};
    \node[gauge, label=below:$2$] (2r) at (10,0){};
    \node[gauge, label=below:$1$] (1r) at (11,0){};
    \node[flavour, label=above:$1$] (1fl) at (5,1){};
    \node[flavour, label=above:$1$] (1fr) at (7,1){};
    \draw[-] (1l)--(2l)--(3l)--(cdotsl)--(n1l)--(n2l)--(cdotsm)--(n1r)--(n2r)--(cdotsr)--(2r)--(1r) (1fl)--(n2l) (1fr)--(n1r);
    \draw [decorate, decoration = {brace, raise=15pt, amplitude=5pt}] (7.25,0) --  (4.75,0) node[pos=0.5,below=20pt,black]{$k+1$};
    \end{tikzpicture}
    \caption{}
    \label{quiv:ONPlus_QQS_mirror}
\end{subfigure}
\begin{subfigure}[t]{0.95\textwidth}
\centering
\begin{tikzpicture}
    \draw[-] (0,-1)--(0,1) (-1,-1)--(-1,1) (-2,-1)--(-2,1) (-3,-1)--(-3,1) (-4,-1)--(-4,1) (-5,-1)--(-5,1) (-6,-1)--(-6,1) (-7,-1)--(-7,1) (-8,-1)--(-8,1) (-9,-1)--(-9,1) (-10,-1)--(-10,1) (-11.5,-1)--(-11.5,1) (-13,-1)--(-13,1);
    \node[O5circle, label=above:$\mathrm{O5}^+$]at  (-14.5,0){};
    \draw[-] (0,0)--(-1,0)node[midway, above]{$1$}--(-2,0)node[midway, above]{$2$};
    \draw[-] (-2,0)--(-2.2,0) (-2.8,0)--(-3,0);
    \node at (-2.5,0){$\cdots$};
    \draw[-] (-3,0)--(-4,0)node[midway, above]{$N$};
    \node[D5] at (-3.5,0.75){};
    \draw[-] (-4,0)--(-5,0)node[midway, above]{$N$};
    \draw[-] (-5,0)--(-5.2,0) (-5.8,0)--(-6,0);
    \node at (-5.5,0){$\cdots$};
    \draw[-] (-6,0)--(-7,0)node[midway,above]{$N$}--(-8,0)node[midway,above]{$N$}--(-9,0)node[midway, above]{$N-1$};
    \node[D5] at (-7.5,0.75){};
    \draw[-] (-9,0)--(-9.2,0) (-9.8,0)--(-10,0);
    \node at (-9.5,0){$\cdots$};
    \draw[-] (-10,0)--(-11.5,0)node[midway, above]{$2n+2$}--(-13,0)node[midway,above]{$2n+1$}--(-14.3,0)node[midway, above]{$2n$};
    \draw [decorate, decoration = {brace, raise=5pt, amplitude=5pt}] (-8,1) --  (-3,1) node[pos=0.5,above=10pt,black]{$k+2$};
    \end{tikzpicture}
    \caption{}
    \label{fig:ONPlusMirrorBrane}
\end{subfigure}
\caption{\Figref{fig:ONPlus_deriv_quiv} results from the gauging a long factor of $\urm(2n)$ for the SQCD in \eqref{quiv:SQCD_3dMirror}. It has a brane realisation using D3 branes, D5 branes, NS5 branes and an $\textrm{ON}^{+}$ orientifold plane in \Figref{fig:ONPlus_deriv_brane}, which under S-duality is exchanged for \Figref{fig:ONPlusMirrorBrane} and the 3d-mirror theory in \Figref{quiv:ONPlus_QQS_mirror}. Note that the S-duality operation between \Figref{fig:ONPlus_deriv_brane} and \Figref{fig:ONPlusMirrorBrane} exchanges the $\textrm{ON}^{+}$ plane for an $\textrm{O5}^{+}$ plane.}
\label{fig:ONPlus_QQS}
\end{figure}

Comparing the $3d$ mirror quiver of $\urm(N)$ with $2N+k$ flavours with \Quiver{quiv:ONPlus_QQS_mirror} suggests that the magnetic realisation of gauging a long $\urm(2n)$ factor corresponds to a quotient quiver subtraction of \Quiver{quiv:ONPlus_quotient_quiver} followed by an \textit{orthogonalisation} of a gauge group.

\begin{equation}
    \raisebox{-0.5\height}{\begin{tikzpicture}
        \node[] (A) at (-3,-0.1){\text{Long}\;$\urm(2n)\; \text{Quotient Quiver}$};
        \node[gauge, label=below:$1$] (1) at (1,0){};
        \node[gauge, label=below:$2$] (2) at (2,0){};
        \node[] (cdots) at (3,0){$\cdots$};
        \node[gauge, label=below:$2n-1$] (nm1) at (4,0){};
        \draw[-] (1)--(2)--(cdots)--(nm1);
        \end{tikzpicture}}
\label{quiv:ONPlus_quotient_quiver}
\end{equation}

\paragraph{Conjecture of the Rules}
The quotient quiver subtraction algorithm for gauging a long $\urm(2n)$ factor is conjectured to be the following.
\begin{enumerate}
    \item Take a framed target quiver with a long leg of unitary gauge nodes of length at least $2n$, shown below.
    \begin{equation}
        \raisebox{-0.5\height}{\begin{tikzpicture}
        \node[gauge, label=below:$1$] (1) at (0,0){};
        \node[gauge, label=below:$2$] (2) at (1,0){};
        \node[] (cdots) (cdots) at (2,0){$\cdots$};
        \node[gauge, label=below:$2n-1$] (2r) at (3,0){};
        \node[gauge, label=below:$2n$] (2rp1) at (4,0){};
        \node[gauge] (Q) at (5,0){$Q$};
        \draw[-] (1)--(2)--(cdots)--(2r)--(2rp1);
        \draw[teal,very thick] (2rp1)--(Q);
    \end{tikzpicture}}
    \end{equation}
    \item Align the quotient quiver \eqref{quiv:ONPlus_quotient_quiver} along the long leg of the target quiver and perform a rank-wise subtraction.
    \begin{equation}\raisebox{-0.5\height}{\begin{tikzpicture}
    \node[font=\bfseries] at (2.5,1) {Subtraction};
    \node[gauge, label=below:$1$] (1) at (0,0){};
    \node[gauge, label=below:$2$] (2) at (1,0){};
    \node (cdots) at (2,0) {$\cdots$};
    \node[gauge, label=below:$2n-1$] (2r) at (3,0){};
    \node[gauge, label=below:$2n$] (2rp1) at (4,0){};
    \node[gauge] (Q) at (5,0) {$Q$};
    \draw[-] (1)--(2)--(cdots)--(2r)--(2rp1);
    \draw[teal, very thick] (2rp1)--(Q);

    \node[font=\Large] at (-0.8,-1.35) {$-$};
    \node[gauge, label=below:$1$] (1sub) at (0,-1){};
    \node[gauge, label=below:$2$] (2sub) at (1,-1){};
    \node (cdotssub) at (2,-1) {$\cdots$};
    \node[gauge, label=below:$2n-1$] (2rsub) at (3,-1){};
    \draw[-] (1sub)--(2sub)--(cdotssub)--(2rsub);
    
    \draw (-0.15,-1.8) -- (5.15,-1.8);
    
    \node[font=\Large] at (-0.8,-2.5) {$=$};
    \node[gauge, label=below:$2n$] (2rp1r) at (1,-2.5){};
    \node[gauge] (Qr) at (2,-2.5) {$Q$};
    \draw[teal, very thick] (2rp1r)--(Qr);
    \end{tikzpicture}}
    \end{equation}
    \item Orthogonalise the $\urm(2n)\rightarrow\sorm(2n)$. This removes a further $n$ moduli.

    \begin{equation}\raisebox{-0.5\height}{\begin{tikzpicture}
    \node[font=\bfseries] at (5.25,1) {Orthogonalisation};
    \node[gauge, label=below:$2n$] (2rp1r) at (1,0){};
    \node[gauge] (Qr) at (2,0) {$Q$};
    \draw[teal, very thick] (2rp1r)--(Qr);

    \draw[->, thick]
    (2.5,0) -- (8.0,0)
    node[midway, fill=white, inner sep=3pt, align=center]
    {\small\textit{Replace}
     $\urm(2n)\;\text{\emph{with}}\;\sorm(2n)$};

    \node[gauger, label=below:$D_n$] (Cnr) at (9.5,0){};
    \node[gauge] (Qrr) at (10.5,0){$Q$};
    \draw[teal, very thick] (Cnr)--(Qrr);
\end{tikzpicture}}
\end{equation}
\end{enumerate}
In terms of the change in dimensions on the magnetic side, the combined action of the subtraction of \Quiver{quiv:ONPlus_quotient_quiver} and the orthogonalisation of a $\urm(2n)\rightarrow\sorm(2n)$ corresponds to a reduction in the Coulomb branch dimension by $2n^2$. This matches the reduction in dimension expected from gauging $\urm(n)\times\urm(n)$ and not that from gauging a $\urm(2n)$ with bifundamental matter, where there is a difference by a factor of two.


\subsection{Gauging a Long $\urm(2n+1)$ Factor}
This case is largely similar to the previous case except that it uses an $\widetilde{\mathrm{ON}^+}$. The starting point for this construction is the $\urm(N)$ SQCD with $2N+k$ flavours. The meaning of gauging a long $\urm(2n+1)$ factor refers to the following operation \begin{equation}
    \raisebox{-0.5\height}{
\begin{tikzpicture}

\node[gauge, label=below:$N$] (L1) at (0,0){};
\node[flavour, label=above:$2N+k$] (L2) at (0,1){};

\draw[-] (L1)--(L2);

\draw[->, very thick]
    (1.0,0) -- (3.0,0);

\node[gauge, label=below:$2n+1$] (2n) at (4,0){};
\node[gauge, label=below:$N$] (N) at (5,0){};
\node[flavour, label=above:$2N+k-2n-1$] (flav) at (5,1){};

\draw[transform canvas={yshift=1.3pt}](2n)--(N);
\draw[transform canvas={yshift=-1.3pt}](2n)--(N);
\draw[-] (4.4,-0.2)--(4.6,0)--(4.4,0.2) (N)--(flav);
\end{tikzpicture}
}
\end{equation}
The effect of gauging a long $\urm(2n+1)$ factor is summarised in \Figref{fig:ONPlusTilde_QQS}. Here the S-dual orientifold to the $\widetilde{\mathrm{ON}^+}$ plane is the $\widetilde{\mathrm{O5}^+}$ plane which gives a $B$-type gauge node in the magnetic quiver. Note that this $\widetilde{\mathrm{O5}^+}$ requires the rank of the gauged factor to be odd.
\begin{figure}[h!]
\centering
    \begin{subfigure}{0.49\textwidth}
    \centering
    \begin{tikzpicture}
    \node[gauge, label=below:$2n+1$] (2n) at (4,0){};
\node[gauge, label=below:$N$] (N) at (5,0){};
\node[flavour, label=above:$2N+k-2n-1$] (flav) at (5,1){};
\draw[transform canvas={yshift=1.3pt}](2n)--(N);
\draw[transform canvas={yshift=-1.3pt}](2n)--(N);
\draw[-] (4.4,-0.2)--(4.6,0)--(4.4,0.2) (N)--(flav);
    \end{tikzpicture}
\caption{}
\label{fig:ONPlusTilde_deriv_quiv}
\end{subfigure}
\begin{subfigure}{0.49\textwidth}
    \centering
    \begin{tikzpicture}
         \draw[-](0,0)--(0,2) (-2,0)--(-2,2);
        \draw[dashed] (-4,0)--(-4,2);
        \draw[-](0,1)--(-2,1)node[midway, below]{$N$}-- (-4,1)node[midway, below]{$2n+1$};
        
         \node[D5] at (-0.5,1.75){};
        \node[D5] at (-1.5,1.75){};
        \node at (-1,1.75){$\cdots$};
        \draw [decorate, decoration = {brace, raise=10pt, 
         amplitude=5pt}] (-1.5,1.75) --  (-0.5,1.75) node[pos=0.5,above=15pt,black]{$2N+k-2n-1$};  
        \node at (-4,2.5) {$\widetilde{\text{ON}^{+}}$};
    \end{tikzpicture}
\caption{}
\label{fig:ONPlusTilde_deriv_brane}
\end{subfigure}
\begin{subfigure}{0.9\textwidth}
\centering
    \begin{tikzpicture}
    \node[gauger, label=below:$B_n$] (1l) at (0,0){};
    \node[gauge, label=below:$2n+2$] (2l) at (1,0){};
    \node[gauge, label=above:$2n+3$] (3l) at (2,0){};
    \node[] (cdotsl) at (3,0){$\cdots$};
    \node[gauge, label=below:$N-1$] (n1l) at (4,0){};
    \node[gauge, label=below:$N$] (n2l) at (5,0){};
    \node[] (cdotsm) at (6,0){$\cdots$};
    \node[gauge, label=below:$N$] (n1r) at (7,0){};
    \node[gauge, label=below:$N-1$] (n2r) at (8,0){};
    \node[] (cdotsr) at (9,0){$\cdots$};
    \node[gauge, label=below:$2$] (2r) at (10,0){};
    \node[gauge, label=below:$1$] (1r) at (11,0){};
    \node[flavour, label=above:$1$] (1fl) at (5,1){};
    \node[flavour, label=above:$1$] (1fr) at (7,1){};
    \draw[-] (1l)--(2l)--(3l)--(cdotsl)--(n1l)--(n2l)--(cdotsm)--(n1r)--(n2r)--(cdotsr)--(2r)--(1r) (1fl)--(n2l) (1fr)--(n1r);
    \draw [decorate, decoration = {brace, raise=15pt, amplitude=5pt}] (7.25,0) --  (4.75,0) node[pos=0.5,below=20pt,black]{$k+1$};
    \end{tikzpicture}
    \caption{}
    \label{quiv:ONPlusTilde_QQS_mirror}
\end{subfigure}
\begin{subfigure}[t]{0.95\textwidth}
\centering
\begin{tikzpicture}
    \draw[-] (0,-1)--(0,1) (-1,-1)--(-1,1) (-2,-1)--(-2,1) (-3,-1)--(-3,1) (-4,-1)--(-4,1) (-5,-1)--(-5,1) (-6,-1)--(-6,1) (-7,-1)--(-7,1) (-8,-1)--(-8,1) (-9,-1)--(-9,1) (-10,-1)--(-10,1) (-11.5,-1)--(-11.5,1) (-13,-1)--(-13,1);
    \node[O5circle, label=above:$\widetilde{\mathrm{O5}^+}$]at  (-14.5,0){};
    \draw[-] (0,0)--(-1,0)node[midway, above]{$1$}--(-2,0)node[midway, above]{$2$};
    \draw[-] (-2,0)--(-2.2,0) (-2.8,0)--(-3,0);
    \node at (-2.5,0){$\cdots$};
    \draw[-] (-3,0)--(-4,0)node[midway, above]{$N$};
    \node[D5] at (-3.5,0.75){};
    \draw[-] (-4,0)--(-5,0)node[midway, above]{$N$};
    \draw[-] (-5,0)--(-5.2,0) (-5.8,0)--(-6,0);
    \node at (-5.5,0){$\cdots$};
    \draw[-] (-6,0)--(-7,0)node[midway,above]{$N$}--(-8,0)node[midway,above]{$N$}--(-9,0)node[midway, above]{$N-1$};
    \node[D5] at (-7.5,0.75){};
    \draw[-] (-9,0)--(-9.2,0) (-9.8,0)--(-10,0);
    \node at (-9.5,0){$\cdots$};
    \draw[-] (-10,0)--(-11.5,0)node[midway, above]{$2n+3$}--(-13,0)node[midway,above]{$2n+2$}--(-14.3,0)node[midway, above]{$2n+1$};
    \draw [decorate, decoration = {brace, raise=5pt, amplitude=5pt}] (-8,1) --  (-3,1) node[pos=0.5,above=10pt,black]{$k+2$};
    \end{tikzpicture}
    \caption{}
    \label{fig:ONPlusTildeMirrorBrane}
\end{subfigure}
\caption{\Figref{fig:ONPlusTilde_deriv_quiv} results from the gauging of a long $\urm(2n+1)$ factor of the SQCD in \eqref{quiv:SQCD_3dMirror}. It has a brane realisation using D3 branes, D5 branes, NS5 branes and an $\widetilde{\textrm{ON}^{+}}$ orientifold plane in \Figref{fig:ONPlusTilde_deriv_brane}, which under S-duality is exchanged for \Figref{fig:ONPlusTildeMirrorBrane} and the 3d-mirror theory in \Figref{quiv:ONPlusTilde_QQS_mirror}. Note that the S-duality operation between \Figref{fig:ONPlusTilde_deriv_brane} and \Figref{fig:ONPlusTildeMirrorBrane} exchanges the $\widetilde{\textrm{ON}^{+}}$ plane for an $\widetilde{\textrm{O5}^{+}}$ plane.}
\label{fig:ONPlusTilde_QQS}
\end{figure}

Comparing the $3d$ mirror quiver of $\urm(N)$ with $2N+k$ flavours with \Quiver{quiv:ONPlus_QQS_mirror} suggests that the magnetic realisation of gauging a long $\urm(2n+1)$ factor corresponds to a quotient quiver subtraction of \Quiver{quiv:ONPlus_quotient_quiver} followed by an \textit{orthogonalisation} of a gauge group.

\begin{equation}
    \raisebox{-0.5\height}{\begin{tikzpicture}
        \node[] (A) at (-3,-0.1){\text{Long}\;$\urm(2n+1)\; \text{Quotient Quiver}$};
        \node[gauge, label=below:$1$] (1) at (1,0){};
        \node[gauge, label=below:$2$] (2) at (2,0){};
        \node[] (cdots) at (3,0){$\cdots$};
        \node[gauge, label=below:$2n$] (nm1) at (4,0){};
        \draw[-] (1)--(2)--(cdots)--(nm1);
        \end{tikzpicture}}
\label{quiv:ONPlusTilde_quotient_quiver}
\end{equation}

\paragraph{Conjecture of the Rules}
The quotient quiver subtraction algorithm for gauging a long $\urm(2n+1)$ factor is conjectured to be the following.
\begin{enumerate}
    \item Take a framed target quiver with a long leg of unitary gauge nodes of length at least $2n+1$, shown below.
    \begin{equation}
        \raisebox{-0.5\height}{\begin{tikzpicture}
        \node[gauge, label=below:$1$] (1) at (0,0){};
        \node[gauge, label=below:$2$] (2) at (1,0){};
        \node[] (cdots) (cdots) at (2,0){$\cdots$};
        \node[gauge, label=below:$2n$] (2r) at (3,0){};
        \node[gauge, label=below:$2n+1$] (2rp1) at (4,0){};
        \node[gauge] (Q) at (5,0){$Q$};
        \draw[-] (1)--(2)--(cdots)--(2r)--(2rp1);
        \draw[teal,very thick] (2rp1)--(Q);
    \end{tikzpicture}}
    \end{equation}
    \item Align the quotient quiver \eqref{quiv:ONPlusTilde_quotient_quiver} along the long leg of the target quiver and perform a rank-wise subtraction.
    \begin{equation}\raisebox{-0.5\height}{\begin{tikzpicture}
    \node[font=\bfseries] at (2.5,1) {Subtraction};
    \node[gauge, label=below:$1$] (1) at (0,0){};
    \node[gauge, label=below:$2$] (2) at (1,0){};
    \node (cdots) at (2,0) {$\cdots$};
    \node[gauge, label=below:$2n$] (2r) at (3,0){};
    \node[gauge, label=below:$2n+1$] (2rp1) at (4,0){};
    \node[gauge] (Q) at (5,0) {$Q$};
    \draw[-] (1)--(2)--(cdots)--(2r)--(2rp1);
    \draw[teal, very thick] (2rp1)--(Q);

    \node[font=\Large] at (-0.8,-1.35) {$-$};
    \node[gauge, label=below:$1$] (1sub) at (0,-1){};
    \node[gauge, label=below:$2$] (2sub) at (1,-1){};
    \node (cdotssub) at (2,-1) {$\cdots$};
    \node[gauge, label=below:$2n$] (2rsub) at (3,-1){};
    \draw[-] (1sub)--(2sub)--(cdotssub)--(2rsub);
    
    \draw (-0.15,-1.8) -- (5.15,-1.8);
    
    \node[font=\Large] at (-0.8,-2.5) {$=$};
    \node[gauge, label=below:$2n+1$] (2rp1r) at (1,-2.5){};
    \node[gauge] (Qr) at (2,-2.5) {$Q$};
    \draw[teal, very thick] (2rp1r)--(Qr);
    \end{tikzpicture}}
    \end{equation}
    \item Orthogonalise the $\urm(2n+1)\rightarrow\sorm(2n+1)$. This removes a further $n+1$ moduli.

    \begin{equation}\raisebox{-0.5\height}{\begin{tikzpicture}
    \node[font=\bfseries] at (5.25,1) {Orthogonalisation};
    \node[gauge, label=below:$2n+1$] (2rp1r) at (1,0){};
    \node[gauge] (Qr) at (2,0) {$Q$};
    \draw[teal, very thick] (2rp1r)--(Qr);

    \draw[->, thick]
    (2.5,0) -- (9.0,0)
    node[midway, fill=white, inner sep=3pt, align=center]
    {\small\textit{Replace}
     $\urm(2n+1)\;\text{\emph{with}}\;\sorm(2n+1)$};

    \node[gauger, label=below:$B_n$] (Cnr) at (9.5,0){};
    \node[gauge] (Qrr) at (10.5,0){$Q$};
    \draw[teal, very thick] (Cnr)--(Qrr);
\end{tikzpicture}}
\end{equation}
\end{enumerate}
In terms of the change in dimensions on the magnetic side, the combined action of the subtraction of \Quiver{quiv:ONPlusTilde_quotient_quiver} and the orthogonalisation of a $\urm(2n+1)\rightarrow\sorm(2n+1)$ corresponds to a reduction in the Coulomb branch dimension by $2n^2+2n+1$. This matches the reduction in dimension expected from gauging $\urm(n)\times\urm(n+1)$ and not that from gauging a $\urm(2n+1)$ with bifundamental matter, where there is a difference by $2n(2n+1)$.

\subsection{Gauging a Short $\urm(n)$ Factor}
This is the third and final case to consider. As before, the starting point for this construction is the $\urm(N)$ SQCD with $2N+k$ flavours. The meaning of gauging a short $\urm(n)$ factor refers to the following operation \begin{equation}
    \raisebox{-0.5\height}{
\begin{tikzpicture}

\node[gauge, label=below:$N$] (L1) at (0,0){};
\node[flavour, label=above:$2N+k$] (L2) at (0,1){};

\draw[-] (L1)--(L2);

\draw[->, very thick]
    (1.0,0) -- (3.0,0);

\node[gauge, label=below:$n$] (2n) at (4,0){};
\node[gauge, label=below:$N$] (N) at (5,0){};
\node[flavour, label=above:$2N+k-2n$] (flav) at (5,1){};

\draw[transform canvas={yshift=1.3pt}](2n)--(N);
\draw[transform canvas={yshift=-1.3pt}](2n)--(N);
\draw[-] (4.6,-0.2)--(4.4,0)--(4.6,0.2) (N)--(flav);
\end{tikzpicture}
}\label{eq:shortUn}
\end{equation}
The effect of gauging a short $\urm(n)$ factor is summarised in \Figref{fig:ONMinusTilde_QQS}. Here the S-dual orientifold to the $\widetilde{\mathrm{ON}^-}$ plane is the $\widetilde{\mathrm{O5}^-}$ plane which gives an $C$-type gauge node in the magnetic quiver with a half hypermultiplet.
\begin{figure}[h!]
\centering
    \begin{subfigure}{0.49\textwidth}
    \centering
    \begin{tikzpicture}
    \node[gauge, label=below:$n$] (2n) at (4,0){};
\node[gauge, label=below:$N$] (N) at (5,0){};
\node[flavour, label=above:$2N+k-2n$] (flav) at (5,1){};
\draw[transform canvas={yshift=1.3pt}](2n)--(N);
\draw[transform canvas={yshift=-1.3pt}](2n)--(N);
\draw[-] (4.6,-0.2)--(4.4,0)--(4.6,0.2) (N)--(flav);
    \end{tikzpicture}
\caption{}
\label{fig:ONMinusTilde_deriv_quiv}
\end{subfigure}
\begin{subfigure}{0.49\textwidth}
    \centering
    \begin{tikzpicture}
         \draw[-](0,0)--(0,2) (-2,0)--(-2,2);
        \draw[dashed] (-4,0)--(-4,2);
        \draw[-](0,1)--(-2,1)node[midway, below]{$N$}-- (-4,1)node[midway, below]{$2n$};
        
         \node[D5] at (-0.5,1.75){};
        \node[D5] at (-1.5,1.75){};
        \node at (-1,1.75){$\cdots$};
        \draw [decorate, decoration = {brace, raise=10pt, 
         amplitude=5pt}] (-1.5,1.75) --  (-0.5,1.75) node[pos=0.5,above=15pt,black]{$2N+k-2n$};  
        \node at (-4,2.5) {$\widetilde{\text{ON}^{-}}$};
    \end{tikzpicture}
\caption{}
\label{fig:ONMinusTilde_deriv_brane}
\end{subfigure}
\begin{subfigure}{0.9\textwidth}
\centering
    \begin{tikzpicture}
    \node[gaugeb, label=below:$C_n$] (1l) at (0,0){};
    \node[flavourr, label=above:$\orm(1)$] (f) at (0,1){};
    \node[gauge, label=below:$2n+1$] (2l) at (1,0){};
    \node[gauge, label=above:$2n+2$] (3l) at (2,0){};
    \node[] (cdotsl) at (3,0){$\cdots$};
    \node[gauge, label=below:$N-1$] (n1l) at (4,0){};
    \node[gauge, label=below:$N$] (n2l) at (5,0){};
    \node[] (cdotsm) at (6,0){$\cdots$};
    \node[gauge, label=below:$N$] (n1r) at (7,0){};
    \node[gauge, label=below:$N-1$] (n2r) at (8,0){};
    \node[] (cdotsr) at (9,0){$\cdots$};
    \node[gauge, label=below:$2$] (2r) at (10,0){};
    \node[gauge, label=below:$1$] (1r) at (11,0){};
    \node[flavour, label=above:$1$] (1fl) at (5,1){};
    \node[flavour, label=above:$1$] (1fr) at (7,1){};
    \draw[-] (f)--(1l)--(2l)--(3l)--(cdotsl)--(n1l)--(n2l)--(cdotsm)--(n1r)--(n2r)--(cdotsr)--(2r)--(1r) (1fl)--(n2l) (1fr)--(n1r);
    \draw [decorate, decoration = {brace, raise=15pt, amplitude=5pt}] (7.25,0) --  (4.75,0) node[pos=0.5,below=20pt,black]{$k+1$};
    \end{tikzpicture}
    \caption{}
    \label{quiv:ONMinusTilde_QQS_mirror}
\end{subfigure}
\begin{subfigure}[t]{0.95\textwidth}
\centering
\begin{tikzpicture}
    \draw[-] (0,-1)--(0,1) (-1,-1)--(-1,1) (-2,-1)--(-2,1) (-3,-1)--(-3,1) (-4,-1)--(-4,1) (-5,-1)--(-5,1) (-6,-1)--(-6,1) (-7,-1)--(-7,1) (-8,-1)--(-8,1) (-9,-1)--(-9,1) (-10,-1)--(-10,1) (-11.5,-1)--(-11.5,1) (-13,-1)--(-13,1);
    \node[O5circle, label=above:$\widetilde{\mathrm{O5}^-}$]at  (-14.5,0){};
    \draw[-] (0,0)--(-1,0)node[midway, above]{$1$}--(-2,0)node[midway, above]{$2$};
    \draw[-] (-2,0)--(-2.2,0) (-2.8,0)--(-3,0);
    \node at (-2.5,0){$\cdots$};
    \draw[-] (-3,0)--(-4,0)node[midway, above]{$N$};
    \node[D5] at (-3.5,0.75){};
    \draw[-] (-4,0)--(-5,0)node[midway, above]{$N$};
    \draw[-] (-5,0)--(-5.2,0) (-5.8,0)--(-6,0);
    \node at (-5.5,0){$\cdots$};
    \draw[-] (-6,0)--(-7,0)node[midway,above]{$N$}--(-8,0)node[midway,above]{$N$}--(-9,0)node[midway, above]{$N-1$};
    \node[D5] at (-7.5,0.75){};
    \draw[-] (-9,0)--(-9.2,0) (-9.8,0)--(-10,0);
    \node at (-9.5,0){$\cdots$};
    \draw[-] (-10,0)--(-11.5,0)node[midway, above]{$2n+2$}--(-13,0)node[midway,above]{$2n+1$}--(-14.3,0)node[midway, above]{$2n$};
    \draw [decorate, decoration = {brace, raise=5pt, amplitude=5pt}] (-8,1) --  (-3,1) node[pos=0.5,above=10pt,black]{$k+2$};
    \end{tikzpicture}
    \caption{}
    \label{fig:ONMinusTildeMirrorBrane}
\end{subfigure}
\caption{\Figref{fig:ONMinusTilde_deriv_quiv} results from the gauging of a short $\urm(n)$ factor of the SQCD in \eqref{quiv:SQCD_3dMirror}. It has a brane realisation using D3 branes, D5 branes, NS5 branes and an $\widetilde{\textrm{ON}^{-}}$ orientifold plane in \Figref{fig:ONMinusTilde_deriv_brane}, which under S-duality is exchanged for \Figref{fig:ONMinusTildeMirrorBrane} and the 3d-mirror theory in \Figref{quiv:ONMinusTilde_QQS_mirror}. Note that the S-duality operation between \Figref{fig:ONMinusTilde_deriv_brane} and \Figref{fig:ONMinusTildeMirrorBrane} exchanges the $\widetilde{\textrm{ON}^{-}}$ plane for an $\widetilde{\textrm{O5}^{-}}$ plane.}
\label{fig:ONMinusTilde_QQS}
\end{figure}

Comparing the $3d$ mirror quiver of $\urm(N)$ with $2N+k$ flavours with \Quiver{quiv:ONMinusTilde_QQS_mirror} suggests that the magnetic realisation of gauging a short $\urm(n)$ corresponds to a quotient quiver subtraction of \Quiver{quiv:ONMinusTilde_quotient_quiver} followed by an \textit{orthogonalisation} of a gauge group.

\begin{equation}
    \raisebox{-0.5\height}{\begin{tikzpicture}
        \node[] (A) at (-3,-0.1){\text{Short}\;$\urm(n)\; \text{Quotient Quiver}$};
        \node[gauge, label=below:$1$] (1) at (1,0){};
        \node[gauge, label=below:$2$] (2) at (2,0){};
        \node[] (cdots) at (3,0){$\cdots$};
        \node[gauge, label=below:$2n-1$] (nm1) at (4,0){};
        \draw[-] (1)--(2)--(cdots)--(nm1);
        \end{tikzpicture}}
\label{quiv:ONMinusTilde_quotient_quiver}
\end{equation}

\paragraph{Conjecture of the Rules}
The quotient quiver subtraction algorithm for gauging $\urm(n)$ on the short side is conjectured to be the following.
\begin{enumerate}
    \item Take a framed target quiver with a long leg of unitary gauge nodes of length at least $2n$, shown below.
    \begin{equation}
        \raisebox{-0.5\height}{\begin{tikzpicture}
        \node[gauge, label=below:$1$] (1) at (0,0){};
        \node[gauge, label=below:$2$] (2) at (1,0){};
        \node[] (cdots) (cdots) at (2,0){$\cdots$};
        \node[gauge, label=below:$2n-1$] (2r) at (3,0){};
        \node[gauge, label=below:$2n$] (2rp1) at (4,0){};
        \node[gauge] (Q) at (5,0){$Q$};
        \draw[-] (1)--(2)--(cdots)--(2r)--(2rp1);
        \draw[teal,very thick] (2rp1)--(Q);
    \end{tikzpicture}}
    \end{equation}
    \item Align the quotient quiver \eqref{quiv:ONPlusTilde_quotient_quiver} along the long leg of the target quiver and perform a rank-wise subtraction.
    \begin{equation}\raisebox{-0.5\height}{\begin{tikzpicture}
    \node[font=\bfseries] at (2.5,1) {Subtraction};
    \node[gauge, label=below:$1$] (1) at (0,0){};
    \node[gauge, label=below:$2$] (2) at (1,0){};
    \node (cdots) at (2,0) {$\cdots$};
    \node[gauge, label=below:$2n-1$] (2r) at (3,0){};
    \node[gauge, label=below:$2n$] (2rp1) at (4,0){};
    \node[gauge] (Q) at (5,0) {$Q$};
    \draw[-] (1)--(2)--(cdots)--(2r)--(2rp1);
    \draw[teal, very thick] (2rp1)--(Q);

    \node[font=\Large] at (-0.8,-1.35) {$-$};
    \node[gauge, label=below:$1$] (1sub) at (0,-1){};
    \node[gauge, label=below:$2$] (2sub) at (1,-1){};
    \node (cdotssub) at (2,-1) {$\cdots$};
    \node[gauge, label=below:$2n-1$] (2rsub) at (3,-1){};
    \draw[-] (1sub)--(2sub)--(cdotssub)--(2rsub);
    
    \draw (-0.15,-1.8) -- (5.15,-1.8);
    
    \node[font=\Large] at (-0.8,-2.5) {$=$};
    \node[gauge, label=below:$2n$] (2rp1r) at (1,-2.5){};
    \node[gauge] (Qr) at (2,-2.5) {$Q$};
    \draw[teal, very thick] (2rp1r)--(Qr);
    \end{tikzpicture}}
    \end{equation}
    \item Symplectise the $\urm(2n)\rightarrow\sprm(n)$ and add a half-hypermultiplet. This removes a further $n+1$ moduli.

    \begin{equation}\raisebox{-0.5\height}{\begin{tikzpicture}
    \node[font=\bfseries] at (5.25,1) {Orthogonalisation};
    \node[gauge, label=below:$2n$] (2rp1r) at (1,0){};
    \node[gauge] (Qr) at (2,0) {$Q$};
    \draw[teal, very thick] (2rp1r)--(Qr);

    \draw[->, thick]
    (2.5,0) -- (9.0,0)
    node[midway, fill=white, inner sep=3pt, align=center]
    {\small\textit{Replace}
     $\urm(2n)\;\text{\emph{with}}\;\sprm(n)$\\\small\textit{and add}\;$\orm(1)$\;\textit{flavour}};

    \node[flavourr, label=above:$\orm(1)$] (1f) at (9.5,1){};
    \node[gaugeb, label=below:$C_n$] (Cnr) at (9.5,0){};
    
    \node[gauge] (Qrr) at (10.5,0){$Q$};
    \draw[-] (1f)--(Cnr);
    \draw[teal, very thick] (Cnr)--(Qrr);
\end{tikzpicture}}
\end{equation}
\end{enumerate}
In terms of the change in dimensions on the magnetic side, the combined action of the subtraction of \Quiver{quiv:ONMinusTilde_quotient_quiver} and the symplectisation of a $\urm(2n)\rightarrow\sprm(n)$ corresponds to a reduction in the Coulomb branch dimension by $2n^2$. This matches the reduction in dimension expected from gauging $\urm(n)\times\urm(n)$ and not that from gauging a $\urm(n)$ with bifundamental matter, which is smaller by a factor of two.

The operation of gauging a short $\urm(n)$ factor, as described in \eqref{eq:shortUn}, can be thought of as first gauging a $\urm(n)\times\urm(n)$ and then folding. On the magnetic quiver, the $\urm(n)\times\urm(n)$ quotient quiver subtraction is the same as the short $\urm(n)$ quotient quiver subtraction procedure except that the $\sprm(n)$ has $D_1$ flavour in the former and $\orm(1)$ flavour in the latter. Perhaps a realisation of folding an electric quiver in the magnetic quiver is in this replacement of $D_1$ with $\orm(1)$.


\bibliographystyle{JHEP}
\bibliography{references.bib}

@article{Henderson_2014,
   title={Diagram automorphisms of quiver varieties},
   volume={267},
   ISSN={0001-8708},
   url={http://dx.doi.org/10.1016/j.aim.2014.08.007},
   DOI={10.1016/j.aim.2014.08.007},
   journal={Advances in Mathematics},
   publisher={Elsevier BV},
   author={Henderson, Anthony and Licata, Anthony},
   year={2014},
   month=Dec, pages={225–276} }

@article{Cabrera:2019izd,
    author = "Cabrera, Santiago and Hanany, Amihay and Sperling, Marcus",
    title = "{Magnetic quivers, Higgs branches, and 6d $\mathcal{N}=(1,0)$ theories}",
    eprint = "1904.12293",
    archivePrefix = "arXiv",
    primaryClass = "hep-th",
    reportNumber = "Imperial/TP/19/AH/01",
    doi = "10.1007/JHEP06(2019)071",
    journal = "JHEP",
    volume = "06",
    pages = "071",
    year = "2019",
    note = "[Erratum: JHEP 07, 137 (2019)]"
}

@article{Dasgupta:1996ij,
    author = "Dasgupta, Keshav and Mukhi, Sunil",
    title = "{F theory at constant coupling}",
    eprint = "hep-th/9606044",
    archivePrefix = "arXiv",
    reportNumber = "TIFR-TH-96-30",
    doi = "10.1016/0370-2693(96)00875-1",
    journal = "Phys. Lett. B",
    volume = "385",
    pages = "125--131",
    year = "1996"
}

@article{Cabrera:2017njm,
    author = "Cabrera, Santiago and Hanany, Amihay",
    title = "{Branes and the Kraft-Procesi transition: classical case}",
    eprint = "1711.02378",
    archivePrefix = "arXiv",
    primaryClass = "hep-th",
    doi = "10.1007/JHEP04(2018)127",
    journal = "JHEP",
    volume = "04",
    pages = "127",
    year = "2018"
}

@article{Hanany:2024fqf,
    author = "Hanany, Amihay and Kalveks, Rudolph and Kumaran, Guhesh",
    title = "{Quiver Polymerisation}",
    eprint = "2406.11561",
    archivePrefix = "arXiv",
    primaryClass = "hep-th",
    month = "6",
    year = "2024"
}

@article{Hanany:2022itc,
    author = "Hanany, Amihay and Sperling, Marcus",
    title = "{Magnetic quivers and negatively charged branes}",
    eprint = "2208.07270",
    archivePrefix = "arXiv",
    primaryClass = "hep-th",
    reportNumber = "Imperial/TP/22/AH/02",
    doi = "10.1007/JHEP11(2022)010",
    journal = "JHEP",
    volume = "11",
    pages = "010",
    year = "2022"
}

@article{Bourget:2021qpx,
    author = "Bourget, Antoine and Dancer, Andrew and Grimminger, Julius F. and Hanany, Amihay and Zhong, Zhenghao",
    title = "{Partial implosions and quivers}",
    eprint = "2112.10825",
    archivePrefix = "arXiv",
    primaryClass = "hep-th",
    doi = "10.1007/JHEP07(2022)049",
    journal = "JHEP",
    volume = "07",
    pages = "049",
    year = "2022"
}

@article{Cabrera:2018ann,
    author = "Cabrera, Santiago and Hanany, Amihay",
    title = "{Quiver Subtractions}",
    eprint = "1803.11205",
    archivePrefix = "arXiv",
    primaryClass = "hep-th",
    doi = "10.1007/JHEP09(2018)008",
    journal = "JHEP",
    volume = "09",
    pages = "008",
    year = "2018"
}

@article{Kobak1996CLASSICALNO,
  title={CLASSICAL NILPOTENT ORBITS AS HYPERK{\"A}HLER QUOTIENTS},
  author={Piotr Z. Kobak and Andrew Swann},
  journal={International Journal of Mathematics},
  year={1996},
  volume={07},
  pages={193-210},
  url={https://api.semanticscholar.org/CorpusID:16460643}
}

@article{Hanany:1996ie,
    author = "Hanany, Amihay and Witten, Edward",
    title = "{Type IIB superstrings, BPS monopoles, and three-dimensional gauge dynamics}",
    eprint = "hep-th/9611230",
    archivePrefix = "arXiv",
    reportNumber = "IASSNS-HEP-96-121",
    doi = "10.1016/S0550-3213(97)00157-0",
    journal = "Nucl. Phys. B",
    volume = "492",
    pages = "152--190",
    year = "1997"
}

@article{Gaiotto:2008ak,
    author = "Gaiotto, Davide and Witten, Edward",
    title = "{S-Duality of Boundary Conditions In N=4 Super Yang-Mills Theory}",
    eprint = "0807.3720",
    archivePrefix = "arXiv",
    primaryClass = "hep-th",
    doi = "10.4310/ATMP.2009.v13.n3.a5",
    journal = "Adv. Theor. Math. Phys.",
    volume = "13",
    number = "3",
    pages = "721--896",
    year = "2009"
}

@article{Cabrera:2018ldc,
    author = "Cabrera, Santiago and Hanany, Amihay and Kalveks, Rudolph",
    title = "{Quiver Theories and Formulae for Slodowy Slices of Classical Algebras}",
    eprint = "1807.02521",
    archivePrefix = "arXiv",
    primaryClass = "hep-th",
    doi = "10.1016/j.nuclphysb.2018.12.022",
    journal = "Nucl. Phys. B",
    volume = "939",
    pages = "308--357",
    year = "2019"
}

@article{Bourget:2022tmw,
    author = "Bourget, Antoine and Grimminger, Julius F.",
    title = "{Fibrations and Hasse diagrams for 6d SCFTs}",
    eprint = "2209.15016",
    archivePrefix = "arXiv",
    primaryClass = "hep-th",
    doi = "10.1007/JHEP12(2022)159",
    journal = "JHEP",
    volume = "12",
    pages = "159",
    year = "2022"
}

@article{Bourget:2022ehw,
    author = "Bourget, Antoine and Grimminger, Julius F. and Hanany, Amihay and Zhong, Zhenghao",
    title = "{The Hasse diagram of the moduli space of instantons}",
    eprint = "2202.01218",
    archivePrefix = "arXiv",
    primaryClass = "hep-th",
    doi = "10.1007/JHEP08(2022)283",
    journal = "JHEP",
    volume = "08",
    pages = "283",
    year = "2022"
}

@article{Bourget:2020xdz,
    author = "Bourget, Antoine and Grimminger, Julius F. and Hanany, Amihay and Kalveks, Rudolph and Sperling, Marcus and Zhong, Zhenghao",
    title = "{Magnetic Lattices for Orthosymplectic Quivers}",
    eprint = "2007.04667",
    archivePrefix = "arXiv",
    primaryClass = "hep-th",
    reportNumber = "Imperial/TP/20/AH/07",
    doi = "10.1007/JHEP12(2020)092",
    journal = "JHEP",
    volume = "12",
    pages = "092",
    year = "2020"
}

@article{Chacaltana:2010ks,
    author = "Chacaltana, Oscar and Distler, Jacques",
    title = "{Tinkertoys for Gaiotto Duality}",
    eprint = "1008.5203",
    archivePrefix = "arXiv",
    primaryClass = "hep-th",
    reportNumber = "UTTG-11-10, TCC-020-10",
    doi = "10.1007/JHEP11(2010)099",
    journal = "JHEP",
    volume = "11",
    pages = "099",
    year = "2010"
}

@article{Chacaltana:2012ch,
    author = "Chacaltana, Oscar and Distler, Jacques and Tachikawa, Yuji",
    title = "{Gaiotto duality for the twisted $A_{2n-1}$ series}",
    eprint = "1212.3952",
    archivePrefix = "arXiv",
    primaryClass = "hep-th",
    reportNumber = "ICTP-SAIFR-2012-010, IPMU-12-0226, UT-12-43, UTTG-24-12, TCC-023-12",
    doi = "10.1007/JHEP05(2015)075",
    journal = "JHEP",
    volume = "05",
    pages = "075",
    year = "2015"
}

@article{Bourget:2019aer,
    author = "Bourget, Antoine and Cabrera, Santiago and Grimminger, Julius F. and Hanany, Amihay and Sperling, Marcus and Zajac, Anton and Zhong, Zhenghao",
    title = "{The Higgs mechanism \textemdash{} Hasse diagrams for symplectic singularities}",
    eprint = "1908.04245",
    archivePrefix = "arXiv",
    primaryClass = "hep-th",
    reportNumber = "Imperial/TP/19/AH/02",
    doi = "10.1007/JHEP01(2020)157",
    journal = "JHEP",
    volume = "01",
    pages = "157",
    year = "2020"
}

@article{Bourget:2024mgn,
    author = "Bourget, Antoine and Sperling, Marcus and Zhong, Zhenghao",
    title = "{Higgs branch RG-flows via Decay and Fission}",
    eprint = "2401.08757",
    archivePrefix = "arXiv",
    primaryClass = "hep-th",
    reportNumber = "UWThPh2024-4",
    month = "1",
    year = "2024"
}

@article{Mansi:2023faa,
    author = "Mansi, Lorenzo and Sperling, Marcus",
    title = "{Unravelling T-Duality: Magnetic Quivers in Rank-zero Little String Theories}",
    eprint = "2312.12510",
    archivePrefix = "arXiv",
    primaryClass = "hep-th",
    reportNumber = "DESY-23-215",
    month = "12",
    year = "2023"
}

@article{Bourget:2021siw,
    author = "Bourget, Antoine and Grimminger, Julius F. and Hanany, Amihay and Sperling, Marcus and Zhong, Zhenghao",
    title = "{Branes, Quivers, and the Affine Grassmannian}",
    eprint = "2102.06190",
    archivePrefix = "arXiv",
    primaryClass = "hep-th",
    doi = "10.2969/aspm/08810331",
    journal = "Adv. Stud. Pure Math.",
    volume = "88",
    pages = "331--435",
    year = "2023"
}

@article{Lawrie:2024wan,
    author = "Lawrie, Craig and Mansi, Lorenzo and Sperling, Marcus and Zhong, Zhenghao",
    title = "{A Pathway to Decay and Fission of Orthosymplectic Quiver Theories}",
    eprint = "2412.15202",
    archivePrefix = "arXiv",
    primaryClass = "hep-th",
    reportNumber = "DESY-24-185, UWThPh 2024-26",
    month = "12",
    year = "2024"
}

@article{Bourget:2023dkj,
    author = "Bourget, Antoine and Sperling, Marcus and Zhong, Zhenghao",
    title = "{Decay and Fission of Magnetic Quivers}",
    eprint = "2312.05304",
    archivePrefix = "arXiv",
    primaryClass = "hep-th",
    month = "12",
    year = "2023"
}

@article{Lawrie:2023uiu,
    author = "Lawrie, Craig and Mansi, Lorenzo",
    title = "{The Higgs Branch of Heterotic LSTs: Hasse Diagrams and Generalized Symmetries}",
    eprint = "2312.05306",
    archivePrefix = "arXiv",
    primaryClass = "hep-th",
    reportNumber = "DESY-23-199",
    month = "12",
    year = "2023"
}

@article{Bourget:2023uhe,
    author = "Bourget, Antoine and Giacomelli, Simone and Grimminger, Julius F.",
    title = "{FI-flows of 3d $N=4$ Theories}",
    eprint = "2302.03698",
    archivePrefix = "arXiv",
    primaryClass = "hep-th",
    doi = "10.1007/JHEP04(2023)015",
    journal = "JHEP",
    volume = "04",
    pages = "015",
    year = "2023"
}

@article{Bourget:2021csg,
    author = "Bourget, Antoine and Grimminger, Julius F. and Martone, Mario and Zafrir, Gabi",
    title = "{Magnetic quivers for rank 2 theories}",
    eprint = "2110.11365",
    archivePrefix = "arXiv",
    primaryClass = "hep-th",
    doi = "10.1007/JHEP03(2022)208",
    journal = "JHEP",
    volume = "03",
    pages = "208",
    year = "2022"
}

@article{Lawrie:2024zon,
    author = "Lawrie, Craig and Mansi, Lorenzo",
    title = "{Higgs branch of 6D (1, 0) SCFTs and little string theories with Dynkin DE-type SUSY enhancement}",
    eprint = "2406.02670",
    archivePrefix = "arXiv",
    primaryClass = "hep-th",
    reportNumber = "DESY-24-079",
    doi = "10.1103/PhysRevD.110.066014",
    journal = "Phys. Rev. D",
    volume = "110",
    number = "6",
    pages = "066014",
    year = "2024"
}

@article{Bourget:2020asf,
    author = "Bourget, Antoine and Grimminger, Julius F. and Hanany, Amihay and Sperling, Marcus and Zafrir, Gabi and Zhong, Zhenghao",
    title = "{Magnetic quivers for rank 1 theories}",
    eprint = "2006.16994",
    archivePrefix = "arXiv",
    primaryClass = "hep-th",
    doi = "10.1007/JHEP09(2020)189",
    journal = "JHEP",
    volume = "09",
    pages = "189",
    year = "2020"
}

@article{Bourget:2020mez,
    author = "Bourget, Antoine and Giacomelli, Simone and Grimminger, Julius F. and Hanany, Amihay and Sperling, Marcus and Zhong, Zhenghao",
    title = "{S-fold magnetic quivers}",
    eprint = "2010.05889",
    archivePrefix = "arXiv",
    primaryClass = "hep-th",
    doi = "10.1007/JHEP02(2021)054",
    journal = "JHEP",
    volume = "02",
    pages = "054",
    year = "2021"
}

@article{Giacomelli:2024dbd,
    author = "Giacomelli, Simone and Savelli, Raffaele and Zoccarato, Gianluca",
    title = "{$ \mathcal{N} $ = 2 Orbi-S-Folds}",
    eprint = "2405.00101",
    archivePrefix = "arXiv",
    primaryClass = "hep-th",
    doi = "10.1007/JHEP01(2025)059",
    journal = "JHEP",
    volume = "01",
    pages = "059",
    year = "2025"
}

@article{Giacomelli:2024ycb,
    author = "Giacomelli, Simone and Harding, William and Mekareeya, Noppadol and Mininno, Alessandro",
    title = "{All class $ \mathcal{S} $ theories of type-A originate from orbi-instantons}",
    eprint = "2411.03425",
    archivePrefix = "arXiv",
    primaryClass = "hep-th",
    reportNumber = "ZMP-HH/24-25",
    doi = "10.1007/JHEP02(2025)185",
    journal = "JHEP",
    volume = "02",
    pages = "185",
    year = "2025"
}

@article{vanBeest:2021xyt,
    author = "van Beest, Marieke and Giacomelli, Simone",
    title = "{Connecting 5d Higgs branches via Fayet-Iliopoulos deformations}",
    eprint = "2110.02872",
    archivePrefix = "arXiv",
    primaryClass = "hep-th",
    doi = "10.1007/JHEP12(2021)202",
    journal = "JHEP",
    volume = "12",
    pages = "202",
    year = "2021"
}

@article{Cabrera:2019dob,
    author = "Cabrera, Santiago and Hanany, Amihay and Sperling, Marcus",
    title = "{Magnetic quivers, Higgs branches, and 6d $ \mathcal{N} $ = (1, 0) theories \textemdash{} orthogonal and symplectic gauge groups}",
    eprint = "1912.02773",
    archivePrefix = "arXiv",
    primaryClass = "hep-th",
    reportNumber = "Imperial/TP/19/AH/05",
    doi = "10.1007/JHEP02(2020)184",
    journal = "JHEP",
    volume = "02",
    pages = "184",
    year = "2020"
}

@article{Hanany:2023tvn,
    author = "Hanany, Amihay and Kalveks, Rudolph and Kumaran, Guhesh",
    title = "{Quotient quiver subtraction}",
    eprint = "2308.05853",
    archivePrefix = "arXiv",
    primaryClass = "hep-th",
    doi = "10.1016/j.nuclphysb.2024.116731",
    journal = "Nucl. Phys. B",
    volume = "1009",
    pages = "116731",
    year = "2024"
}

@article{Intriligator:1996ex,
    author = "Intriligator, Kenneth A. and Seiberg, N.",
    title = "{Mirror symmetry in three-dimensional gauge theories}",
    eprint = "hep-th/9607207",
    archivePrefix = "arXiv",
    reportNumber = "RU-96-63, IASSNS-HEP-96-80",
    doi = "10.1016/0370-2693(96)01088-X",
    journal = "Phys. Lett. B",
    volume = "387",
    pages = "513--519",
    year = "1996"
}

@article{Argyres:1996eh,
    author = "Argyres, Philip C. and Plesser, M. Ronen and Seiberg, Nathan",
    title = "{The Moduli space of vacua of N=2 SUSY QCD and duality in N=1 SUSY QCD}",
    eprint = "hep-th/9603042",
    archivePrefix = "arXiv",
    reportNumber = "RU-96-07, WIS-96-1-PH",
    doi = "10.1016/0550-3213(96)00210-6",
    journal = "Nucl. Phys. B",
    volume = "471",
    pages = "159--194",
    year = "1996"
}

@article{Dancer:2024lra,
    author = "Dancer, Andrew and Grimminger, Julius F. and Martens, Johan and Zhong, Zhenghao",
    title = "{Complex Symplectic Contractions and 3d Mirrors}",
    eprint = "2406.09626",
    archivePrefix = "arXiv",
    primaryClass = "hep-th",
    month = "6",
    year = "2024"
}

@article{Aharony:1997bx,
    author = "Aharony, Ofer and Hanany, Amihay and Intriligator, Kenneth A. and Seiberg, N. and Strassler, M. J.",
    title = "{Aspects of N=2 supersymmetric gauge theories in three-dimensions}",
    eprint = "hep-th/9703110",
    archivePrefix = "arXiv",
    reportNumber = "RU-97-10A, IASSNS-HEP-97-18",
    doi = "10.1016/S0550-3213(97)00323-4",
    journal = "Nucl. Phys. B",
    volume = "499",
    pages = "67--99",
    year = "1997"
}

@inproceedings{Seiberg:1996nz,
    author = "Seiberg, Nathan and Witten, Edward",
    title = "{Gauge dynamics and compactification to three-dimensions}",
    booktitle = "{Conference on the Mathematical Beauty of Physics (In Memory of C. Itzykson)}",
    eprint = "hep-th/9607163",
    archivePrefix = "arXiv",
    reportNumber = "IASSNS-HEP-96-78, RU-96-55",
    pages = "333--366",
    month = "6",
    year = "1996"
}

@article{Bennett:2024llh,
    author = "Bennett, Sam and Hanany, Amihay and Kumaran, Guhesh",
    title = "{Orthosymplectic quotient quiver subtraction}",
    eprint = "2409.15419",
    archivePrefix = "arXiv",
    primaryClass = "hep-th",
    reportNumber = "Imperial/TP/24/AH/02, Imperial/TP/24/AH/02",
    doi = "10.1007/JHEP12(2024)063",
    journal = "JHEP",
    volume = "12",
    pages = "063",
    year = "2024"
}

@article{Bennett:2025edk,
    author = "Bennett, Sam and Hanany, Amihay and Kumaran, Guhesh",
    title = "{Orthosymplectic Quotient Quiver Subtraction II: Framed Quivers}",
    eprint = "2503.19954",
    archivePrefix = "arXiv",
    primaryClass = "hep-th",
    reportNumber = "Imperial/TP/25/AH/03",
    month = "3",
    year = "2025"
}

@article{Bennett:2024loi,
    author = "Bennett, Sam and Hanany, Amihay and Kumaran, Guhesh and Li, Chunhao and Liu, Deshuo and Sperling, Marcus",
    title = "{Quiver subtraction on the Higgs branch}",
    eprint = "2409.16356",
    archivePrefix = "arXiv",
    primaryClass = "hep-th",
    reportNumber = "Imperial/TP/24/AH/03, Imperial/TP/24/AH/03",
    doi = "10.1016/j.nuclphysb.2025.116917",
    journal = "Nucl. Phys. B",
    volume = "1016",
    pages = "116917",
    year = "2025"
}

@article{Argyres:2007tq,
    author = "Argyres, Philip C. and Wittig, John R.",
    title = "{Infinite coupling duals of N=2 gauge theories and new rank 1 superconformal field theories}",
    eprint = "0712.2028",
    archivePrefix = "arXiv",
    primaryClass = "hep-th",
    doi = "10.1088/1126-6708/2008/01/074",
    journal = "JHEP",
    volume = "01",
    pages = "074",
    year = "2008"
}

@article{Bennett:2026gtm,
    author = "Bennett, Sam and Hanany, Amihay and Kumaran, Guhesh",
    title = "{Quotient Quiver Subtraction -- Classical Groups}",
    eprint = "2603.08774",
    archivePrefix = "arXiv",
    primaryClass = "hep-th",
    month = "3",
    year = "2026"
}

@article{Bennett:2025zor,
    author = "Bennett, Sam and Hanany, Amihay",
    title = "{$\mathrm{O}$/$\mathrm{SO}$ Gauge Groups, $BC$ Quivers and $O3$ Planes}",
    eprint = "2503.05443",
    archivePrefix = "arXiv",
    primaryClass = "hep-th",
    reportNumber = "Imperial/TP/22/AH/01",
    month = "3",
    year = "2025"
}

@article{Harding:2025vov,
    author = "Harding, William and Mekareeya, Noppadol and Zhong, Zhenghao",
    title = "{Orthosymplectic quivers: indices, Hilbert series, and generalised symmetries}",
    eprint = "2505.03875",
    archivePrefix = "arXiv",
    primaryClass = "hep-th",
    doi = "10.1007/JHEP09(2025)212",
    journal = "JHEP",
    volume = "09",
    pages = "212",
    year = "2025"
}

@article{Cabrera:2017ucb,
    author = "Cabrera, Santiago and Hanany, Amihay and Zhong, Zhenghao",
    title = "{Nilpotent orbits and the Coulomb branch of $T^\sigma (G)$ theories: special orthogonal vs orthogonal gauge group factors}",
    eprint = "1707.06941",
    archivePrefix = "arXiv",
    primaryClass = "hep-th",
    doi = "10.1007/JHEP11(2017)079",
    journal = "JHEP",
    volume = "11",
    pages = "079",
    year = "2017"
}

@article{Sela_2019,
   title={Symmetry enhancement in 4d Spin(n) gauge theories and compactification from 6d},
   volume={2019},
   ISSN={1029-8479},
   url={http://dx.doi.org/10.1007/JHEP12(2019)052},
   DOI={10.1007/jhep12(2019)052},
   number={12},
   journal={Journal of High Energy Physics},
   publisher={Springer Science and Business Media LLC},
   author={Sela, Orr and Zafrir, Gabi},
   year={2019},
   month=Dec }

@article{Minahan_1997,
   title={Superconformal fixed points with En global symmetry},
   volume={489},
   ISSN={0550-3213},
   url={http://dx.doi.org/10.1016/S0550-3213(97)00039-4},
   DOI={10.1016/s0550-3213(97)00039-4},
   number={1-2},
   journal={Nuclear Physics B},
   publisher={Elsevier BV},
   author={Minahan, Joseph A. and Nemeschansky, Dennis},
   year={1997},
   month=Mar, pages={24–46} }

@article {braverman2018coulombbranches3dmathcal,
    AUTHOR = {Braverman, Alexander and Finkelberg, Michael and Nakajima,
              Hiraku},
     TITLE = {Coulomb branches of {$3d$} {$\mathcal{N}=4$} quiver gauge theories
              and slices in the affine Grassmannian},
      NOTE = {With two appendices by Braverman, Finkelberg, Joel Kamnitzer,
              Ryosuke Kodera, Nakajima, Ben Webster and Alex Weekes},
   JOURNAL = {Adv. Theor. Math. Phys.},
  FJOURNAL = {Advances in Theoretical and Mathematical Physics},
    VOLUME = {23},
      YEAR = {2019},
    NUMBER = {1},
     PAGES = {75--166},
      ISSN = {1095-0761,1095-0753},
   MRCLASS = {57R57 (14M15 16G20 17B81 81T13)},
  MRNUMBER = {4020310},
       DOI = {10.4310/ATMP.2019.v23.n1.a3},
       URL = {https://doi.org/10.4310/ATMP.2019.v23.n1.a3},
}

@article{Baume_2022,
   title={Two 6D origins of 4D SCFTs: Class $\mathcal{S}$ and 6D (1,0) on a torus},
   volume={106},
   ISSN={2470-0029},
   url={http://dx.doi.org/10.1103/PhysRevD.106.086003},
   DOI={10.1103/physrevd.106.086003},
   number={8},
   journal={Physical Review D},
   publisher={American Physical Society (APS)},
   author={Baume, Florent and Kang, Monica Jinwoo and Lawrie, Craig},
   year={2022},
   month=Oct }
\end{document}